\documentclass[12pt]{article}

\usepackage[a4paper, total={6in, 8in}]{geometry}
\usepackage{authblk}
\usepackage[utf8]{inputenc} 
\usepackage[T1]{fontenc}    
\usepackage{hyperref}       
\usepackage{url}            
\usepackage{booktabs}       
\usepackage{amsfonts}       
\usepackage{nicefrac}       
\usepackage{bbm}
\usepackage{subcaption}

\usepackage[table]{xcolor}
\usepackage{amsmath,amssymb,amsthm}

\usepackage{graphicx}
\usepackage{natbib}
\usepackage{color}

\newcommand{\LL}[1]{\mathcal{L}\left(#1\right)}
\newcommand{\LLh}[1]{\hat{\mathcal{L}}\left(#1\right)}

\newcommand{\eqdef}{\mathrel{\mathop=}:}

\DeclareMathOperator*{\argmax}{argmax}

\numberwithin{equation}{section}

\title{Co-existence of Trend and Value\\ in Financial Markets:\\ Estimating an Extended Chiarella Model
}
 \author{Adam A. Majewski, Stefano Ciliberti and Jean-Philippe Bouchaud\thanks{All authors would like to thank Andr{\'e} Breedt, Yves Lemp{\'e}ri{\`e}re, Marco Leoni, Iacopo Mastromatteo, Adam Rej, Philip Seager and Emmanuel S{\'e}ri{\'e} for insightful comments and fruitful discussions. Moreover, we are grateful to participants of Imperial ETH Workshop on Mathematical Finance 2018 in Zurich and seminar in Capital Fund Management for helpful remarks.}}

\affil{ Capital Fund Management\\ 
23 rue de l’Universit\'{e}\\ 
Paris 75007, France\\ 
}

\begin{document}
\maketitle
\begin{abstract} 
Trend and Value are pervasive anomalies, common to all financial markets. We address the problem of their co-existence and interaction within the framework of Heterogeneous Agent Based Models (HABM). More specifically, we extend the \cite{Chiarella1992} model by adding noise traders and a non-linear demand of fundamentalists. We use Bayesian filtering techniques to calibrate the model on time series of prices across a variety of asset classes since 1800. The fundamental value is an output of the calibration, and does not require the use of an external pricing model. Our extended model reproduces many empirical observations, including the non-monotonic relation between past trends and future returns. The destabilizing activity of trend-followers leads to a qualitative change of mispricing distribution, from unimodal to bimodal, meaning that some markets tend to be over- (or under-) valued for long periods of time. \end{abstract}

\newpage
\begin{flushright} 
\textit{''Nowadays people know the price of everything and the value of nothing.''} 
\end{flushright} 
\vspace{-5mm} 
\begin{flushright} 
Oscar Wilde, The Picture of Dorian Gray 
\end{flushright} 
\vspace{5mm}

\section{Introduction}

Among the most challenging anomalies for the efficient market hypothesis are the so-called `Trend' and `Value' effects, that pervade all financial markets (see e.g. \cite{AMP2013}). Trend means that positive (negative) returns over a short to medium period are likely to be followed by positive (negative) short-term return. Value means that assets with prices lower (higher) than their `fundamental value' tend to have positive (negative) future returns. While the former anomaly essentially implies positive autocorrelations of short to medium-term returns (weeks to months), the latter implies a negative correlation of long term returns (corresponding to price mean-reversion on a multi-year time scale). Both anomalies have been extensively validated statistically in the empirical literature for various types of assets.\footnote{The first ones to advocate for value investing were \cite{Graham1934}. Later on, the value anomaly was investigated by many researchers, among which \cite{BWT1985}, \cite{RRL1985},
\cite{Summers1986}, \cite{CS1988}, \cite{FF1992}, \cite{LSV1994}, \cite{BCLMSS2017}. The trend effect was documented by \cite{Jegadeesh1990}, \cite{Lehmann1990}, \cite{JT1993}, \cite{Carhart1997}, \cite{MOP2012}, \cite{LDSPB2014} and others. Value and trend effects together were statistically validated by \cite{FF2012} and \cite{AMP2013}.} The present paper studies the qualitative and quantitative nature of these two effects and their interaction, within the framework of an Agent-Based inspired model.

Having acknowledged the existence of a trend effect, it is interesting to investigate the structure of the relation between past and future returns. In two recent papers, \cite{LDSPB2014} and \cite{BCLMSS2017} present empirical evidence that this relation is non-linear and surprisingly non-monotonic. The trend effect saturates for larger trend signals, and even appears to {\it invert} for very large trend signals. The first goal of our study is to introduce a mechanism that leads to such an effect. The intuitive explanation of this phenomenon is that when trend signals are very strong it is very likely that the price is far away from the fundamental value. Fundamentalists (i.e. investors believing in value) then become more active, causing price mean-reversion, overriding the influence of chartists (or trend-followers). It is therefore natural to approach this problem within the framework of heterogeneous agent-based market models.

The cornerstone of classical financial theory is the existence of a representative investor with rational expectations. This paradigm has been questioned and rejected by many scholars who have provided evidence that investors are in fact heterogeneous and at best boundedly rational (see e.g. \cite{Shiller1987}, \cite{Thaler1993}, \cite{DHS1998}, \cite{Thaler2005}, \cite{Kahneman2011}, \cite{LMT2017} and many others). In order to meet these empirical findings, the HABM literature assumes the existence of several types of investors with simple investment heuristics, that interact with each other via market prices. For reviews on HABMs see \cite{Hommes2006}, \cite{Lebaron2006}, \cite{CDH2009} and \cite{DH2018}. 

In this paper we assume that market clearing is done by a price impact mechanism. In that case, the accumulated demand functions of investors are crucial for determining the price dynamics. Two elements seem necessary to obtain the non-monotonic shape of the trend effect: $i$) a bounded demand of trend-followers that are active for small and medium trend signals, before saturating for large signals; and $ii$) a demand of fundamentalists that grows with increasing mispricing. The model of \citep{Chiarella1992} encapsulates these features, and is therefore able to reproduce the desired shape of the trend effect. We modify the original model of \citep{Chiarella1992} by allowing the fundamental value to have a long term drift, and by adding a third group of agents, which represent noise traders. This leads to a more realistic price dynamics.

The second goal of our paper is to model and measure the value effect. A detailed analysis reveals the non-linear nature of this effect. It seems natural that profits from value investing grow proportionally to the difference between price and fundamental value. However, a model with such a linear demand of fundamentalists fails to reproduce some aspect of the data. On the other hand a non-linear demand of fundamentalists implies a non-linear shape of the value effect, similar to the one observed empirically.

Finally, one would like to know how often, and by how much, markets are over/under-valued. In his famous paper, \cite{Black1986} suggested that prices are typically `a factor 2' away from value, a conclusion bolstered by the analysis of \cite{BCLMSS2017}, who report 50 \% typical mispricings and a time scale of several years for markets to self-correct. Recently, \cite{SW2017} presented evidence that the distribution of price distortion of S\&P 500 is actually bimodal, with a local minimum of probability distribution around zero. That is a very surprising discovery suggesting that market is more often overvalued or undervalued than close to fundamental value. We confirm that our model-implied price distortion has a bimodal distribution for certain assets. It is worth stressing that the proposed agent-based model indeed exhibits a phenomenological bifurcation. The qualitative change of mispricing distribution, from unimodal to bimodal, emerges when the destabilizing activity of trend-followers exceeds the trading activity of fundamentalists (see a related discussion in \cite{BBDG2018}, Chapter 20).


We estimate our HABM on a rich and diverse dataset (stock indices, commodities, FX rates and government bonds) going back to 1800. The main challenge in estimating HABM is that fundamental value is not an observable quantity. HABM models are often calibrated by first estimating fundamental value using an additional economic model (for example the model of \cite{Gordon1962} for stock markets) and then estimating the parameters of the HABM model by standard econometric methods like OLS or maximum likelihood (see \cite{BH2007}, \cite{CHZ2014} and others). Another approach to HABM estimation is via the method of simulated moments, which searches for the parameter setting such that simulated moments and other statistics match the ones observed in the data \citep{FW2011, FW2016, Barde2016,GL2016}. Recently, independently from our own research, a new approach has been proposed by \cite{Lux2017} and \cite{BMS2018}.\footnote{Another interesting approach to HABM estimation is via modern machine learning techniques, see \cite{LRS2017}.} Both papers treat the unobserved quantities as hidden variables that are estimated using particle filtering. The parameters of the system are determined by a numerical black-box procedure maximizing the likelihood given the filtered-out hidden variables.

Our estimation approach falls into the last category. We treat the fundamental value of the asset as the hidden variable and we filter it out using Bayesian filtering techniques. Contrary to \cite{Lux2017} and \cite{BMS2018}, who use computationally expensive particle filtering, we apply a classical Kalman filter to the model with a linear value demand function and an Unscented Kalman Filter to the non-linear version of the model. Moreover, for the linear demand function we apply the Expectation-Maximization (EM) algorithm that provides closed-form formulas for iterative procedure maximizing likelihood of the model. The advantage of Bayesian filtering methods over a method of simulated moments is both higher precision of parameter estimates and the possibility of identification of the hidden variables trajectories that are of crucial conceptual importance in most agent-based models. We also find the Bayesian filtering approach more consistent and more universal than estimating first a fundamental value using an additional economic model and then estimating the HABM: our approach does not need any additional assumption on the fundamental value process and can thus be applied to all asset classes.

The rest of the paper is organized as follows. In Section \ref{sec:literature} we review the literature related to trend and value anomalies, and HABMs. Section~\ref{sec:HABM} introduces our version of the Chiarella HABM, with a linear demand of fundamentalists. In the same section we provide some basic properties of the model (Subsection \ref{sec:oscillations}) and we describe the estimation methodology (Subsection~\ref{sec:Estimation}). Subsection~\ref{sec:TrendValueEffects} describes the model-implied trend and value effects and compares them with the ones observed in the data. Section \ref{sec:HABMNL} extends the model by considering a non-linear demand function of fundamentalists. In Section~\ref{sec:StationaryDistribution} we analyze the statistical properties of the stationary distribution of mispricing. Section~\ref{sec:Conclusions} concludes.

\section{Related Literature}\label{sec:literature}
\subsection{Interpretations of the Trend and Value Anomalies}

The economic rationale for the trend and value anomalies is one of the most debated problems in modern finance theory. The classical branch of finance literature based on the efficient market hypothesis has serious difficulties in explaining the existence and nature of these two anomalies, especially trend which is often denied or swept under the rug.\footnote{The momentum factor is so at odds with Fama's view of the world that it is not included in the most recent 5 factor version of the Fama-French model! See \cite{FF2016}.} Even if in principle one cannot reject the existence of a risk-based asset pricing theory that would accommodate the trend effect, there is no clear identification of the risk that could be responsible for excess returns of the trend-following strategy. Indeed, macroeconomic variables such as the business cycle, or consumption, have been rejected as potential risk variables to explain it \citep{GJS2003, AMP2013}.\footnote{Of note, there is also some evidence of macroeconomic risks, like industrial production, explaining a trend effect \citep{LZ2008}. Assuming that those results are not spurious due to extensive data mining, the proposed rational asset pricing models can only explain the trend effect for single-name equities. Since one observes the presence of the anomaly over all asset classes, we feel that a more general approach is needed.} \cite{MOP2012} and \cite{DNDLBP2018} show that the trend-following strategy performs best when the price experiences large up and down moves, which rules out the possibility that the trend effect is a compensation for crash risk or tail events. The most plausible evidence of a link between risk and trend-following returns is illiquidity risk \citep{PS2003, Sadka2006, MOP2012}. However, even in that case, the transaction cost factor can only explain a small fraction of abnormal returns of trend-followers. Finally, \cite{LDNSPB2017} show that the trend-following strategy has a {\it positive skewness}, which is at odds with the interpretation of trend excess returns as a risk premium of sorts (see also \cite{DNDLBP2018}).



The problems encountered by classical theories to provide a satisfactory explanation for the trend and value anomalies have stimulated alternative approaches to modelling financial markets in the past three decades. One attempt is to relax the assumption of a representative investor by introducing heterogeneity of market participants. For example, \cite{AC2017} propose a theory based on information percolation in an heterogeneous market setting. In the model of \cite{BEK2017} the trend effect is generated by the crowding behavior of investors. 

\cite{VW2013} propose an interpretation of trend and value anomalies based on delegated management. According to this theory investors try to learn about a portfolio manager’s skill from his past performance and effectively chase returns. In this scenario, fund managers are prone to moral hazard -- even if they are aware of a bubble, they prefer to herd with the mass of investors as it is bringing short-term profits. Going with the flow is safe -- for example, no manager was fired for holding tech stocks during the dot-com bubble in late 90s. On the other hand, going against the flock of investors can be financially damaging in the short-term. Since value strategies pay off slowly, unevenly and only over long horizons, impatient investors are likely to withdraw their funds from managers not following the trends. As a result, managers have an incentive to herd and fund flows push prices away from fair values, thus inducing short-term momentum. This approach is in its infancy and it is hard to judge its ability to explain the mechanism beyond the two effects since it has not been tested on financial data. Note that this mechanism relies, {\it in fine}, on strong behavioral biases affecting final investors.

More generally, as already mentioned in the introduction, there are many reasons to think that investors do not act rationally. 
Behavioral finance literature presents a long list of cognitive biases that may lead to irrational decisions. Conservatism bias, defined as the slow updating of beliefs in the face of new information \citep{Edwards1968}, may lead to investors' underreaction. A similar cognitive bias, also leading to underreaction, is anchoring \citep{TK1974}, the tendency of people to assess probabilities by starting with an implicitly suggested reference point (the `anchor') and making small incremental adjustments based on new evidence to reach their final estimate heavily biased by the initial anchor. \cite{KT1973} find that people are prone to recency bias -- a tendency to assign more importance to more recent observations compared to those further in the past. Market overreaction can be caused by a number of biases: one example is representativeness bias -- the tendency of people to categorize some events as typical or representative and ignore the laws of probability \citep{TK1974}. Another behavior leading to overreaction is self-attribution bias -- taking credit for positive returns of an investment strategy as proof of one’s own skill, but blaming bad luck for losses \citep{Bem1965}.

Two notable examples of behavioral finance models for trend \& value are \cite{BSV1998} and \cite{HS1999}. The first stage in both models is underreaction to arriving economic news, caused by the conservatism bias of investors. In the second phase, both models have some overreaction that drives the prices far away from the fundamental value. Finally price reversal occurs. In the model of \cite{BSV1998} overreaction is a consequence of the representativeness bias among investors, while in \cite{HS1999} part of the investors are trend-following causing the overreaction. Another prominent behavioral finance model is \cite{DHS1998}, which is motivated by different cognitive biases: overconfidence and self-attribution. In this model investors are over-confident about their private signals. When the price changes are in line with their prediction, they attribute it to their own skills and become even more over-confident, creating further short-term trends. In the famous behavioral model of \cite{DSSW1990}, trend effect is attributed to investor sentiment and recency bias in forecasting.

\subsection{Agent-Based Models}

One of the weaknesses of these behavioral finance models is that it is not clear how the biases of individual investors at the micro level translate into price dynamics at the macro level. On the other hand, the link between micro- and macro-levels is the central object of the Agent-Based Model (ABM) approach. For these reasons, agent-based technologies are well suited for testing behavioral theories \citep{Lebaron2006}.

Much like behavioral finance, the ABM approach does not assume that investors are rational. Rather,  financial markets are viewed as complex evolving systems with interacting groups of boundedly rational, heterogeneous agents who follow simple `rule of thumb' strategies \citep{Lebaron2006}. It is a `bottom-up' approach -- the micro interactions of agents are pieced together to give rise to a macro dynamics of the whole economy. It has been shown that ABM are able to replicate and explain a variety of stylized facts of financial markets, like excess volatility, volatility clustering, fat-tails of returns (see for example \cite{LUX1998}, \cite{LM1999}, \cite{GB2003}, \cite{CMZ2013} and \cite{HL2018}).

One type of traders in agent-based models are trend-followers with extrapolative expectations: their beliefs on future prices are formed by extrapolating recent trends in price. For example, \cite{LMT2017} provide quantitative evidence that investors are more prone to extrapolative expectation than to under-reaction bias. Extrapolative expectations are related to recency bias, as in \cite{KT1973}. The self-referential behavior of trend-followers creates trends and moves price away from the fundamental value, as in \cite{WB2007}. When the market price is far away from the fundamental value, a second group of traders, the fundamentalists, step in and take opposite positions, thus creating the price reversal responsible for the value effect (\cite{Chiarella1992,LUX1998}).

The great advantage of ABMs is that they offer a description of the resulting price dynamics. It can be provided as a solution of a nonlinear dynamical system, where bifurcation and chaos theories may be applied to help understand the resulting model dynamics. This approach is called analytical and in this group of ABM one can include the seminal models of \cite{BG1980}, \cite{Chiarella1992}, \cite{BrH1997, BH1998}, \cite{LUX1998}, and later models by \cite{CH2001}, \cite{CDG2002, CDG2006}, \cite{WB2007}. The second approach consists in studying the price dynamic by numerically simulating agents' behavior. This computational approach is considered in a very large number of works, starting from the so-called Santa-Fe models (see \cite{LAP1999} and for a recent review \cite{HL2018}).

Another classification of ABMs is with respect to the price setting mechanism. The most common price setting mechanisms are (i) Walrasian-type market clearing and (ii) price impact. Examples of ABM with Walrasian clearing are \cite{BrH1997, BH1998}. In price impact mechanism models, one typically postulates a linear relation between excess aggregate demand and price changes, which describes the behavior of a Kyle market maker (see \cite{Kyle1985}). Among models with a price impact mechanism, one can cite \cite{BG1980}, \cite{Chiarella1992}, \cite{BC1998}, \cite{GB2003}, \cite{WB2007}.

\section{An Extended Chiarella Model}\label{sec:HABM}
\subsection{Model Set-Up}

We will denote the log-price of an asset at time $t$ as $P_t$. We assume that the price dynamics is governed by a linear price impact mechanism: the price change from $t$ to $t+\Delta$ is linear in the cumulative demand imbalance (total signed volume traded on the market) in the same period, which we denote as $D \left(t, t+\Delta \right)$. This can be written as 
\begin{equation}
P_{t+\Delta} - P_{t}= \lambda D \left(t, t+\Delta \right), 
\end{equation} 
where $\lambda$ is `Kyle's lambda', and is inversely proportional to the liquidity of the market \citep{Kyle1985}.

The aggregated demand of investors depends on the investment strategies of market participants, which we assume are heterogeneous in their investment decisions, as postulated in the early HABMs of \cite{Chiarella1992,LUX1998,LM1999}. The empirical research of \cite{BH2007} on the S\&P 500 index from 1871 until 2003 provides evidence that the two largest groups of investors active in the market are fundamentalists, betting on mean reversion of stock prices towards the fundamental value, and chartists betting that trends will continue. Consistent with these observations, we classify agents as fundamentalists, trend-followers and noise traders -- i.e. agents with trading strategies uncorrelated with the future price change.

\paragraph{Trend-followers}
A commonly used trend signal, $M_t$, is the exponentially weighted moving average of past log-returns with decay rate $\alpha$. Following \cite{Chiarella1992} we assume that the demand function of trend-followers is given by $\tilde{\beta} \,\, \mathrm{tanh}(\gamma M_t)$ with $\gamma>0$. Parameter $\gamma$ describes the saturation of chartists' demand for large signals, induced for example by risk aversion and/or budget constraints. For trend signals larger than $\gamma^{-1}$ trend-followers' demand becomes almost constant.

The overall weight of trend-followers in the market is captured by parameter $\tilde{\beta}$. Consistently with the behavior of trend-followers, we assume that $\tilde{\beta} > 0$, i.e. the demand of chartists is positive (negative) if trend is positive (negative) and it is an increasing function of trend.

\paragraph{Fundamentalists}
The investment strategy of fundamentalists is based on the perceived intrinsic (or fundamental) value of the asset, $V_t$. A value trader will buy an asset whenever it is under-priced according to her beliefs ($V_t - P_t > 0$). She will sell it otherwise. Following \cite{BG1980}, \cite{Chiarella1992} and \cite{BC1998} we assume in this section that the aggregated demand of fundamentalists is linear in mispricing, i.e. given by $\tilde{\kappa} \left( V_t - P_t \right)$, where $\tilde{\kappa}$ describes the overall weight of fundamentalists in the market. A non-linear extension will be considered in Section \ref{sec:HABMNL}.

\paragraph{Noise-traders}
The last group of traders are noise traders, who either have different investment horizon than we are considering (months) or are following investment strategies different than trend-following or value investing. The cumulative demand of noise traders is described by Brownian Motion, $W^{(N)}_t$, multiplied by $\tilde{\sigma}_N$, which describes the size of noise traders in the market.

\paragraph{Model Dynamics}
The total demand of all three groups of investors is then given by \begin{equation}\label{eq:PriceImpact} D \left(t, t+\Delta \right) = \tilde{\kappa} \int\limits_{t}^{t+\Delta}( V_s - P_s )ds + \tilde{\beta} \int\limits_{t}^{t+\Delta}\tanh \left( \gamma M_s \right)ds + \tilde{\sigma}_N \int\limits_{t}^{t+\Delta}dW^{(N)}_s. \end{equation} Furthermore, we assume that the (log-)fundamental value is driven by a diffusion with volatility $\sigma_V$ and average growth $g$. The resulting price dynamics for $\Delta \rightarrow 0$ is described by a stochastic dynamical system: \begin{equation}\label{eq:Model} \begin{split} dP_t &= \kappa( V_t - P_t )dt + \beta \, \tanh \left( \gamma M_t \right)dt + \sigma_NdW^{(N)}_t,\\ dM_{t} &= - \alpha M_t dt + \alpha dP_t,\\ dV_t &= gdt + \sigma_VdW^{(V)}_t, \end{split} \end{equation} where $\kappa$, $\sigma_N$, $\beta$ are equal to $\lambda \tilde{\kappa}$, $\lambda \tilde{\sigma}_N$, $\lambda \tilde{\beta}$, respectively. Observe that for $g=0$ and $\sigma_N = 0$ we recover the model of \cite{Chiarella1992}. The dynamics produced by similar ABMs with linear demand functions was investigated in \cite{BG1980}, and for ABMs with non-linear demand of trend-followers in \cite{BC1998}.

\subsection{The deterministic case: Non-linear oscillations}\label{sec:oscillations}
In this section we briefly review the mathematical properties of model (\ref{eq:Model}). We start our considerations by assuming $\sigma_N = \sigma_V = g = 0$. In that case the system is deterministic and the dynamics of the trend signal is described by a second-order differential equation: \begin{equation}\label{eq:lienard} \frac{d^2 M_t}{dt^2} + \mathcal{G}\left( M_t \right) \frac{d M_t }{dt} + \alpha \kappa M_t = 0, \end{equation} where \begin{equation}\label{eq:damping} \mathcal{G}\left( x \right) := \alpha + \kappa - \alpha \gamma \beta \left( 1 - \mbox{tanh}^2 \left( \gamma x \right) \right). \end{equation} Equation (\ref{eq:lienard}) with a general function $\mathcal{G}$ is called the Li\'{e}nard equation and the famous Van der Pol oscillator is the special case where $\mathcal{G}\left( x \right) = \mu \left( x^2 - 1 \right)$. In theory of non-linear oscillations the function $\mathcal{G}$ is called damping force. If \begin{equation}\label{eq:bifurcation} \alpha + \kappa - \alpha \gamma \beta < 0, \end{equation}
then for small $|M_t|$, the damping $\mathcal{G}(M_t)$ is negative, which will make $|M_t|$ increase with time. For large trend signal in absolute value, the damping force is positive and large, which will make $|M_t|$ decrease with time. Naturally, this phenomenon will cause non-linear oscillations of trend signal for the set of parameters satisfying (\ref{eq:bifurcation}).

\begin{figure}[!h]
\begin{center}
\includegraphics[scale=0.7]{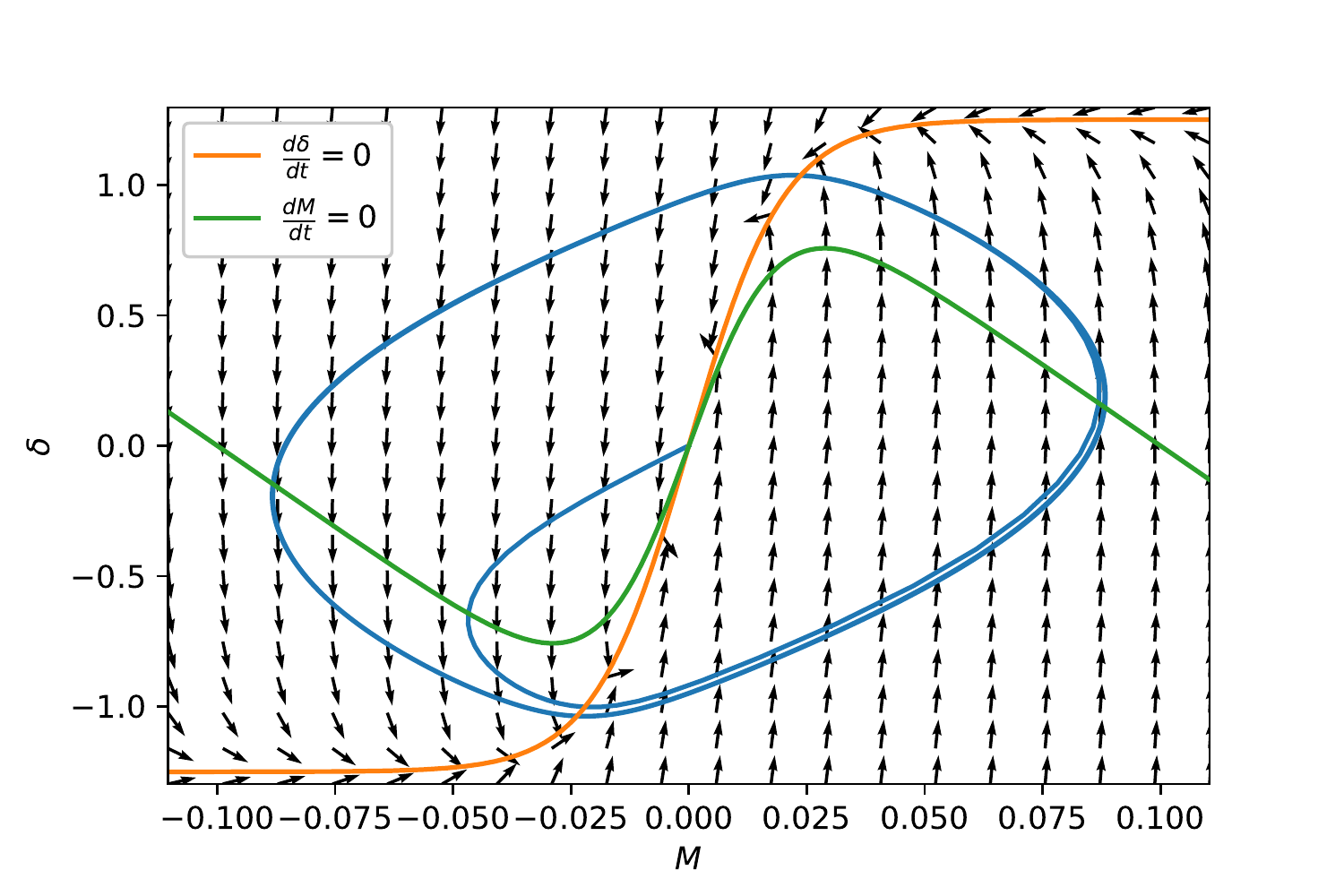} \caption{Limit cycle. $\delta = P - V$ denotes price
distortion.}\label{fig:VF} 
\end{center} 
\end{figure}

It is more subtle to show that (\ref{eq:lienard}) has a unique non-degenerate periodic solution. The Levinson-Smith theorem states that if (\ref{eq:bifurcation}) is satisfied then the dynamic system $\left( M_t,R_t \right)$, where $R_t = dM_t/dt$, has a unique and stable limit cycle surrounding the origin in the phase plane. Consequently, $M_t$ has a unique periodic solution. Otherwise, if (\ref{eq:bifurcation}) is not satisfied, the system $\left( M_t, R_t \right)$ converges to the origin which is the critical point of the system. Figure \ref{fig:VF} shows the limit cycle generated by the system (\ref{eq:lienard}) in the $\delta-M$ plane, where $\delta=P-V$ is the price distortion. There is no closed formula for the shape of the limit cycle. Establishing the properties of limit cycles is still an area of active research. More details about the geometry of limit cycles and the dynamic analysis of the system can be found in \cite{Chiarella1992, CHWZ2008}.


To obtain an economic interpretation of the non-linear oscillations we rewrite the condition (\ref{eq:bifurcation}) as $\kappa < \alpha \left( \gamma \beta - 1 \right)$. Here we immediately observe that the trading activity of trend-followers is the force boosting oscillations, while fundamentalists are damping it. The oscillations are stronger for longer trend horizons ($\alpha$ small), faster saturation of trend-followers demand ($\gamma$ large) and larger weight of trend-followers in the market ($\beta$ large). A larger weight $\kappa$ of fundamentalists weakens the price oscillations, to the point that, if $\kappa \geq \alpha \left( \gamma \beta - 1 \right)$, then the price simply relaxes towards the fundamental equilibrium with no oscillations at all.
 \begin{figure*}[t!]

 \centering
 \begin{subfigure}[t]{0.5\textwidth}
 \centering \includegraphics[scale = 0.5]{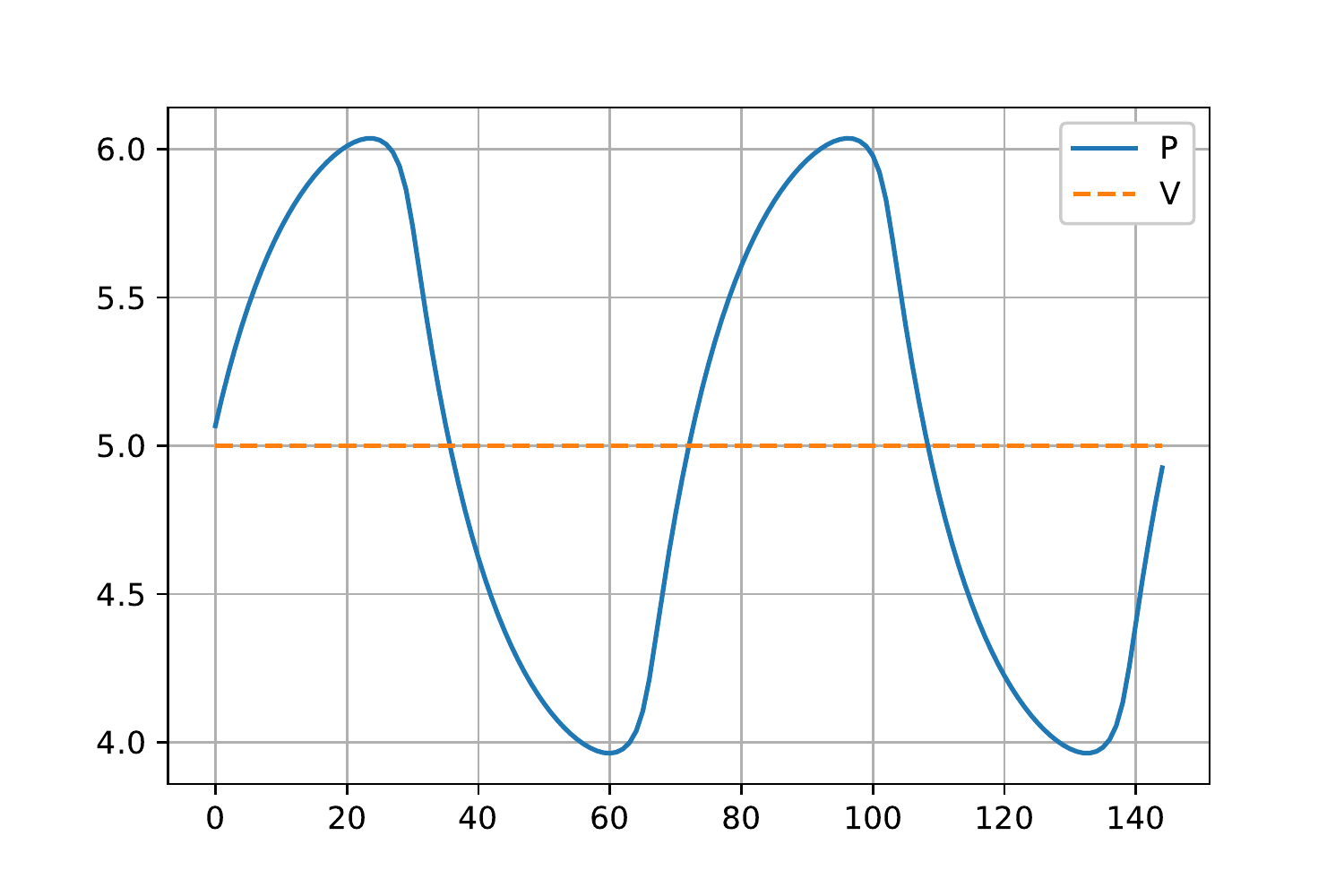}
 \caption{{\footnotesize Price and value in a deterministic setting.}}
 \end{subfigure}%
 ~
 \begin{subfigure}[t]{0.5\textwidth}
 \centering \includegraphics[scale=0.5]{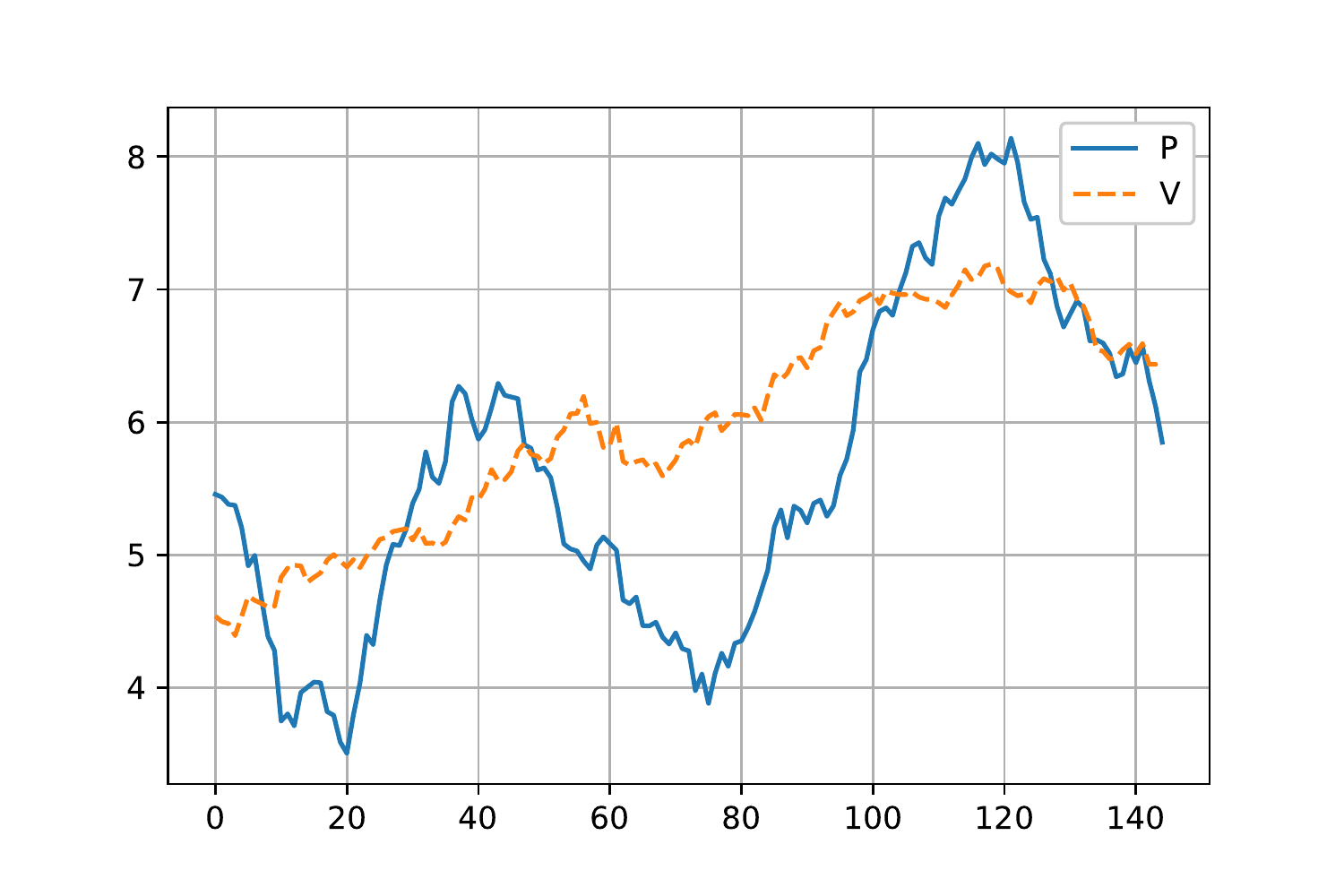}
 \caption{{\footnotesize Price and value in a stochastic setting.}}
 \end{subfigure}\\
\begin{subfigure}[t]{0.5\textwidth}
 \centering \includegraphics[scale = 0.5]{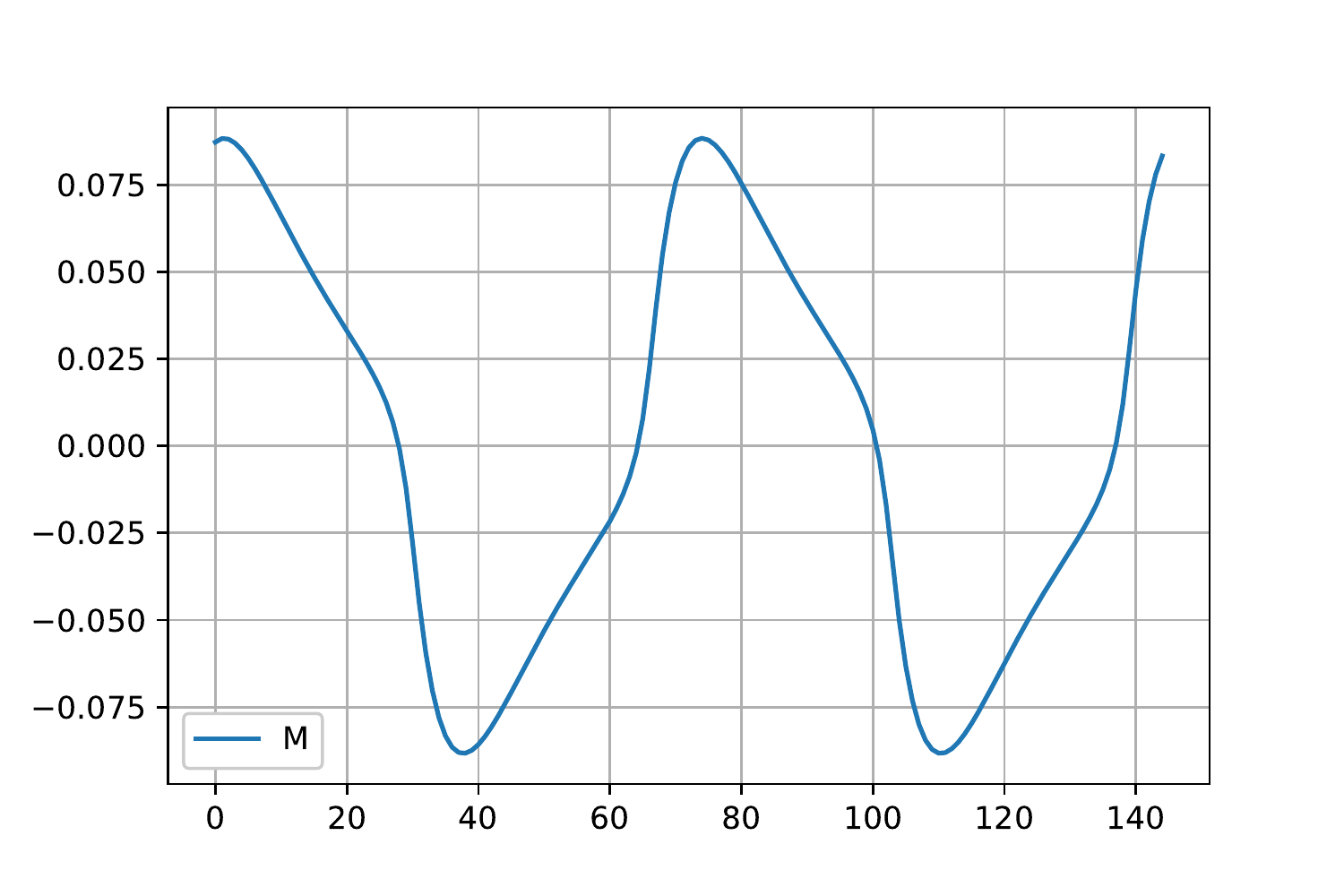}
 \caption{{\footnotesize Trend signal in a deterministic setting.}}
 \end{subfigure}%
 ~
 \begin{subfigure}[t]{0.5\textwidth}
 \centering \includegraphics[scale=0.5]{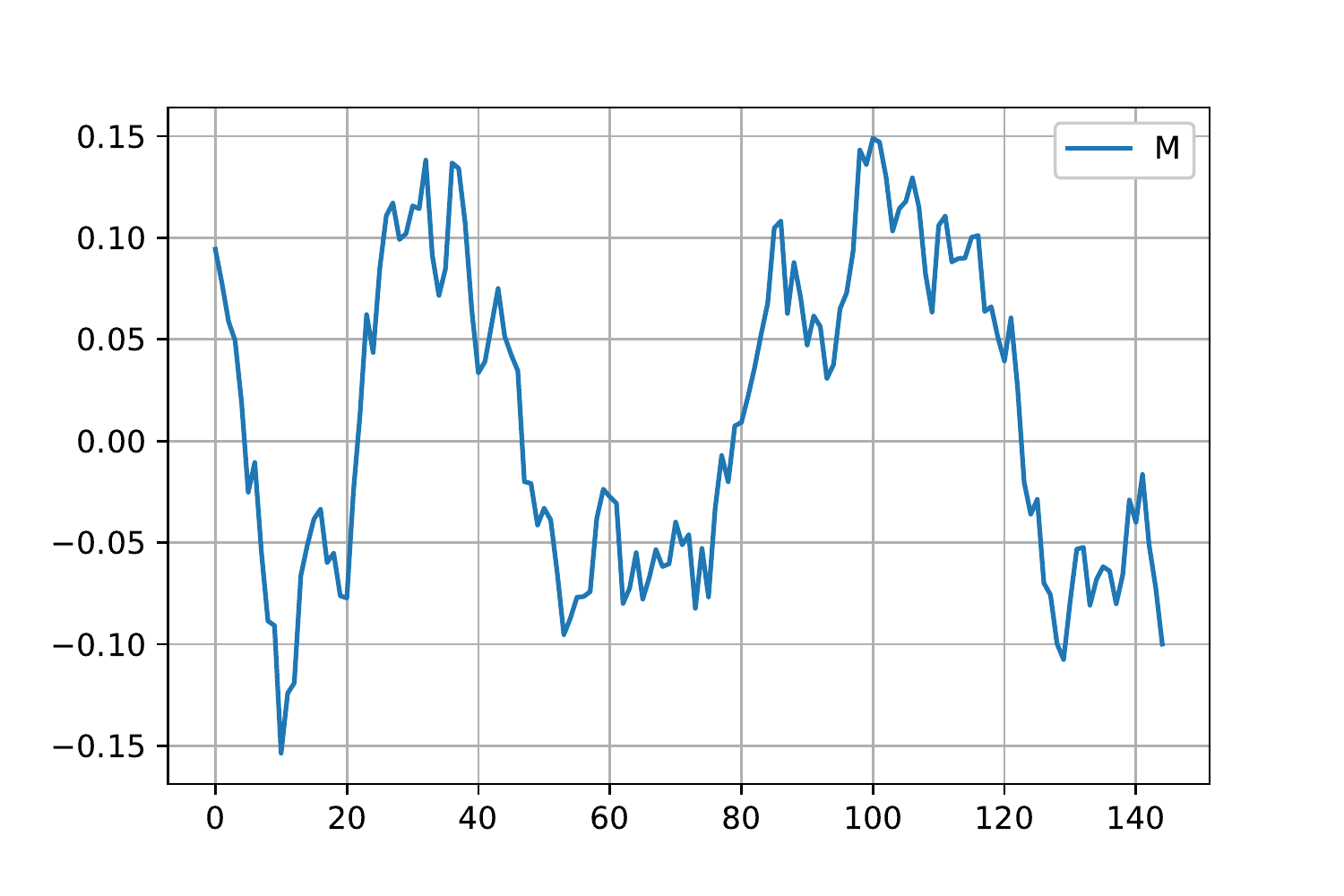}
 \caption{{\footnotesize Trend signal in a stochastic setting.}}
\end{subfigure}
 \caption{{\footnotesize Trajectories of price, fundamental value and trend signal for model \ref{eq:Model} with parameters: $\kappa = 0.08$, $\beta = 0.1$, $\gamma = 50.0$, $v_0 =5.0$, $g=0.0$. On the left: $\sigma_N = \sigma_V = 0.0$, on the right: $\sigma_N = 0.15$ and $\sigma_V = 0.075$. }}\label{fig:trajectories} 
\end{figure*}

Figure \ref{fig:trajectories} shows the trajectories of price ($P_t$), fundamental value ($V_t$) and trend signal ($M_t$) for a set of parameters satisfying condition (\ref{eq:bifurcation}). In the left part we present the evolution of price and trend when value is constant and there are no noise traders. We observe that price is oscillating around the fundamental value. The oscillations are driven by two competing forces: value and trend. When the price is near value the trend forces are pushing the price away from its value. When the price of an asset walks too far from its fundamental value, fundamentalists are enforcing mean-reversion. The situation is similar in the stochastic case but less clear (and more realistic!) due to the noise. In that case we describe the properties of the system by analyzing its stationary distributions in Section \ref{sec:StationaryDistribution}.

\subsection{Model calibration}\label{sec:Estimation}
\subsubsection{Data}\label{sec:Data}

We estimate the model on large pool of monthly spot prices belonging to four types of assets classes: stock indices, commodities, fixed income and currencies. Our aim is to investigate as diverse a universe of assets as possible. At the same time, to increase the robustness of our results, we will choose only the time series with long enough history. Therefore, for indices, bonds and currencies, we consider the following seven countries: Australia, Canada, Germany, Japan, Switzerland, the United Kingdom and the United States. For currency exchange rates we consider only pairs versus the US dollar. Our commodities data consist of crude oil, Henry Hub natural gas, corn, wheat, sugar, live cattle and copper. Our source of spot prices is Global Financial Data. This is the dataset considered in \cite{LDSPB2014}.

We constrain our time series only to periods when the asset are freely traded with high liquidity. For that reason we will limit the time series of currency exchange rates from 1973. From 1945 to 1973 all currencies were pegged to the US dollar inside the Bretton Woods system. We consider only government bonds yields after 1920, as before they were not liquidly traded. The emergence of a stable government debt market coincided with the creation of modern central banks that pursue monetary policy independently from the political authority of the country.

Time series of spot prices of equity indices and commodities do not suffer from similar limitations and the available history of prices is over 200 years. From all commodities series we will only exclude the period of World War II and from Crude Oil we exclude additionally period 1939-1985, when the price of oil was fixed. Equity indices have witnessed large jumps due to political reasons. Since our model is not designed to deal with jumps, we remove World War II period from German and Japanese indices, and the period of  World War I from German and British equity indices (for Germany we also exclude the post First World War period termed `hyperinflation of the Weimar Republic'). Additionally we exclude the period 1973-1974 for Great Britain corresponding to stock market crash after the collapse of the Bretton Woods system.

The time series of equity indices and commodities prices are inflation adjusted. Real price of equity index at time $t$ is obtained by multiplying the nominal price at time $t$ by the ratio between local CPI (Consumer Price Index) at the last observation and CPI at time $t$. To obtain real prices of commodities we use US CPI.

\subsubsection{Estimation: the Expectation-Maximization Algorithm}\label{sec:EMalgorithm}

We will estimate a discrete-time version of model (\ref{eq:Model}), where one time step is equal to one month: 
\begin{equation}\label{eq:DiscreteModel} 
\begin{split} 
p_{t+1} - p_{t} &= \kappa( v_t - p_t ) + \beta \tanh \left( \gamma m_t \right) + \epsilon_{t+1},\\ 
m_{t+1} &= (1 - \alpha)m_t + \alpha (p_{t} - p_{t-1}),\\ 
v_{t+1} &= v_t + g + \eta_{t+1} 
\end{split} 
\end{equation} 
where $\epsilon_t$ and $\eta_t$ are i.i.d. with normal distribution with zero mean and standard deviations, $\sigma_N$ and $\sigma_V$, respectively.

The correlation between the SG CTA index (formerly called Newedge CTA index) and the P\&L of trend-following strategies with different window lengths, reveals that the typical horizon of trend computation for trend-following strategy is around six months. For this reason we will take $\alpha = 1/(1+\tau) = 1/7$, where $\tau = 6$.\footnote{We have also performed the estimation for $\alpha$ equal $1/4$ and $1/10$ corresponding respectively to $3$ and $9$ months, with very similar results.} Since the model estimation will have difficulties pinning down the value of parameters $\beta$ and $\gamma$, we will fix $\gamma^{-1}$ to twice the standard deviation of $m_t$ of the asset. This factor $2$ comes from the fact that saturation in trend effect starts around $2$ (see Figure \ref{fig:trend_effect}).

The main challenge of estimating such agent-based models is the non-observability of the fundamental value. In the economic literature there are several ways to estimate the fundamental value of a company. According to a popular finance theory \citep{Gordon1962,Shiller2000}, the fundamental value of a stock should be equal to the expected value of discounted future dividends that the company will pay to the shareholders. The drawback of this methodology is that one needs to make strong assumptions on the dynamics of these dividends. Moreover, we cannot use this approach to compute the fundamental value of assets that do not pay dividends, for example commodities.


In this paper we propose a different approach in which we estimate {\it both} the fundamental value of an asset and the parameters of the model simultaneously. We note that the dynamical system (\ref{eq:DiscreteModel}) is linear in $v_t$, which we treat as a hidden variable of the system. For convenience and consistency with state-space literature, let us denote by $\tilde{v}_t:=v_{t-1}$ the fundamental value of the asset the day before $t$ and $\tilde{\eta}_{t} := \eta_{t-1}$. We can then rewrite model (\ref{eq:DiscreteModel}) as
\begin{equation}\label{eq:StateSpaceModel} 
\begin{split} 
\tilde{v}_{t+1} &= g + \tilde{v}_t + \tilde{\eta}_{t+1}\\ 
p_{t+1} &= p_t + \kappa \left( \tilde{v}_{t+1} - p_t \right) + \beta u_{t+1} + \epsilon_{t+1}, 
\end{split} 
\end{equation} 
where $u_{t+1} := \tanh \left( \gamma m_t \right)$ is a function of past prices. Treating $u_t$ component as a control term of the system, we immediately see that the fundamental value $\tilde{v}_t$ can be filtered out by means of a classical Kalman filter \citep{Kalman1960}, were the parameters of the system known. In the case where parameters, and hidden states of the model are not known, we can estimate both of these by applying an Expectation-Maximization (EM) algorithm.

The algorithm is maximizing a difficult-to-compute marginal log-likelihood by maximizing the conditional expectation of an easy-to-compute joint log-likelihood, in a two-step iterative procedure. It alternates between one E-step that computes a conditional expectation of joint log-likelihood on posterior distribution over the hidden variables given prices and the parameters, and one M-step which infers the parameters for which the expectation obtained in E-step reaches maximum. The detailed description of the EM algorithm is provided for completeness in Appendix \ref{app:EMalgorithm}. The EM algorithm was introduced by \cite{BaumEtAl1970} for Hidden Markov Models and it was generalized for models with incomplete observations in \cite{DempsterEtAl1977}. The EM algorithm with Kalman filter was applied for estimation of linear dynamics system for the first time in \cite{Chen1981} and \cite{SS1982}.

It is quite remarkable that it is not necessary to have any economic information about the asset (such as the dividends paid by a company) to estimate its underlying fundamental value. It is sufficient to observe the time series of prices and assume that its dynamics is governed by (\ref{eq:DiscreteModel}). We also note that our framework provides a very good estimation accuracy on surrogate data and that it is more computationally efficient than the recent approach by \cite{Lux2017} or \cite{BMS2018}.

\subsubsection{Results}\label{sec:ResultsLDS}

In this section we present the results of our estimation procedure introduced in Section \ref{sec:EMalgorithm}, applied to the time series described in \ref{sec:Data}. To obtain a more robust estimation procedure we will calibrate parameters in two steps. In the first step we calibrate each time series of log-prices to obtain asset specific values of $\sigma_N$, $g$ and the initial fundamental value $v_0$. In the second step we search for a set parameters $\kappa$, $\beta$, $\sigma_V$ common to an asset class (i.e. one set of parameters for stock indices, one for commodities, one for FX rates and one for government bonds) that maximizes the common log-likelihood.

In Table \ref{tab:estimation_EM} (see Appendix \ref{app:estimation_results}) we report the results from the EM estimation of model (\ref{eq:DiscreteModel}) with T-statistics of estimated parameters in Table \ref{tab:tstat_EM} (see Appendix \ref{app:estimation_results}). From this table we observe that for all assets $\sigma_N$ is substantially larger than $\sigma_V$. This confirms that noise traders provide an important contribution to the volatility and that it is crucial to include them in the HABM. This feature can be also viewed as another piece of evidence for the famous `excess volatility puzzle' first reported by \cite{Shiller1980}.

\begin{figure}[h!]
\begin{center}
 \includegraphics[scale=0.7]{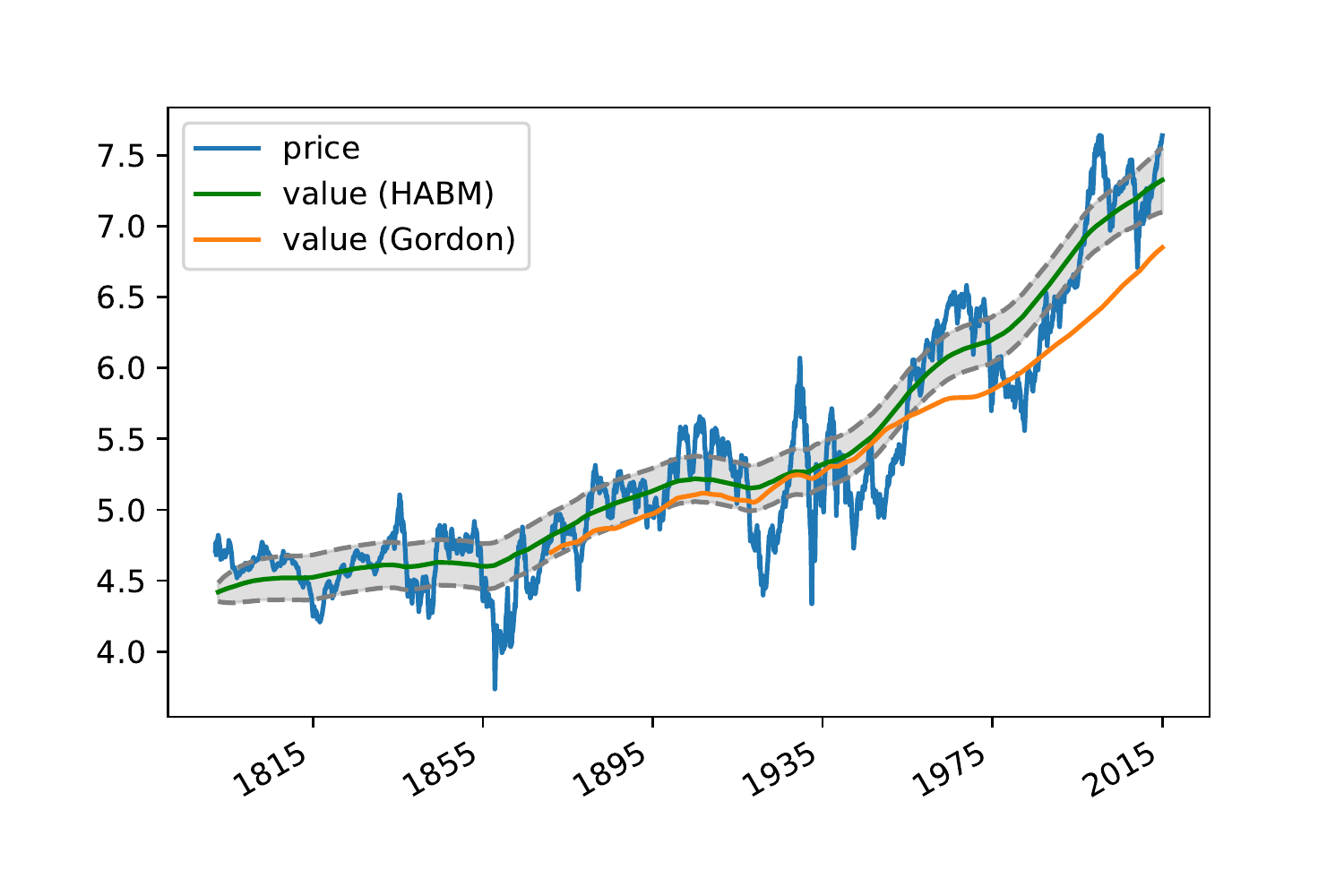} 
\caption{{\footnotesize Log-level of the US stock index, together with the smoothed fundamental value as obtained from model (\ref{eq:Model}) with parameters in Table \ref{tab:estimation_EM}. We also plot value plus/minus standard one deviation of the estimation interval, and the benchmark fundamental value obtained from Gordon model.}}\label{fig:US_LDS} 
\end{center} 
\end{figure}

In Figure \ref{fig:US_LDS} we present the smoothed estimate of fundamental value (using a Kalman smoother) for the US stock index given the estimated parameters in Table \ref{tab:estimation_EM}. We can see that market price is oscillating around the fundamental values of the asset. Since $\alpha + \kappa - \alpha \gamma \beta$ is found to be slightly positive, the driving force of oscillations is the uncertainty of fundamental value and the noise traders' activity. The price stabilizing activity of fundamentalists is not strong enough (relatively to the level of noise and activity of trend-followers) to prevent oscillations.

To compute a fundamental value benchmark for the US index we apply the model of \cite{Gordon1962}. According to this model the fundamental value of an equity asset is the present value of dividends and is computed from the actual subsequent real dividends using a constant real discount rate of $6.8 \%$ per year, equal to the historical average real return on the market since 1871.\footnote{We take the dividend data from Shiller’s dataset, available at http://www.econ.yale.edu/~shiller/.} For dividends after December 2014, it is assumed that these will themselves grow forever from the last observed dividend, with a growth rate of $2.2\%$ percent per year (which is the dividends’ average growth rate since 1871).

From Figure \ref{fig:US_LDS} we can observe that the benchmark fundamental value is 1-$\sigma$ below our own estimate from 1871 till 1950's. Since then, the discrepancy estimates diverge more strongly, with values obtained with Gordon's model suggesting that the S\&P 500 is systematically, and substantially overpriced since 70's. Some authors assume the future dividend growth to be equal to average dividends growth over some short recent period (for example 10 years in \cite{SW2017}). This choice of dividends growth is justified by the increase of dividends growth during the last decades. As a consequence, the divergence between our estimates and the Gordon model can be mitigated. However, we find this approach inconsistent and non-robust, and is another argument in favor of our estimation methodology.
 
\subsection{Trend and value effects}\label{sec:TrendValueEffects}

\begin{figure*}[t!]
\centering
\begin{subfigure}[t]{0.5\textwidth}
\centering \includegraphics[scale = 0.5]{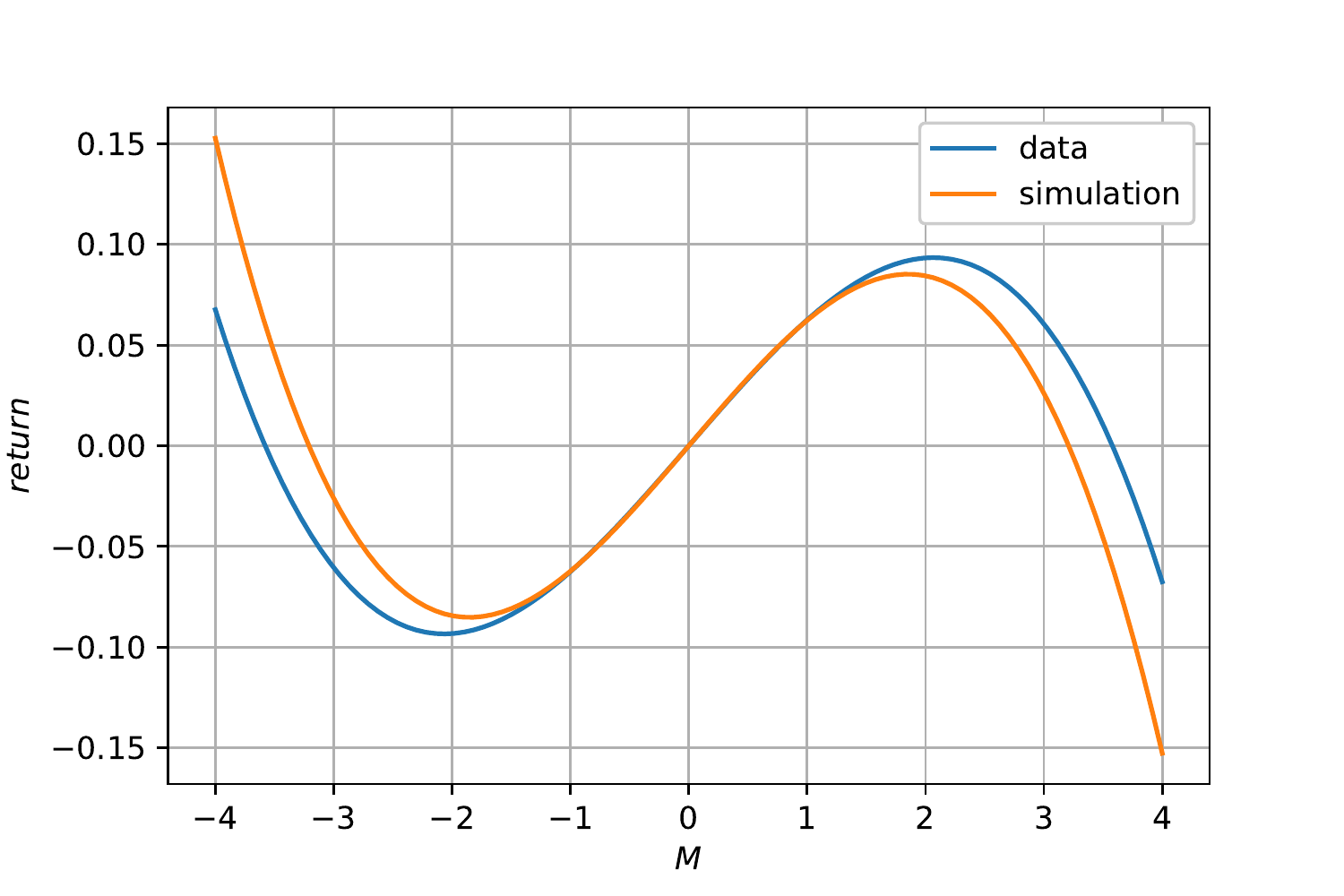}
\caption{{\footnotesize Trend effect }}\label{fig:trend_effect}
\end{subfigure}%
~
\begin{subfigure}[t]{0.5\textwidth}
\centering 
\includegraphics[scale=0.5]{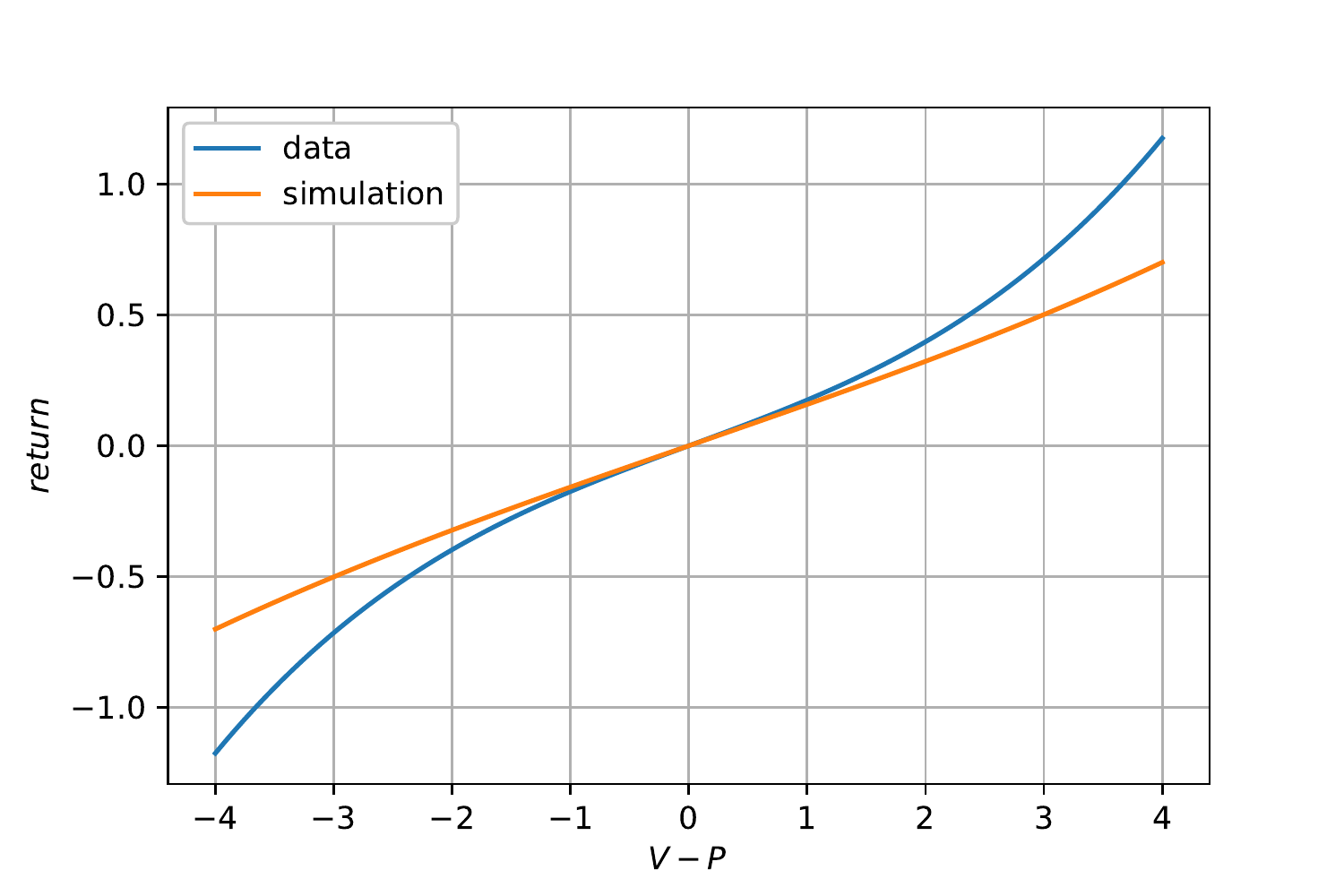}
\caption{{\footnotesize Value effect}}\label{fig:value_effect}
\end{subfigure}
\caption{{\footnotesize Left plot: regression of log-returns on the trend signal $m_t$ for real data (second row of Table \ref{tab:regressions}) and simulated using the calibrated model, with the parameters in Table \ref{tab:estimation_EM}. Right plot: regression of log-returns on the value signal $V - P$ for real data (with $V$ implied by the model) and simulated using the same model.}} 
\end{figure*}

We repeated the analysis of \cite{BCLMSS2017}, who showed that monthly returns $p_{t+1} - p_t$ can be fitted by a cubic polynomial of the past trend signal $m_t$, defined as before. Figure \ref{fig:trend_effect} confirms that the relation between trend signals and future log-returns is non-monotonic, and quite well reproduced by the model with the parameters in Table \ref{tab:estimation_EM}. The coefficients of the regression are reported in the second row of Table \ref{tab:regressions}. The adjusted R-square of the regression is $1.3\%$ and it is substantially higher than for linear regression (row one of Table \ref{tab:regressions}). Negative cubic coefficient suggests that large moves (in absolute value) tend to mean-revert.

A careful analysis of the value effect reveals some non-linearities, not well accounted for by the model. Contrary to the trend effect, the estimation of the value effect is model-dependent since $V$ has to be extracted from the calibration procedure itself. In Figure \ref{fig:value_effect} we plot the value effect using real data and the fundamental value $V$ implied by our HABM, and compare it with the same quantity obtained from a simulation of the model. We see that the restoring force towards $V$ is clearly non-linear and {\it increases} at large mispricing. In order to account for this observation, we will consider in the next section an extended model with a non-linear demand function of fundamentalists.

\begin{figure*}[t!]
\centering
\begin{subfigure}[t]{0.5\textwidth}
\centering \includegraphics[scale = 0.5]{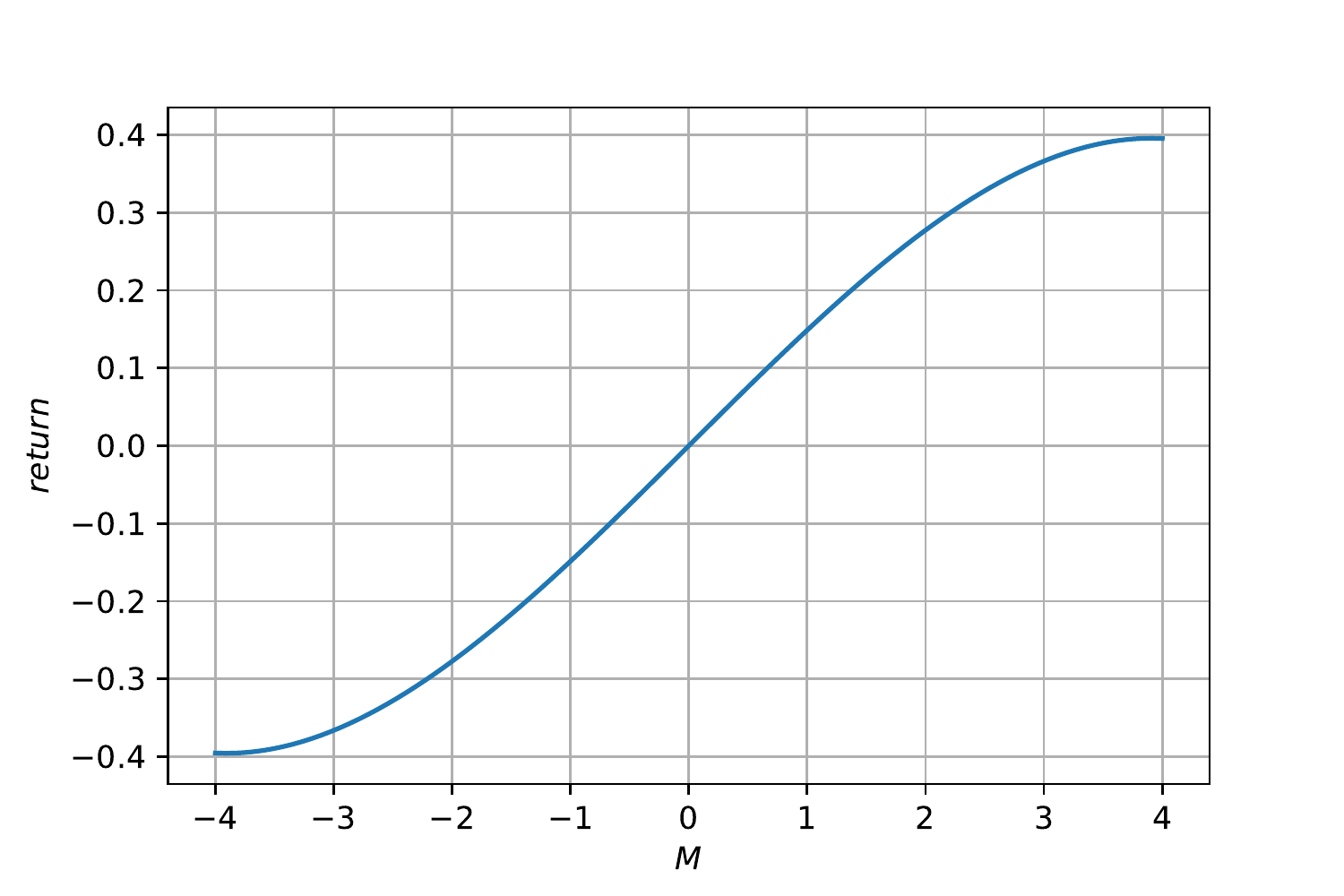}
\caption{{\footnotesize Trend effect controlled for value.}}\label{fig:trend_effect_control}
\end{subfigure}%
 ~
\begin{subfigure}[t]{0.5\textwidth}
\centering \includegraphics[scale=0.5]{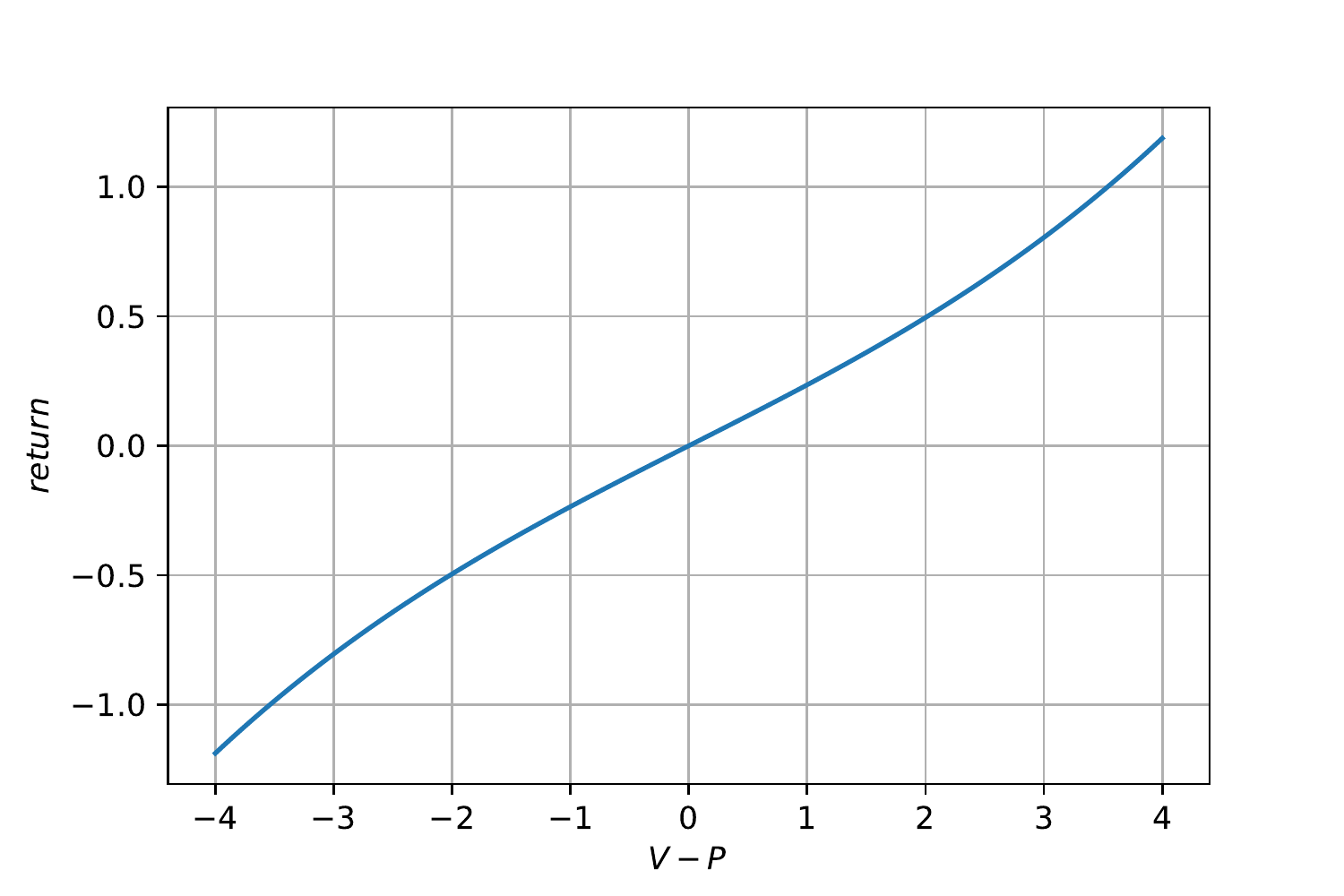}
\caption{{\footnotesize Value effect controlled for trend.}}\label{fig:value_effect_control}
\end{subfigure}
\caption{{\footnotesize On the left: function $F_t(m) = a_1 m + a_3 m^3$, where coefficients $a_1$ and $a_3$ are given in the last rows of Table \ref{tab:regressions}. On the right: function $F_v(V-P) = b_1 \left( V - P \right) + b_3 \left( V - P \right)^3 $, where coefficients $b_1$ and $b_3$ are given in the last rows of Tables \ref{tab:regressions}.}} 
\end{figure*}

The analysis of the trend and value effects together delivers important messages. In Table \ref{tab:regressions} we report the result of regressions of log-returns with several configurations of regressands composed of trend and mispricing. We observe that in the configuration with all regressands, the cubic term $m^3$ becomes smaller in absolute value and the linear component larger, with respect to a regression with only the trend component. Consistently with intuition, the non-linear mean-reversion contribution to trend apparent in Fig. \ref{fig:trend_effect} (Left) disappears when we control for the value effect, as is replaced by a saturation effect -- compare Figures \ref{fig:trend_effect} and \ref{fig:trend_effect_control}. In other terms, when we condition our observations to a price distortion equal to zero, the contrarian behavior for large trend signals disappears.

\begin{table}
\begin{center}
{\footnotesize
\begin{tabular}{lrrrrrrlr}
\toprule
{} &  const &     $m$ &     $m^2$ &     $m^3$ &     $d$ &     $d^3$ &   &     $R^2$ \\
\midrule
\rowcolor{gray!30}  \textbf{coefficient} & -0.000 &  0.024 &    - &    - &    - &    - &    &  0.001 \\
\rowcolor{gray!30}  \textbf{P-value    } &  0.997 &  0.000 &    - &    - &    - &    - &    &    - \\
\textbf{coefficient} & -0.003 &  0.068 &  0.001 & -0.005 &    - &    - &    &  0.013 \\
\textbf{P-value    } &  0.631 &  0.000 &  0.710 &  0.000 &    - &    - &    &    - \\
\rowcolor{gray!30}  \textbf{coefficient} &  0.000 &    - &    - &    - &  0.210 &    - &    &  0.044 \\
\rowcolor{gray!30}  \textbf{P-value    } &  0.996 &    - &    - &    - &  0.000 &    - &    &    - \\
\textbf{coefficient} &  0.001 &    - &    - &    - &  0.167 &  0.008 &    &  0.050 \\
\textbf{P-value    } &  0.820 &    - &    - &    - &  0.000 &  0.000 &    &    - \\
\rowcolor{gray!30}  \textbf{coefficient} & -0.000 &  0.119 &    - &    - &  0.255 &    - &    &  0.056 \\
\rowcolor{gray!30}  \textbf{P-value    } &  0.990 &  0.000 &    - &    - &  0.000 &    - &    &    - \\
\textbf{coefficient} & -0.007 &  0.158 &  0.005 & -0.005 &  0.251 &    - &    &  0.067 \\
\textbf{P-value    } &  0.239 &  0.000 &  0.020 &  0.000 &  0.000 &    - &    &    - \\
\rowcolor{gray!30}  \textbf{coefficient} & -0.006 &  0.152 &  0.005 & -0.003 &  0.231 &  0.004 &    &  0.068 \\
\rowcolor{gray!30}  \textbf{P-value    } &  0.337 &  0.000 &  0.024 &  0.000 &  0.000 &  0.000 &    &    - \\
\bottomrule
\end{tabular}

\caption{{\footnotesize Regressions of log-returns on trend and value components for the model (\ref{eq:Model}). In the last column we present the adjusted R square of the regression.}}\label{tab:regressions}}
\end{center}
\end{table}

Finally, the value effect also changes when controlling for the trend. We can see that the linear component of the value effect becomes larger if we add the trend signal as regressand (compare Figure \ref{fig:value_effect_control} with Figure \ref{fig:value_effect}). This suggests that trend-followers are an important force reducing the expected return of value investors for small price discrepancies, reducing the incentive for such investors to step in. On the other hand, observe that the value effect is stronger for large price distortions, even when controlled for trend.

\section{Non-linear Fundamentalist Demand Function}\label{sec:HABMNL}
\subsection{Model dynamics}

The model proposed in the previous section is not able to reproduce the shape of the value effect observed in financial data. One of the reason might be the inadequacy of a linear demand function of fundamentalists. Demand of fundamentalists depends on many factors like attitude towards risk, perception and, more importantly, uncertainty about fundamental value. It seems natural also to assume that fundamentalists' demand function picks up when mispricing is large, at least within some range, meaning that the demand function is of higher order than linear. Note that this does not necessarily mean that agents are risk-lovers, but rather implies that more agents following value investing start to trade when market misvaluation becomes more evident. If price distortion is small ($|V_t - P_t|$ small), on the other hand, fundamentalists are not even sure about the sign of mispricing, due to a large uncertainty on the unobservable fundamental value. Consequently, we expect fundamentalists' demand function to be rather flat in the region of small price distortion.

Therefore, we propose to further modify the Chiarella model by positing that the fundamentalists' demand function $\tilde{f}\left( V_t - P_t \right)$ contains a cubic term, on top of the linear term considered so far: $\tilde{f}(x) = \tilde{\kappa} x + \tilde{\kappa}_3 x^3$, where $\tilde{\kappa}_3$ is presumably positive. The linear coefficient $\tilde{\kappa}$, on the other hand, might be negative, as a result of behavioral biases such as herding (see the discussion in \cite{BBDG2018}, Chapter 20). We keep the demand function of trend-followers as in the previous section.

The resulting price dynamics is now described by the following generalized stochastic dynamical system: 
\begin{equation}\label{eq:ModelNL} 
\begin{split} 
dP_t &= f( V_t - P_t )dt + \beta \tanh \left( \gamma M_t \right)dt + \sigma_NdW^{(N)}_t,\\ 
dM_{t} &= - \alpha M_t dt + \alpha dP_t,\\ 
dV_t &= gdt + \sigma_VdW^{(V)}_t, 
\end{split} 
\end{equation} 
where $f(x) = \kappa x + \kappa_3 x^3$, where $\kappa_3 = \lambda \tilde{\kappa}_3$, $\kappa = \lambda \tilde{\kappa}$ and $\lambda$ is Kyle's impact parameter. Note that when $\kappa_3=0$ one recovers exactly Eq. (\ref{eq:Model}).

\subsection{Estimation}
\subsubsection{Unscented Kalman Filter}\label{sec:CKF}

Following the same reasoning as in Section \ref{sec:EMalgorithm}, we estimate the discretized version of model (\ref{eq:ModelNL}) \begin{equation}\label{eq:NonlinearModel} \begin{split} \tilde{v}_{t+1} &= \tilde{v}_t + g + \tilde{\eta}_{t+1},\\ p_{t+1} &= p_{t} + f \left( \tilde{v}_{t+1} - p_t \right) + \beta u_{t+1} + \epsilon_{t+1}.\\ \end{split} \end{equation} In the present case, however, the conditional distribution of $\tilde{v}_t$ given the past prices is no longer Gaussian and we cannot use Kalman filters anymore.

One can find several filtering approaches for non-linear dynamic systems in the Machine Learning literature. For example `particle filtering' (also called Sequential Monte Carlo) amounts to simulate the dynamics of the hidden state $\tilde{v}_{t+1}$ a certain number of times (predicting step) and then to adjust in the filtering step the weight of each prediction based on the likelihood of observing log-price $p_{t+1}$, given the fundamental value obtained in simulation. This approach has been used for estimating ABMs in \cite{Lux2017} and \cite{BMS2018}. Another approach, called Extended Kalman Filter (EKF), approximates the distribution of fundamental values by a Gaussian which is then propagated analytically through the first-order linearization of the non-linear demand function of fundamentalists. Both approaches are described in \cite{DK2012} or \cite{Sarkka2013}.

In this paper we will apply a third method called Unscented Kalman Filter (UKF), introduced by \cite{JU1997}. The idea of the filter is to approximate directly the demand of fundamentalists by a Gaussian distribution with mean and variance determined by an Unscented Transform. Details of the UKF method are given in Appendix \ref{app:UKF}. The main advantage of UKF over EKF is a higher accuracy with the same computational complexity. UKF approximates posterior mean and covariance accurately to the 3rd order (Taylor series expansion) for any non-linearity. EKF, in contrast, only achieves first-order accuracy.

All these filtering recipes can be applied together with EM for the identification of non-linear dynamical systems (see \cite{RG2001} for EM combined with EKF and \cite{KSS2014} or \cite{KSS2015} for EM combined with particle filter and UKF). Unfortunately, in the case of non-linear systems we cannot guarantee that the algorithm will converge to the local maximum of log-likelihood as for linear systems. In fact, if our approximation of the conditional distribution of value is too far away from the true distribution (in a Kullback–Leibler sense), the algorithm will maximize a quantity that is substantially different from the log-likelihood of the system.\footnote{See discussion below equation (\ref{eq:convergence}) in Appendix \ref{app:EMalgorithm}.} For this reason, instead of applying the EM algorithm we will optimize our non-linear model with a direct likelihood maximization. Details of estimation procedure are provided in Appendix \ref{app:estNL}.

\subsubsection{Results}
We follow the same two-step estimation procedure as in Section \ref{sec:ResultsLDS}. In the first step we calibrate each time series of log-prices to obtain an asset specific $\sigma_N$, $g$ and the initial fundamental value $v_0$, while in the second step we search for a common set of parameters $\kappa$, $\kappa_3$, $\beta$ for a given asset class. In order to prevent the algorithm to converge to unphysical values, we fix the fundamental volatility $\sigma_V$ to the value obtained for the linear model in Section \ref{sec:ResultsLDS}.

\begin{figure}[!h]
\begin{center}
\includegraphics[scale=0.7]{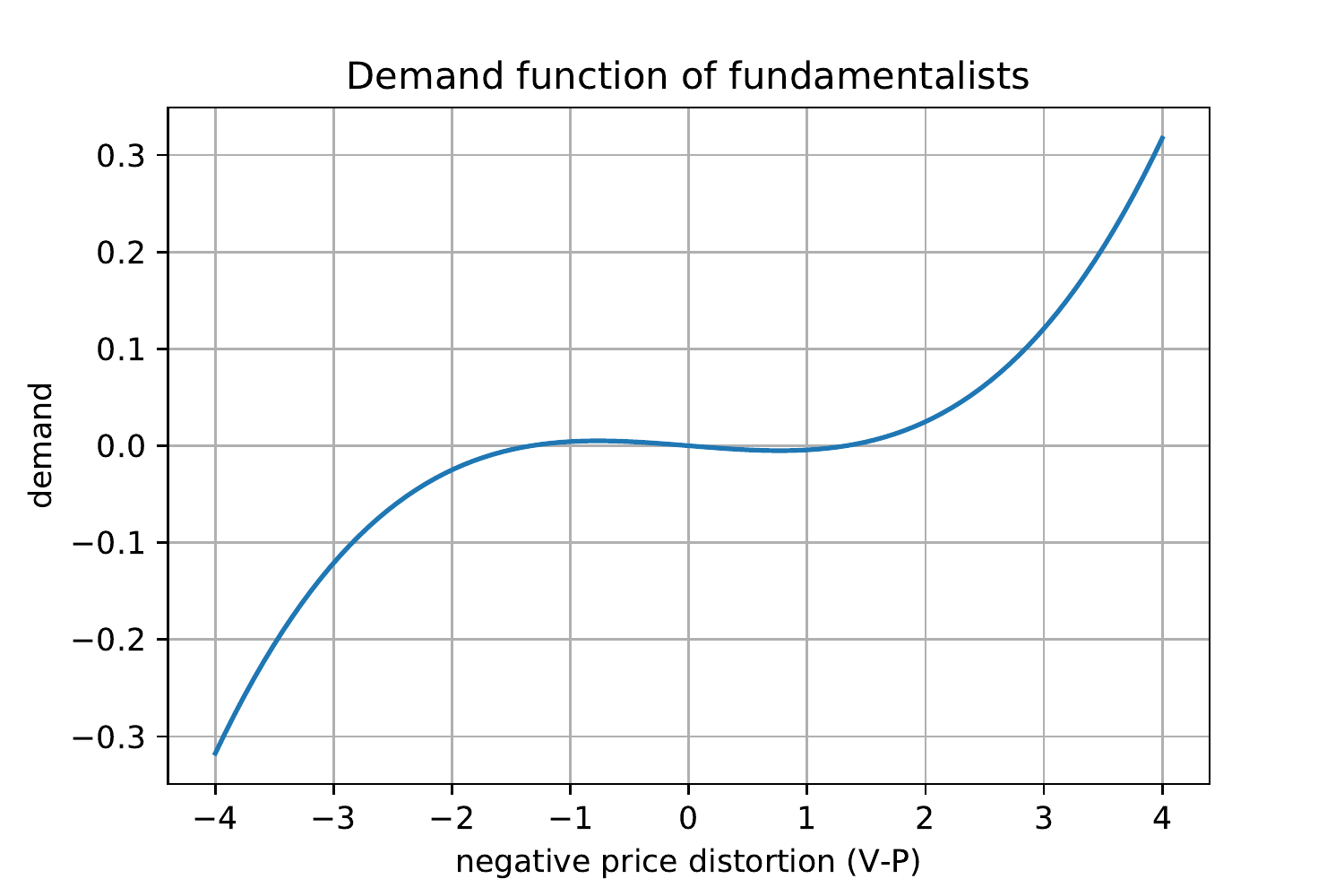} 
\caption{{\footnotesize Plot of $\bar{f}(x) := \frac{1}{N} \sum_{i=1}^{N} f^{(i)}\left( x \sigma_D^{(i)} \right)$, where $f^{(i)}$ is the estimated demand function of fundamentalists for asset $i$ ($f^{(i)}\left( x \right) := \kappa^{(i)} x + \kappa_3^{(i)}x^3$), $\sigma_D^{(i)}$ is the standard deviation of the price distortions of asset $i$, as implied by our non-linear model, and $N$ is the number of assets. Note that fundamental demand is extremely flat when mispricing is small, in agreement with intuition.}}\label{fig:demand} 
\end{center} 
\end{figure}

Table \ref{tab:estimation_direct} (see Appendix \ref{app:estimation_results}) presents the results of the direct maximum-likelihood estimation of model (\ref{eq:NonlinearModel}). In Figure \ref{fig:demand} we plot the demand function of fundamentalists, standardized and averaged over all assets. The linear parameter $\kappa$ is found to be {\it negative} and significant (see T-statistics in Table \ref{tab:tstat_direct} in Appendix \ref{app:estimation_results}). However, Figure \ref{fig:demand} shows that its impact on the full demand function $f(x)$ is small: the fundamental demand is essentially flat when $P \approx V$.\footnote{This suggests that a better two-parameter family of demand functions might be $f_{\mu}(x)=\kappa_{\mu} \mathrm{sign}(x) |x|^{\mu}$ with $\mu > 1$. Model with such a demand of fundamentalists has been proposed by \cite{IS2002} in a different context. We have not investigated this possibility as yet.\label{footnote:OtherDemand}} On the other hand the non-linear parameter $\kappa_3$ is positive and significant. It means, as anticipated, that fundamentalists tend to be very active when their estimate of value is sufficiently far from the market price, whereas they are nearly inactive when value and price are close to each other.

\begin{figure}[!h]
\begin{center}
\includegraphics[scale=0.7]{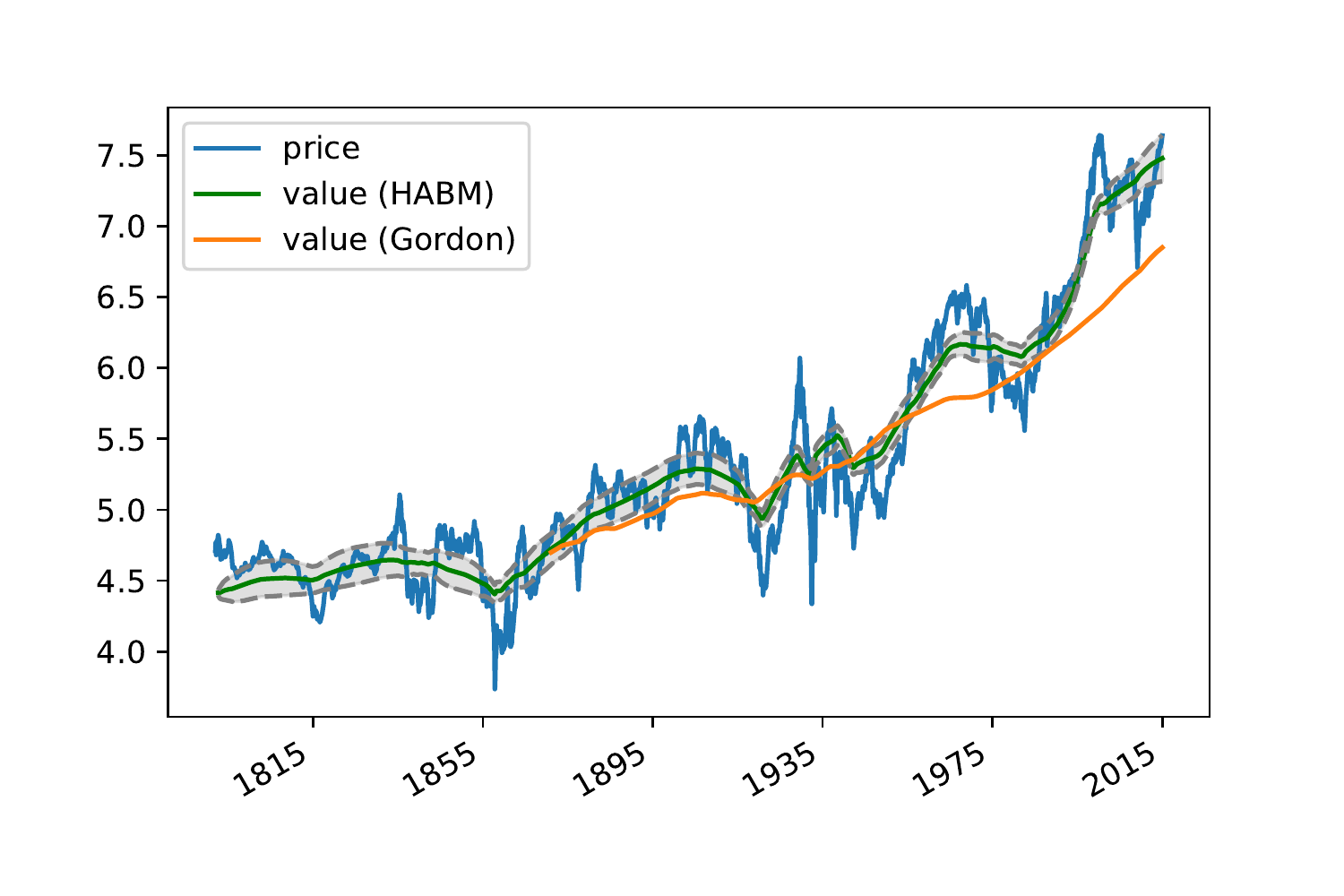} \caption{{\footnotesize Log-level of the US stock index, together with the smoothed fundamental value as inferred from the non-linear model (\ref{eq:ModelNL}) for the parameters given in Table \ref{tab:estimation_direct}. We also plot value plus/minus one standard deviation of the estimation interval, and the benchmark fundamental value obtained from Gordon model.}}\label{fig:US_NLDSP} 
\end{center} 
\end{figure}

In Figure \ref{fig:US_NLDSP} we plot the smoothed estimate of fundamental value of US stock index, using the Unscented Kalman smoother, as inferred from the non-linear model (\ref{eq:NonlinearModel}) for the parameters given in Table \ref{tab:estimation_direct}. Figures \ref{fig:US} - \ref{fig:USBND} in Appendix \ref{app:estimation_results} gives the smoothed estimates of value for several assets, given the estimated parameters of both the linear (\ref{eq:DiscreteModel}) and non-linear (\ref{eq:NonlinearModel}) models. We see that the fundamental value estimated with the extended model tracks the market price better than the one found using the linear version of the model. The reason is that the cubic demand function of fundamentalists acts as a strong mean-reverting force when mispricing becomes large, and therefore prevents such large excursions much more effectively than the linear model. Correspondingly, the most likely path of value is closer to the market price. We believe that this non-linear model makes more economic sense: not only is the idea of a steeply increasing demand enticing, but the final output of the model (in terms of average mispricing) is also more plausible.

\subsection{Trend and value effects}\label{sec:TrendValueEffectsNLDS}

\begin{figure*}[!h]
\centering
\begin{subfigure}[t]{0.5\textwidth}
\centering \includegraphics[scale = 0.5]{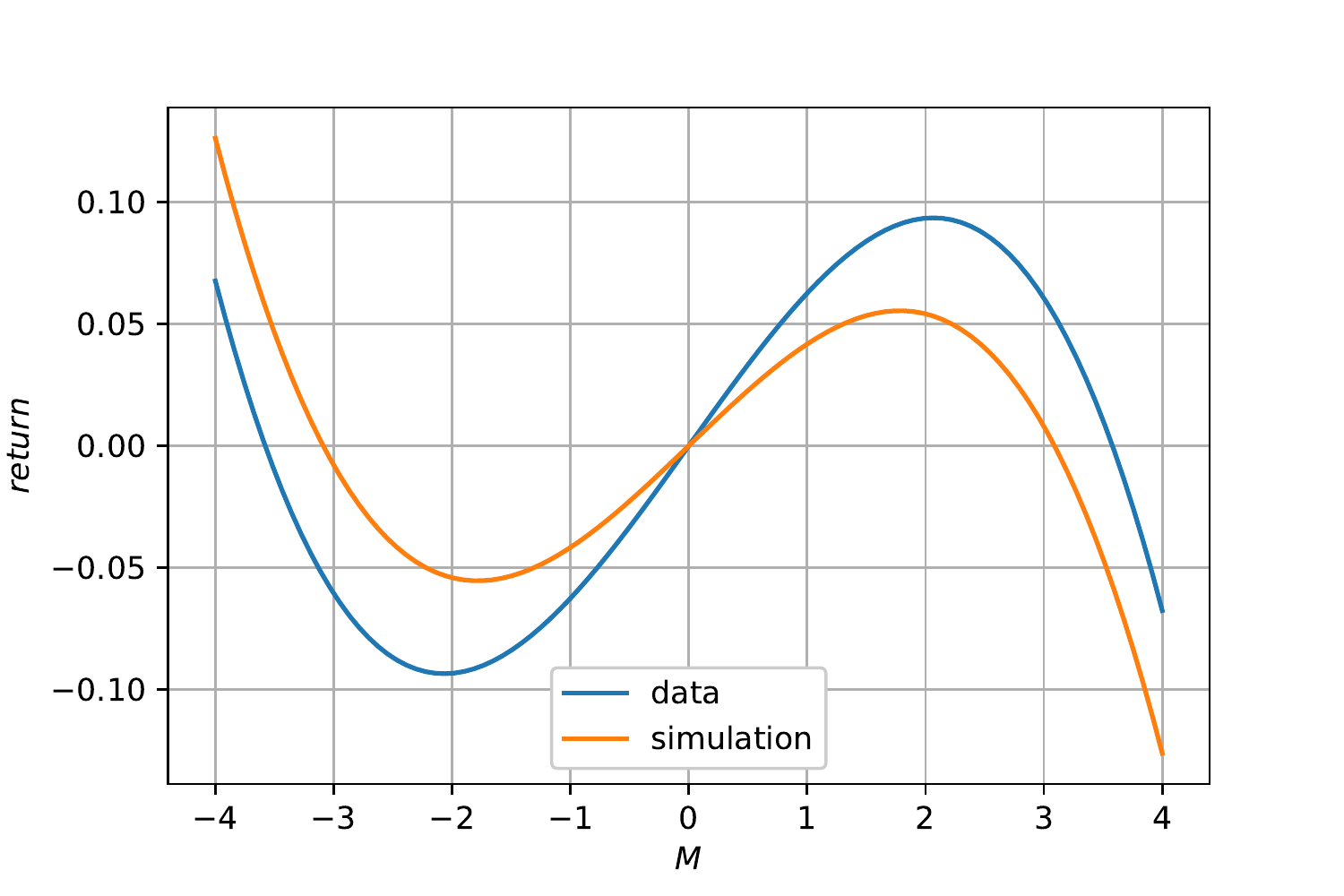}
\caption{{\footnotesize Trend effect}}\label{fig:trend_effectP}
 \end{subfigure}%
 ~
\begin{subfigure}[t]{0.5\textwidth}
\centering \includegraphics[scale=0.5]{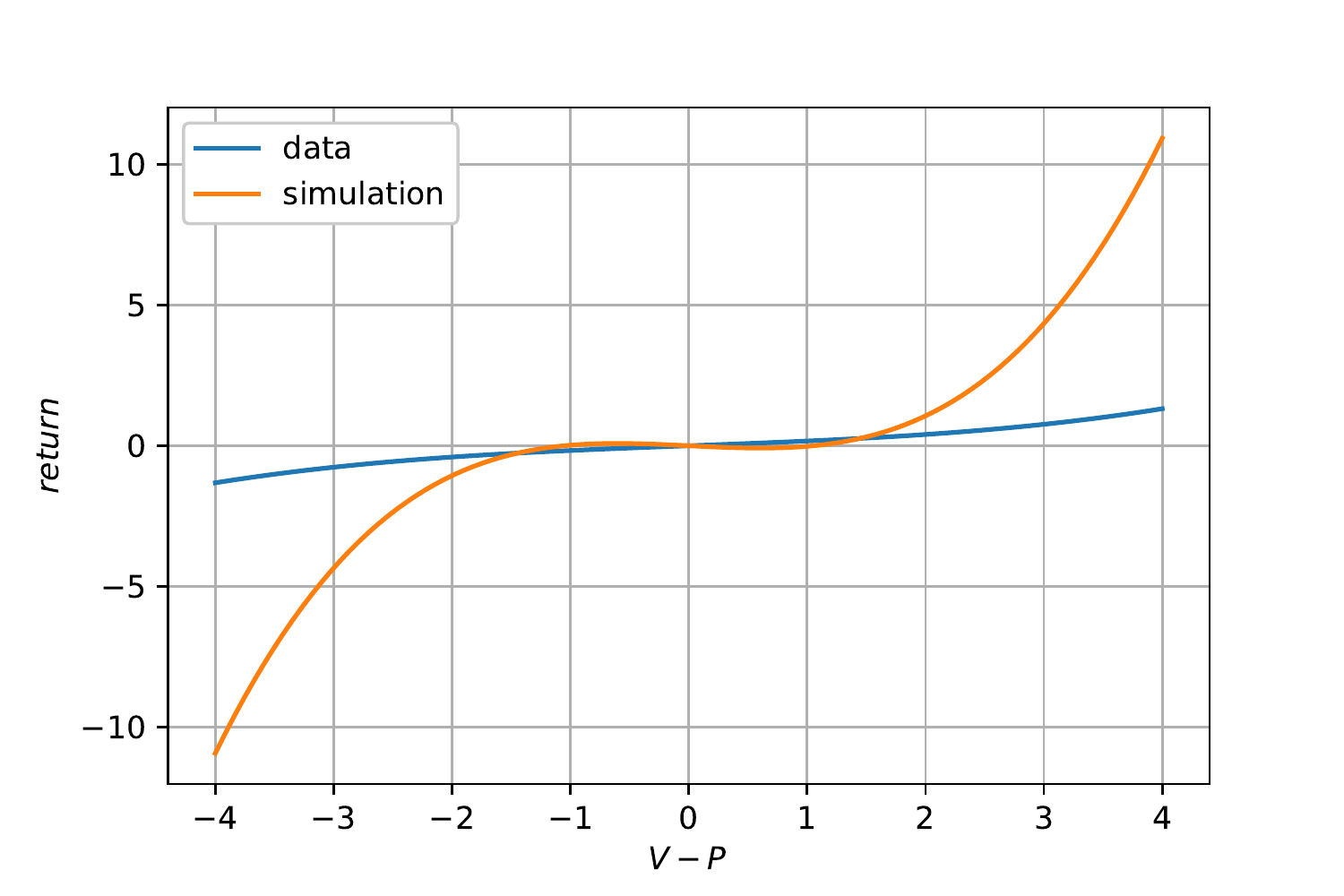}
\caption{{\footnotesize Value effect}}\label{fig:value_effectP}
\end{subfigure}
\caption{{\footnotesize Left plot: regression of log-returns on the trend signal $m_t$ for real data and simulated using the calibrated model. Right plot: regression of log-returns on the value signal $V - P$ for real data (with $V$ implied by the model) and simulated using the same model.}} 
\end{figure*}

\begin{figure*}[!h]
\centering
\begin{subfigure}[t]{0.5\textwidth}
\centering \includegraphics[scale = 0.5]{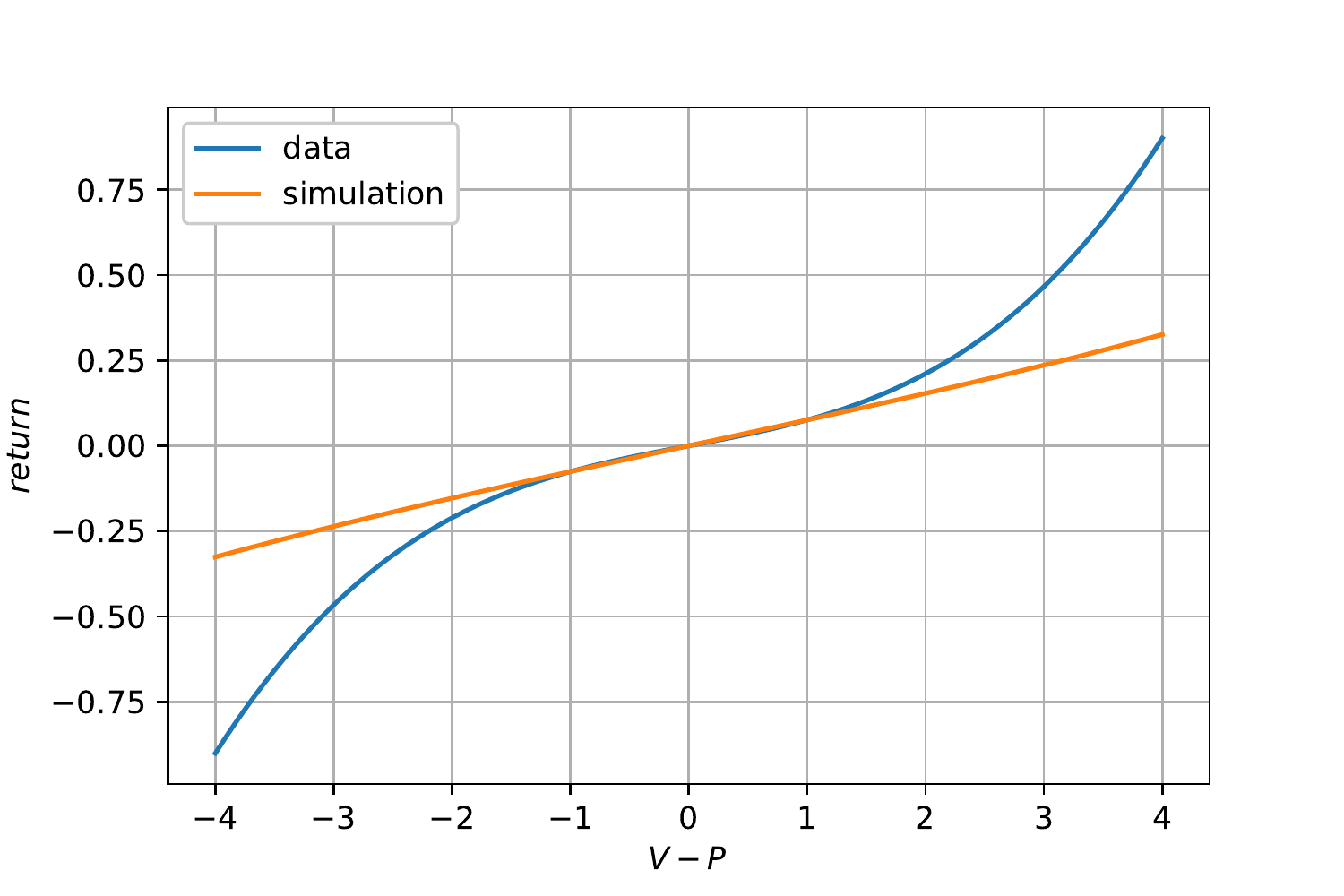}
\caption{{\footnotesize $\kappa_3 = 0$}}\label{fig:value_effect_IDX}
 \end{subfigure}%
 ~
\begin{subfigure}[t]{0.5\textwidth}
\centering \includegraphics[scale=0.5]{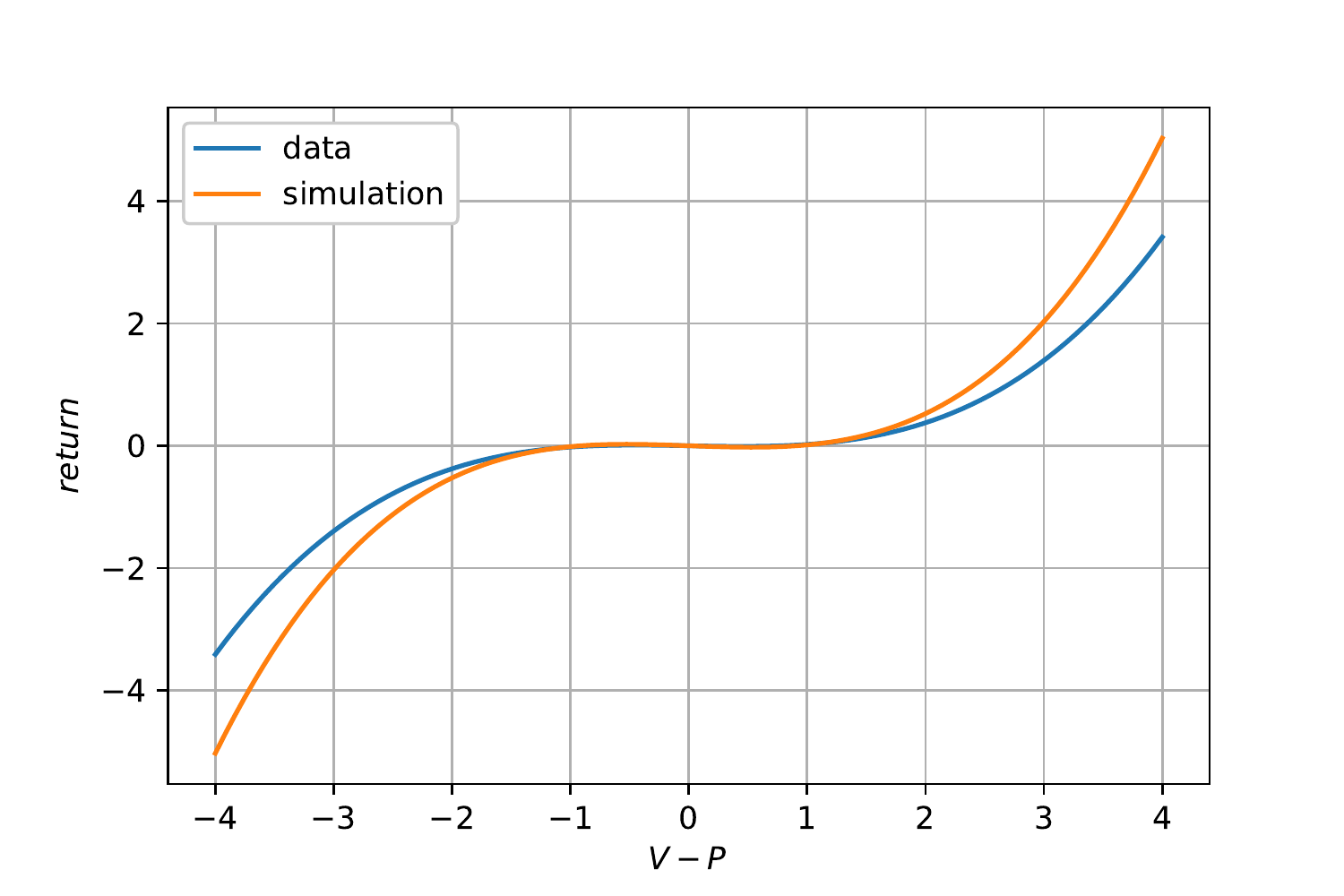}
\caption{{\footnotesize $\kappa_3 \neq 0$}}\label{fig:value_effectP_IDX}
\end{subfigure}
\caption{{\footnotesize Regression of log-returns of indices of equites on the value signal $V - P$ for real data (with $V$ implied by the model) and simulated using the model with linear demand of fundamentalists (left plot) and the model with non-linear demand of fundamentalists (right plot).}} 
\end{figure*}

Figures \ref{fig:trend_effectP}, \ref{fig:value_effectP} are the counterpart, for the non-linear model, of Figures \ref{fig:trend_effect}, \ref{fig:value_effect} in the linear case. The regressions have been obtained using the parameters obtained in Table \ref{tab:estimation_direct}. While the simulated trend effect is very similar to the one observed in the data, we now find that the non-linear model over-estimates the non-linearity of the fundamental demand for large values of $|V-P|$. However, when limited to indices of equities, we observe that non-linear model provides consistent estimated and simulated value effects (see Figure \ref{fig:value_effectP_IDX}), while linear model is not able to reproduce the non-linearity (see Figure \ref{fig:value_effect_IDX}), as we have observed for complete dataset in Section~\ref{sec:TrendValueEffects}. This again suggests that a different specification of the non-linear function $f(x)$ could be more adapted (for example $f_{\mu}(x)$ in footnote \ref{footnote:OtherDemand} with $\mu$ being asset specific).

As in Section~\ref{sec:TrendValueEffects}, we now proceed to the analysis of trend and value together. Comparing Tables \ref{tab:regressions} and \ref{tab:regressionsP}, we notice that the model with a non-linear demand of fundamentalists gives a higher adjusted R-square. This is further evidence that our extended model explains the price dynamics better.

\begin{table}
\begin{center}
{\footnotesize
\begin{tabular}{lrrrrrrlr}
\toprule
{} &  const &     $m$ &     $m^2$ &     $m^3$ &     $d$ &     $d^3$ &   &     $R^2$ \\
\midrule
\rowcolor{gray!30}  \textbf{coefficient} & -0.000 &  0.024 &    - &    - &    - &    - &    &  0.001 \\
\rowcolor{gray!30}  \textbf{P-value    } &  0.997 &  0.000 &    - &    - &    - &    - &    &    - \\
\textbf{coefficient} & -0.003 &  0.068 &  0.001 & -0.005 &    - &    - &    &  0.013 \\
\textbf{P-value    } &  0.631 &  0.000 &  0.710 &  0.000 &    - &    - &    &    - \\
\rowcolor{gray!30}  \textbf{coefficient} &  0.000 &    - &    - &    - &  0.202 &    - &    &  0.041 \\
\rowcolor{gray!30}  \textbf{P-value    } &  0.996 &    - &    - &    - &  0.000 &    - &    &    - \\
\textbf{coefficient} &  0.002 &    - &    - &    - &  0.158 &  0.011 &    &  0.048 \\
\textbf{P-value    } &  0.730 &    - &    - &    - &  0.000 &  0.000 &    &    - \\
\rowcolor{gray!30}  \textbf{coefficient} & -0.000 &  0.127 &    - &    - &  0.253 &    - &    &  0.054 \\
\rowcolor{gray!30}  \textbf{P-value    } &  0.990 &  0.000 &    - &    - &  0.000 &    - &    &    - \\
\textbf{coefficient} & -0.007 &  0.172 &  0.005 & -0.005 &  0.254 &    - &    &  0.067 \\
\textbf{P-value    } &  0.242 &  0.000 &  0.028 &  0.000 &  0.000 &    - &    &    - \\
\rowcolor{gray!30}  \textbf{coefficient} & -0.005 &  0.165 &  0.005 & -0.004 &  0.223 &  0.008 &    &  0.070 \\
\rowcolor{gray!30}  \textbf{P-value    } &  0.347 &  0.000 &  0.021 &  0.000 &  0.000 &  0.000 &    &    - \\
\bottomrule
\end{tabular}
\caption{{\footnotesize Regressions of log-returns on trend and value components for the model (\ref{eq:ModelNL}). In the last column we present the adjusted R square of the regression.}}\label{tab:regressionsP}}
\end{center}
\end{table}

Similar to the linear version of the model, the inversion of the trend effect for large signals morphs into a saturation when we control for value (that is, when we condition it to $P=V$), and the value effect controlled for trend becomes stronger for small price distortion. The resulting graphs are very similar to Figures \ref{fig:trend_effect_control} and \ref{fig:value_effect_control}, so we do not duplicate them here.

\section{The Distribution of Mispricings}\label{sec:StationaryDistribution}

Both the linear and non-linear models studied in the previous sections admit a stationary distribution for the {\it price distortion} (or mispricing) $\delta_t = P_t - V_t$ (remember that both $P$ and $V$ are logarithms). In this section we would like to investigate further the properties of this stationary distribution, both on real data and in simulation. Of particular interest are (i) the variance of this distribution, which gives us an indication of the typical amplitude of mispricings and (ii) the shape of the distribution (unimodal vs. bimodal).

\begin{figure*}[!h]
 \centering
 \begin{subfigure}[t]{0.5\textwidth}
 \centering \includegraphics[scale = 0.5]{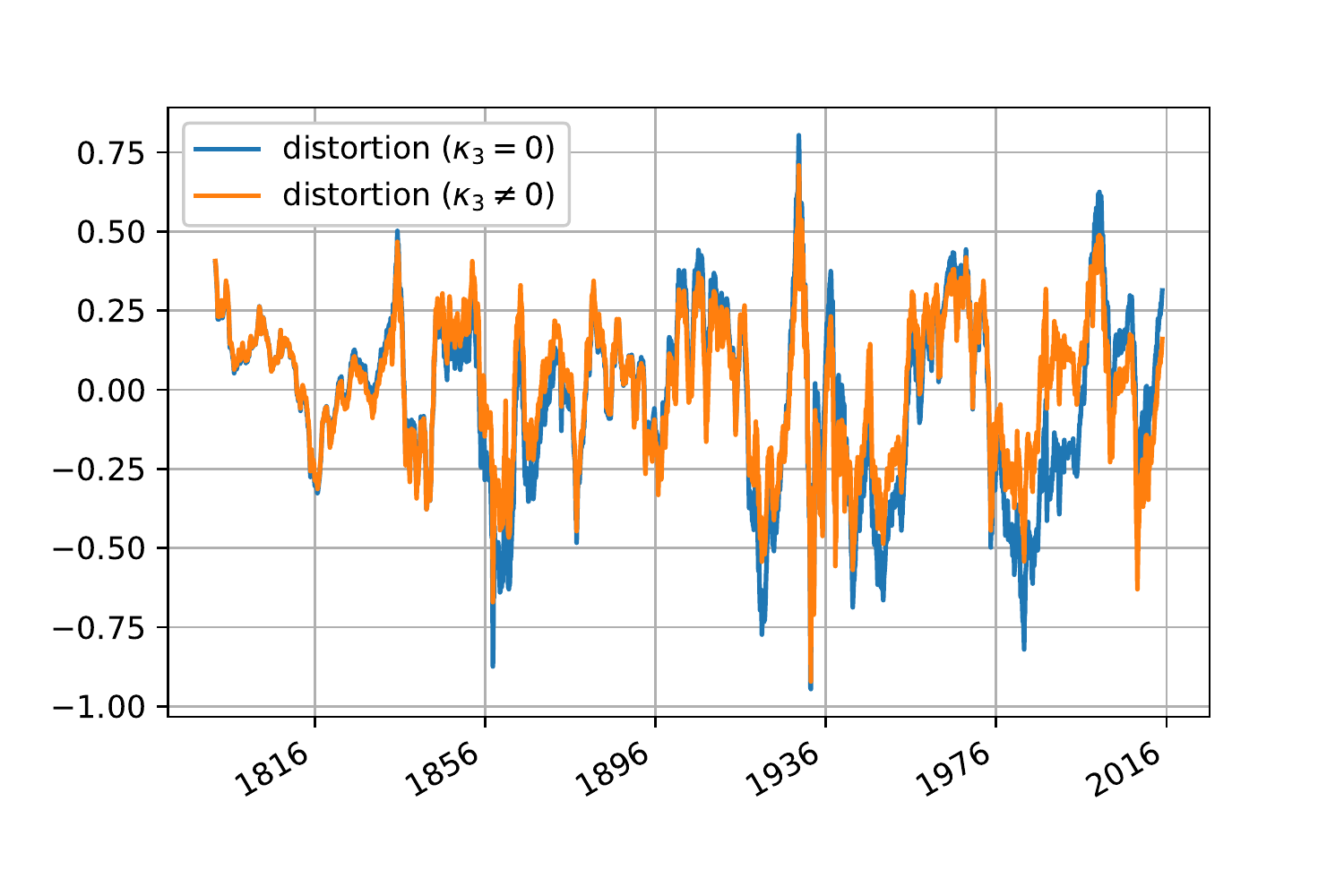}
 \end{subfigure}%
 ~
 \begin{subfigure}[t]{0.5\textwidth}
 \centering \includegraphics[scale=0.5]{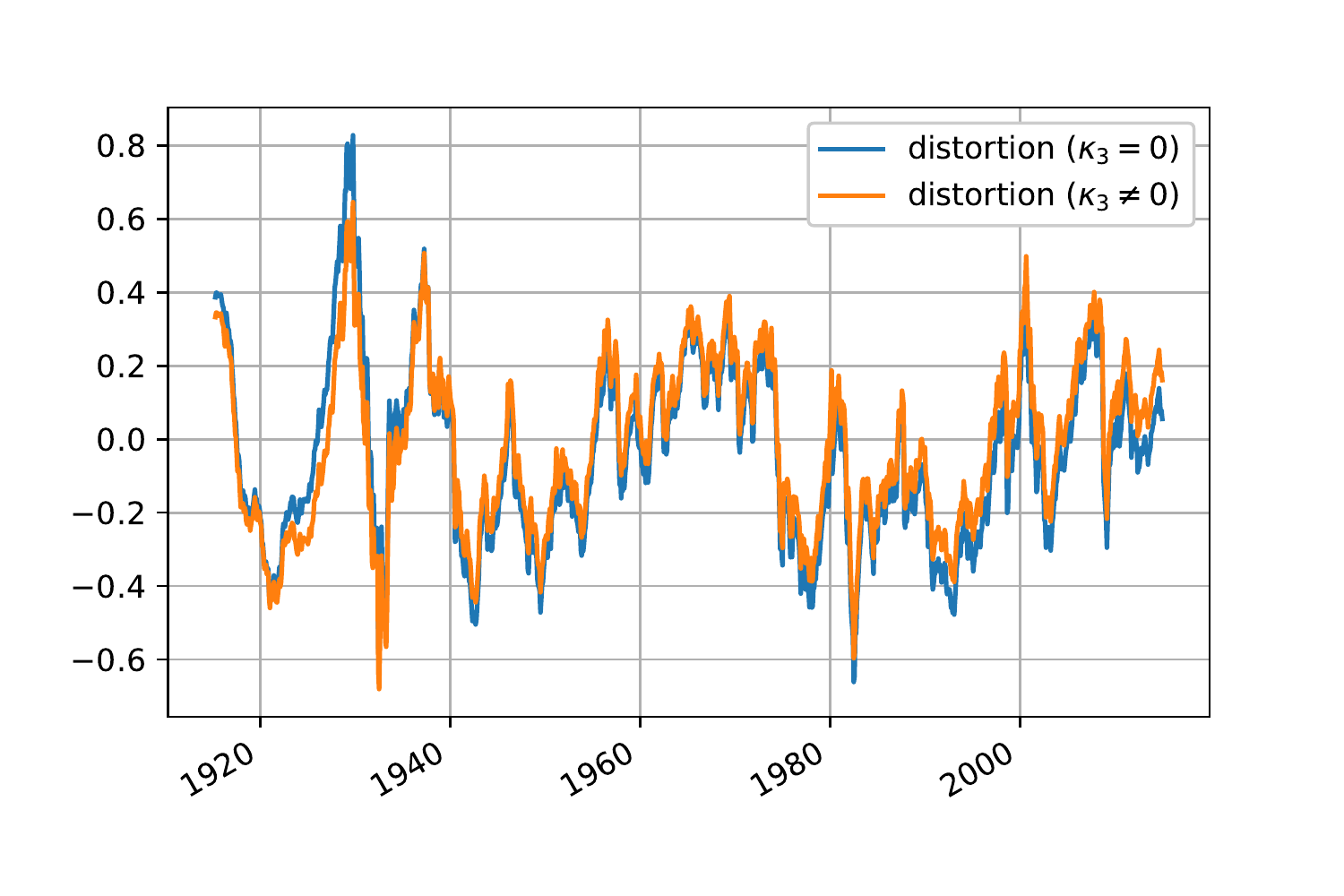}
 \end{subfigure}
 \caption{{\footnotesize (Log-)Price distortion time series for the US stock index (left) and Canadian stock index (right), for both models considered in this paper (linear and non-linear). Note that the mispricings are smaller for the non-linear model, as explained in the previous section.}}
\label{fig:distortion_data} 
\end{figure*}


\cite{Black1986} conjectured that ``price is more than half of value and less than twice value''. This conjecture was recently confirmed empirically by \cite{BCLMSS2017}. In Figure \ref{fig:distortion_data} we plot as an illustration the time series of the log-price distortion for the US and Canadian stock indices. Since the distortion is in natural log scale, a value of $0.5$ on the y-axis corresponds to a mispricing ratio of $e^{0.5} = 1.65$. We find that variance of price distortion implied by the linear model (\ref{eq:DiscreteModel}), averaged over all assets, is equal to $0.233$, corresponding to a root-mean-square distortion of $0.5$, precisely as reported in \cite{BCLMSS2017}. The non-linear model (\ref{eq:NonlinearModel}) leads to a slightly smaller variance of $0.215$, as expected from the discussion of the previous section.
 
\begin{figure*}[!h]
\begin{subfigure}[t]{0.5\textwidth}
\includegraphics[scale = 0.5]{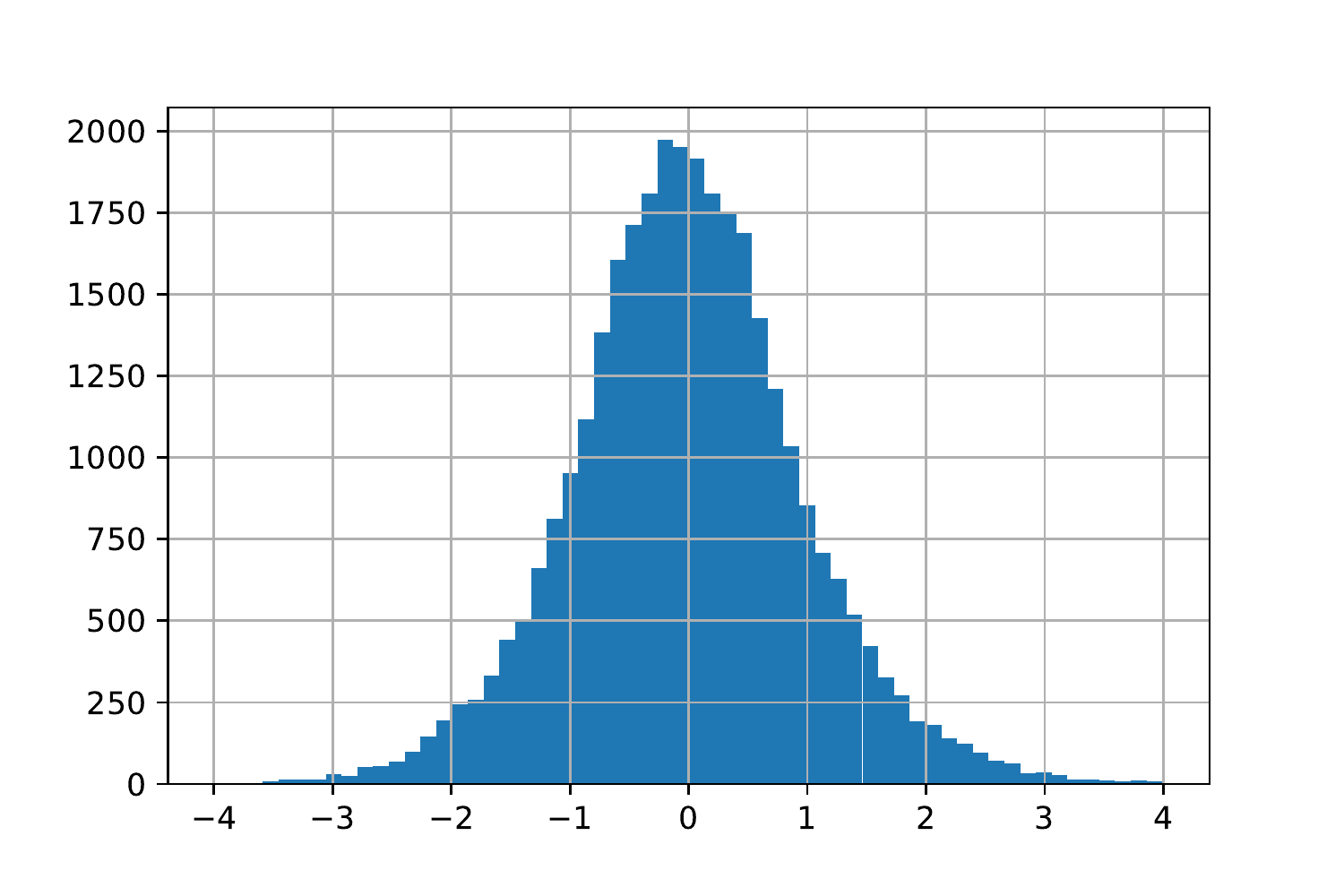}
\end{subfigure}
\begin{subfigure}[t]{0.5\textwidth}
\includegraphics[scale=0.5]{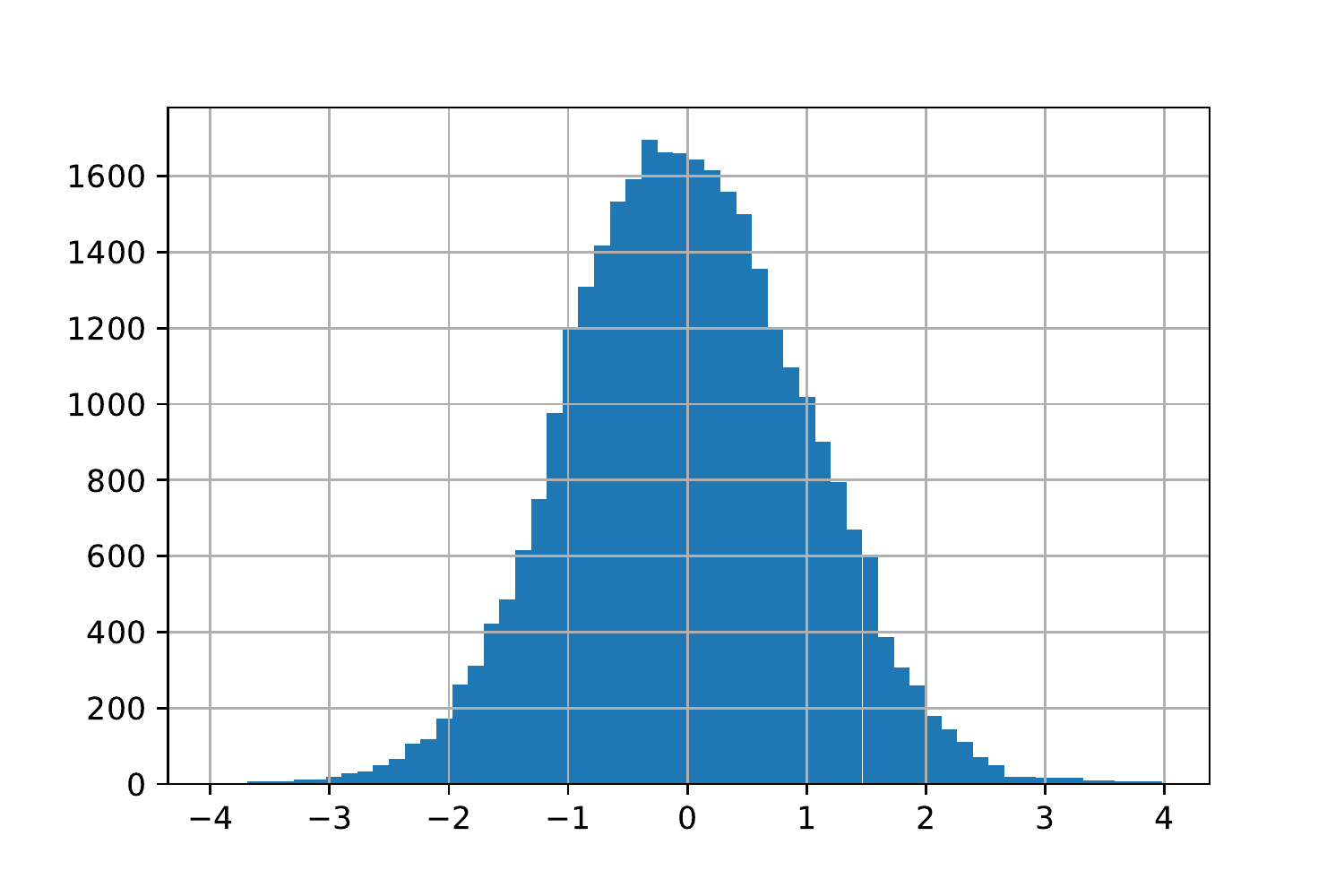}
\end{subfigure}
\caption{{\footnotesize Model-implied histogram of price distortions, averaged over all assets. Left: linear model (\ref{eq:Model}); right: non-linear model (\ref{eq:ModelNL}).}}\label{fig:bimodality_data} 
\end{figure*}

In Figure \ref{fig:bimodality_data} we plot the histogram of price distortions, averaged over all assets, obtained using the two models considered in this paper. We can see that both of them lead to a unimodal shape. That means that the market price is most likely to be near the fundamental value of the asset, albeit with a large dispersion, as noted above.

\begin{figure*}[!h]
 ~
 \begin{subfigure}[t]{0.5\textwidth}
 \centering \includegraphics[scale=0.45]{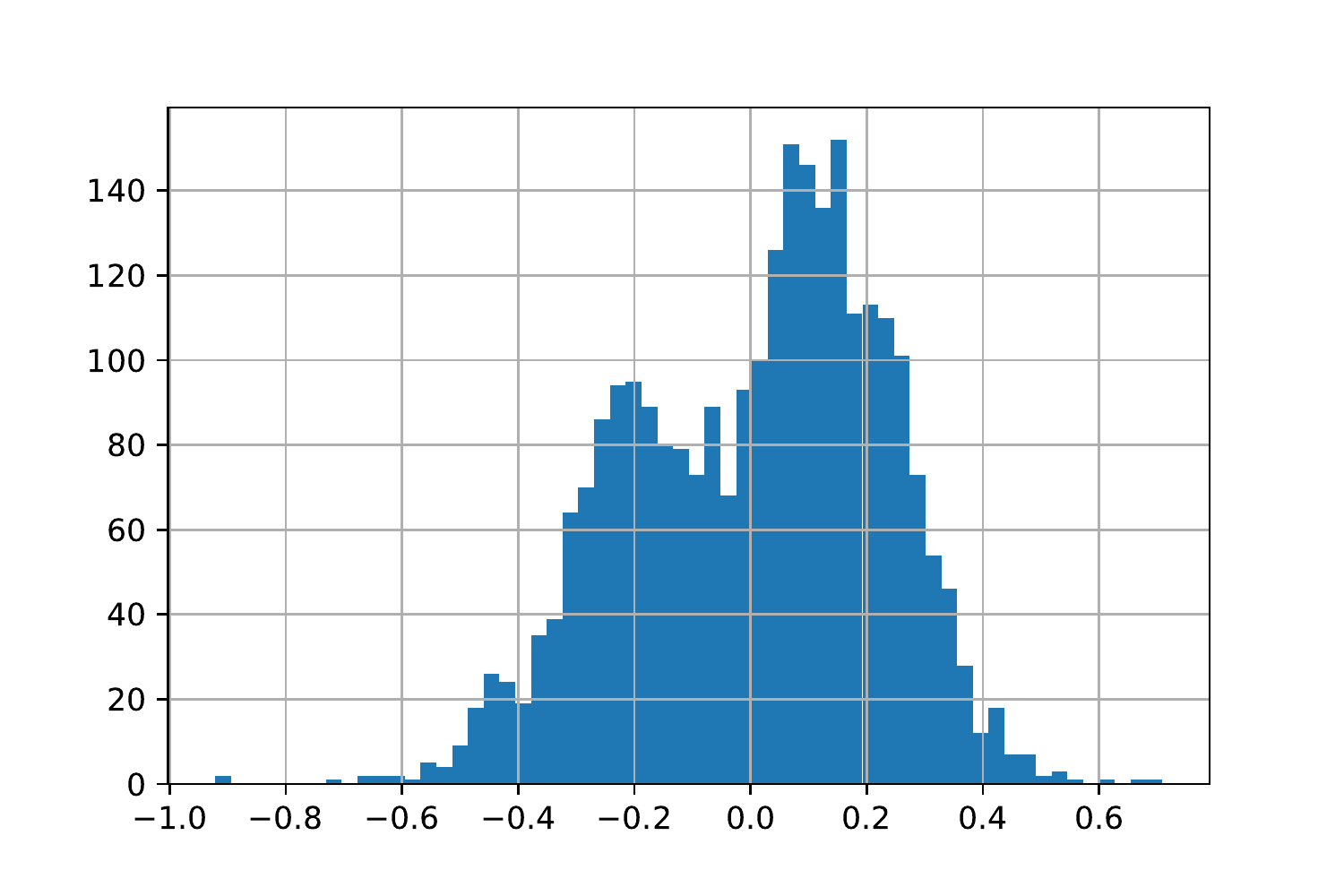}
 \end{subfigure}
 ~
 \begin{subfigure}[t]{0.5\textwidth}
 \centering \includegraphics[scale=0.45]{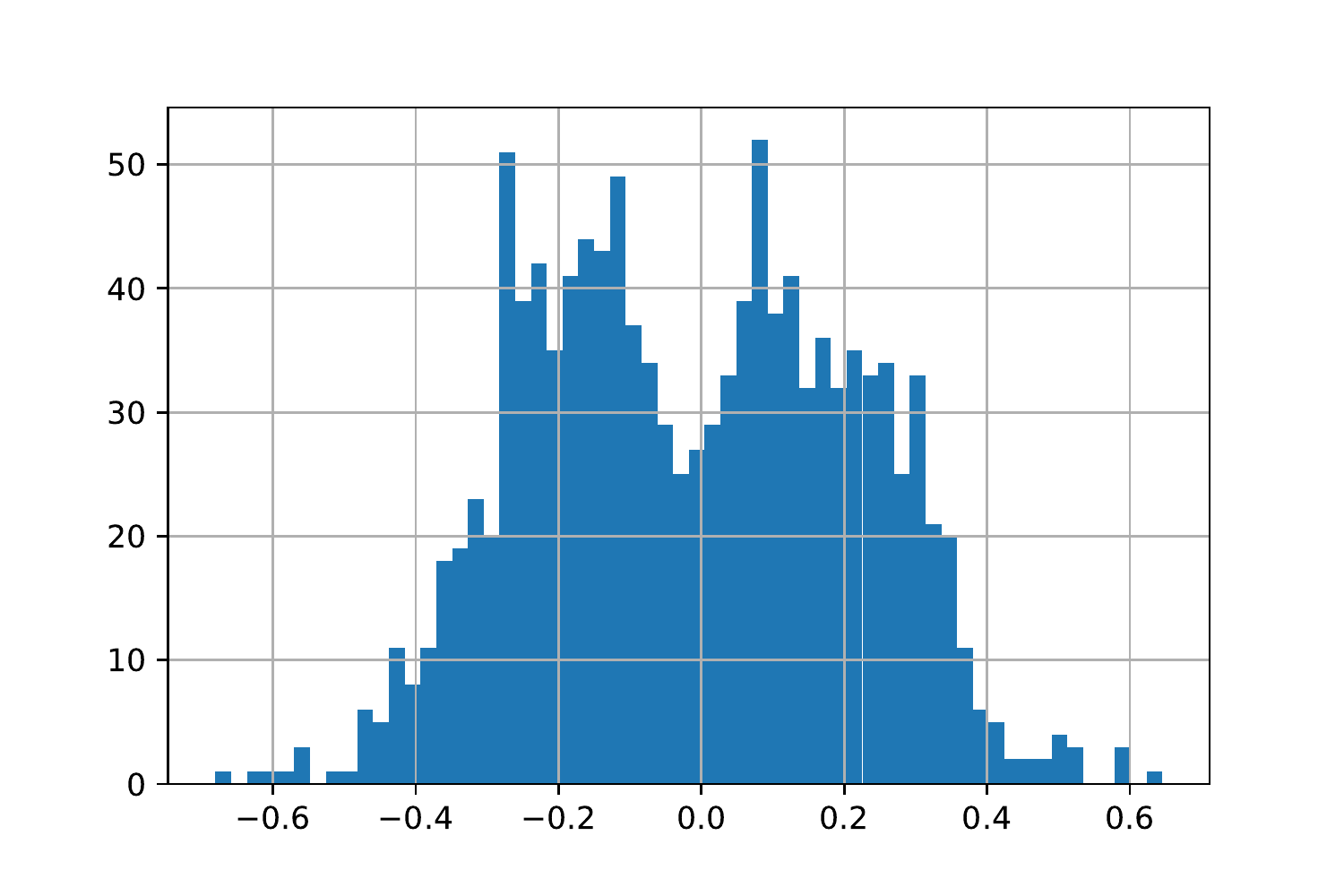}
\end{subfigure}
\caption{{\footnotesize Histogram of the price distortion for US stock index (left) and Canadian stock index (right), using the non-linear model (\ref{eq:ModelNL}).}}\label{fig:bimodality_data_spec} 
\end{figure*}

However, when restricting to certain assets, such as the US stock index or the Canadian stock index, we observe hints of bimodality (see Figure \ref{fig:bimodality_data_spec}). This is a somewhat intriguing observation since it suggests that the market's most likely state is being over-priced or under-priced. This was also recently reported in \cite{SW2017}.

We statistically confirm the bimodality of mispricing distribution by applying \cite{Silverman1981} test.\footnote{We apply the R package \texttt{silvermantest}, which is an implementation of  \cite{Silverman1981} test taking into account modification suggested by \cite{HY2001} in order to prevent it from being too conservative.} The null hypothesis of the test is that the investigated distribution has at most $k$ modes, where $k$ is a parameter of the test. The test rejects the null hypothesis that distribution of mispricing for US equity index has at most one mode ($p$-value is $1.5\%$) while the null hypothesis of the distribution having at most two modes cannot be rejected ($p$-value is $60.4\%$). We receive similar results for the distribution of mispricing of Canadian stock index - $p$-value equals $0.1\%$ for null hypothesis of at most one mode and $p$-value equals $90.6\%$ for null hypothesis of at most two modes.  Consequently, at $5\%$ significance level we reject the hypothesis that both empirical distribution plotted on Figure \ref{fig:bimodality_data_spec} have one mode and we cannot reject the hypothesis that they have at most two modes, which suggests that the distributions of mispricing of US and Canadian index are bimodal.

\begin{figure*}[!h]
\centering
 \begin{subfigure}[t]{0.5\textwidth}
\centering 
\includegraphics[scale = 0.45]{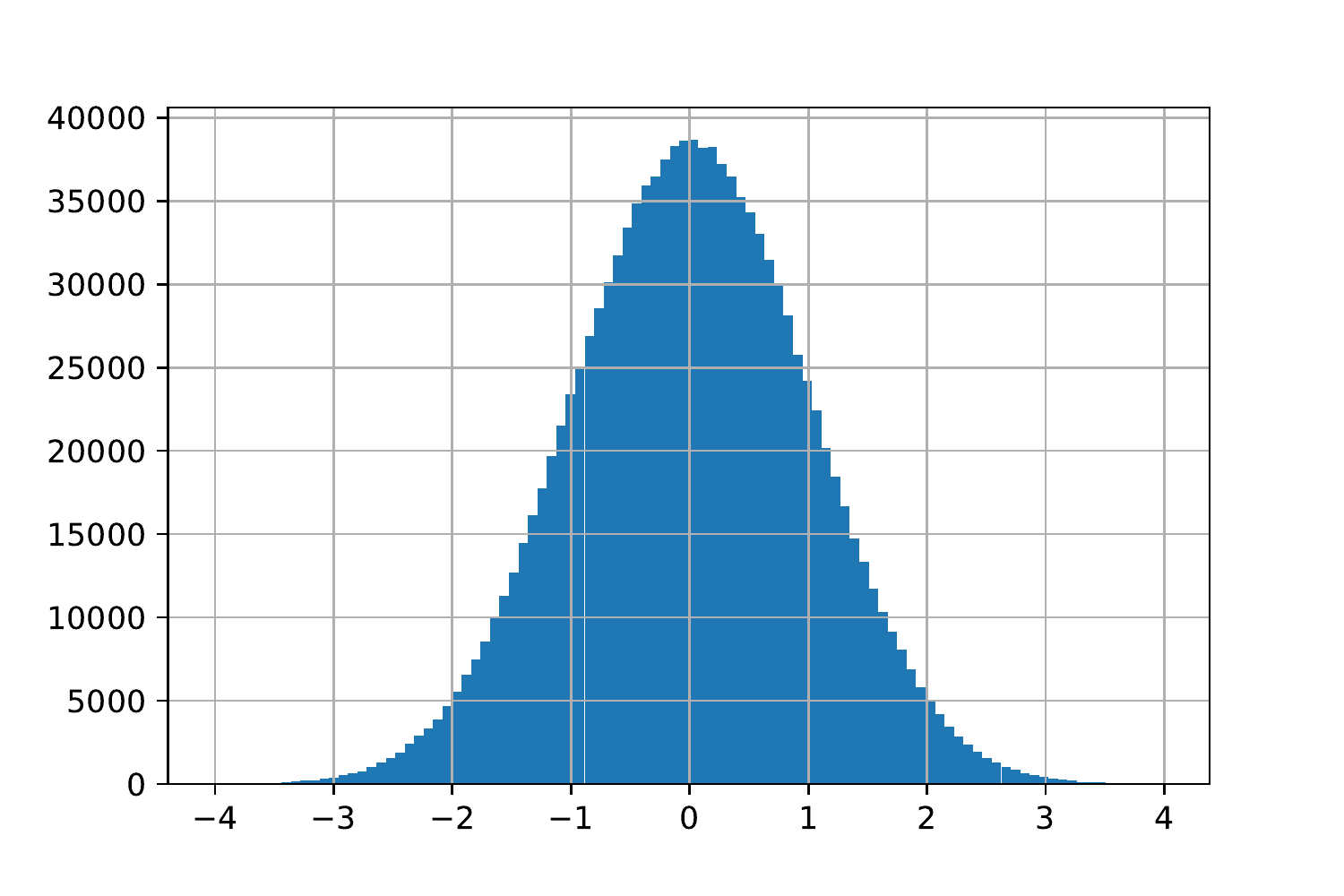}
\end{subfigure}%
 ~
\begin{subfigure}[t]{0.5\textwidth}
\centering 
\includegraphics[scale=0.45]{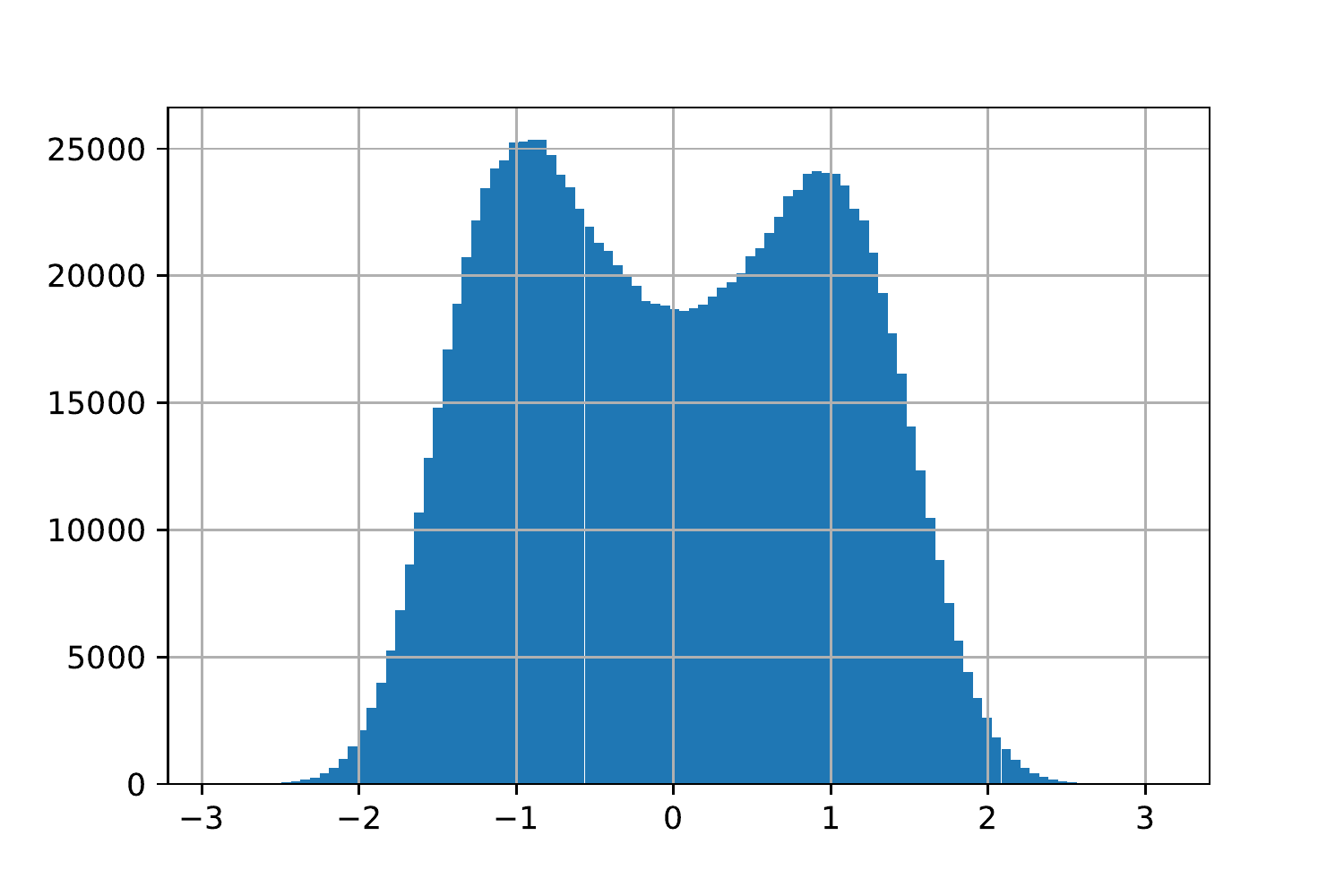}
 \end{subfigure}
 \caption{{\footnotesize Simulated histogram of price distortion. The left plot correspond to model (\ref{eq:Model}) and the right to model (\ref{eq:ModelNL}). We use parameters from Tables \ref{tab:estimation_EM} and \ref{tab:estimation_direct}.}}\label{fig:bimodality_sim} 
\end{figure*}

Note that the stochastic dynamic systems described by (\ref{eq:Model}) or (\ref{eq:ModelNL}) indeed undergo a phenomenological bifurcation (P-bifurcation) in parameter space, which means a qualitative change in the stationary distribution of mispricing from unimodal to bimodal. Since the Fokker-Planck equation associated with those systems does not have a known solution, one has to use approximation methods to find the set of parameters for which the bifurcation occurs.  The result of the analysis of \cite{CHWZ2008} and \cite{CHZ2011} is the following condition for P-bifurcation: when $\alpha + \kappa - \alpha \gamma \beta < 0$ the stationary distribution of $(\delta_t, M_t)$ has a crater shape, otherwise it has a single peak. Observe that it is the same condition as for the existence of limit cycles in the deterministic case (see Section \ref{sec:oscillations}). The crater shape distribution of the two-dimensional system corresponds to a bimodal distribution of mispricings.\footnote{The phase diagram of model (\ref{eq:ModelNL}) is richer and it has not yet been studied analytically. Obviously if $\kappa < 0$ and $\kappa_3 > 0$, then the stationary distribution of mispricing is bimodal, see \cite{BBDG2018}, Chapter 20.} In Figure \ref{fig:bimodality_sim} we plot the histogram of simulated price distortion distribution for both models. Since $\alpha + \kappa - \alpha \gamma \beta > 0$ for all assets, we observe a unimodal distribution for model (\ref{eq:Model}). For model (\ref{eq:ModelNL}) we observe a bimodal distribution due to negative value of $\kappa$.

\begin{figure*}[!h]
\centering
\begin{subfigure}[t]{0.32\textwidth}
\centering 
\includegraphics[scale = 0.32]{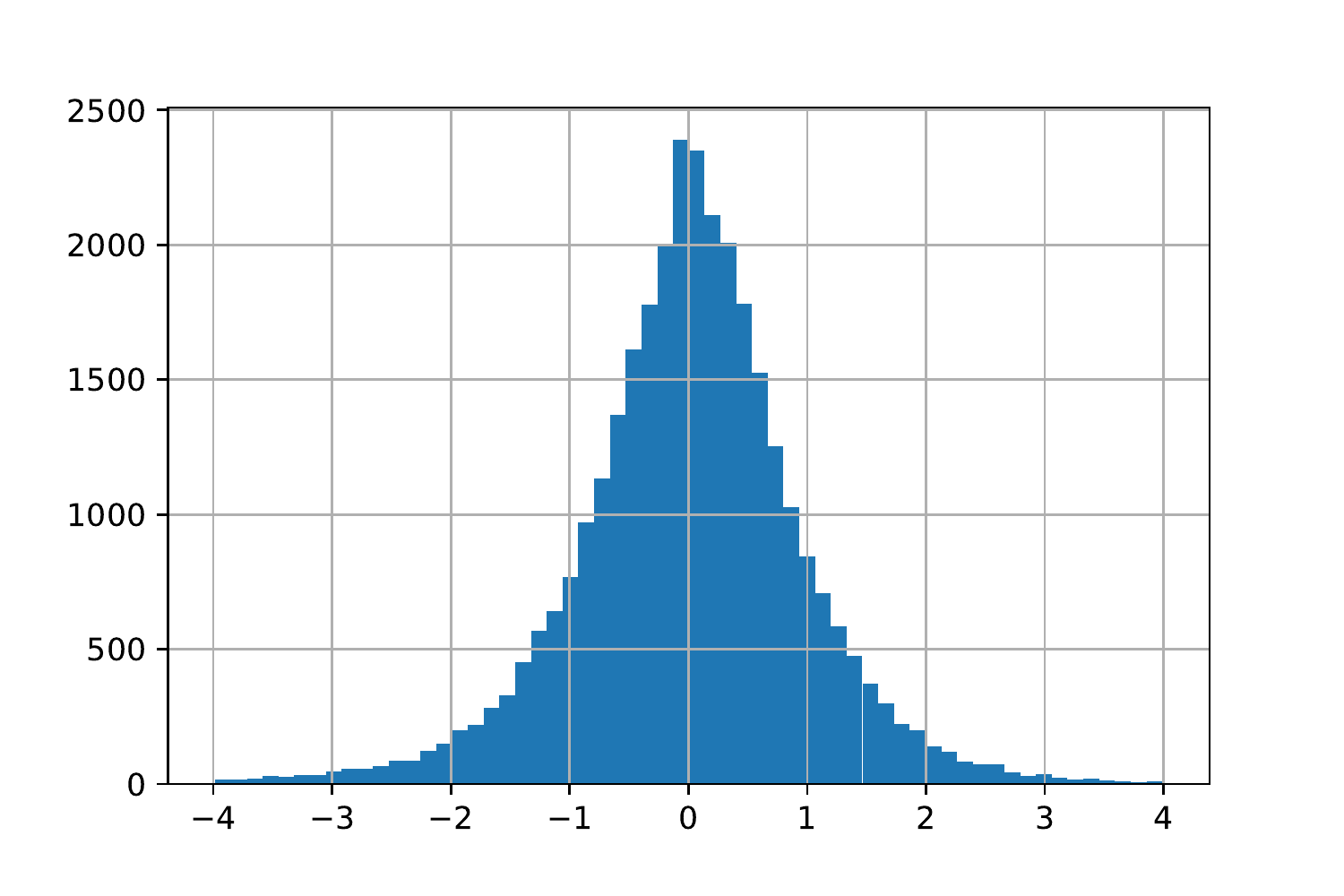}
 \end{subfigure}
 \begin{subfigure}[t]{0.32\textwidth}
 \centering \includegraphics[scale=0.32]{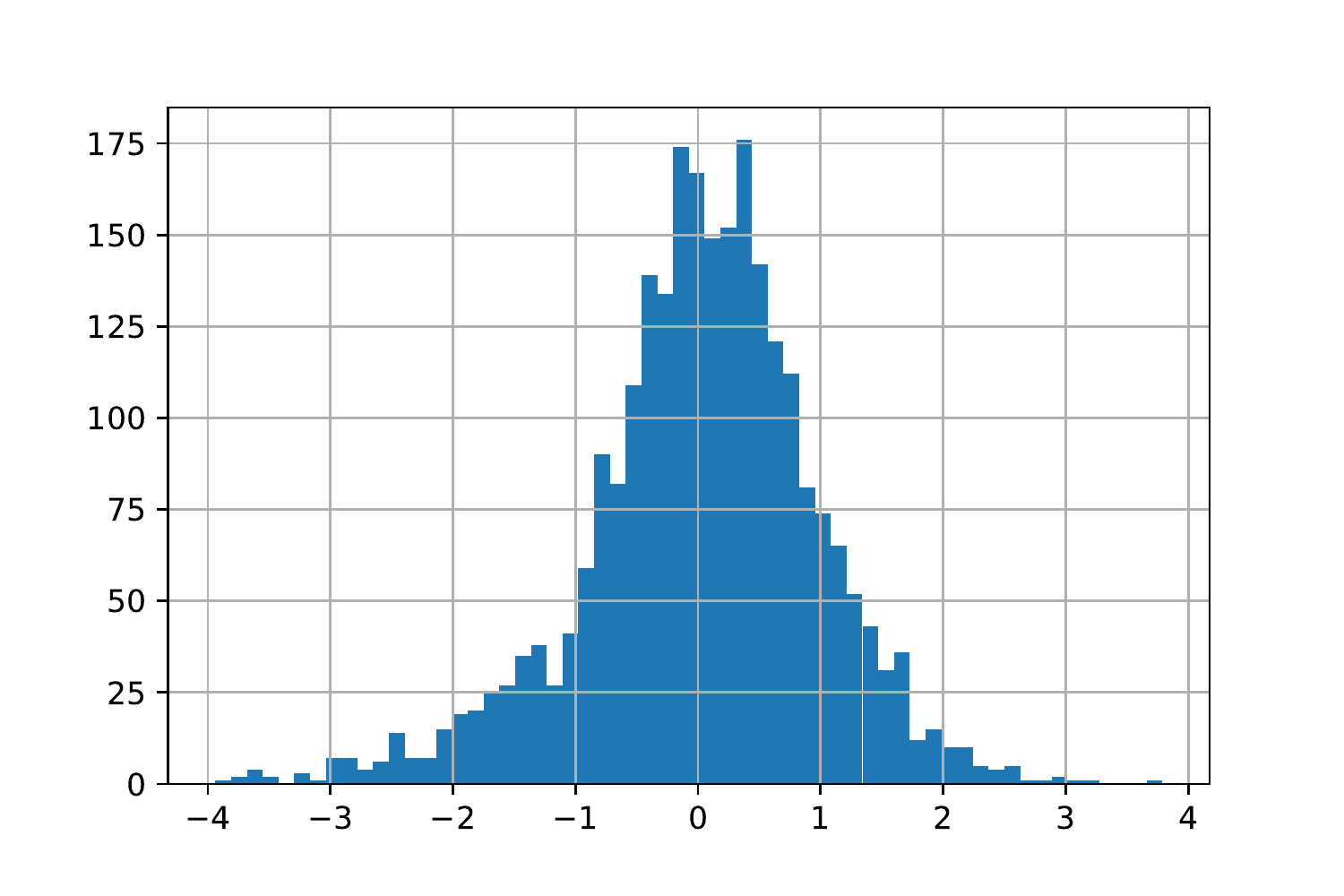}
 \end{subfigure}
 \begin{subfigure}[t]{0.32\textwidth}
 \centering 
\includegraphics[scale=0.32]{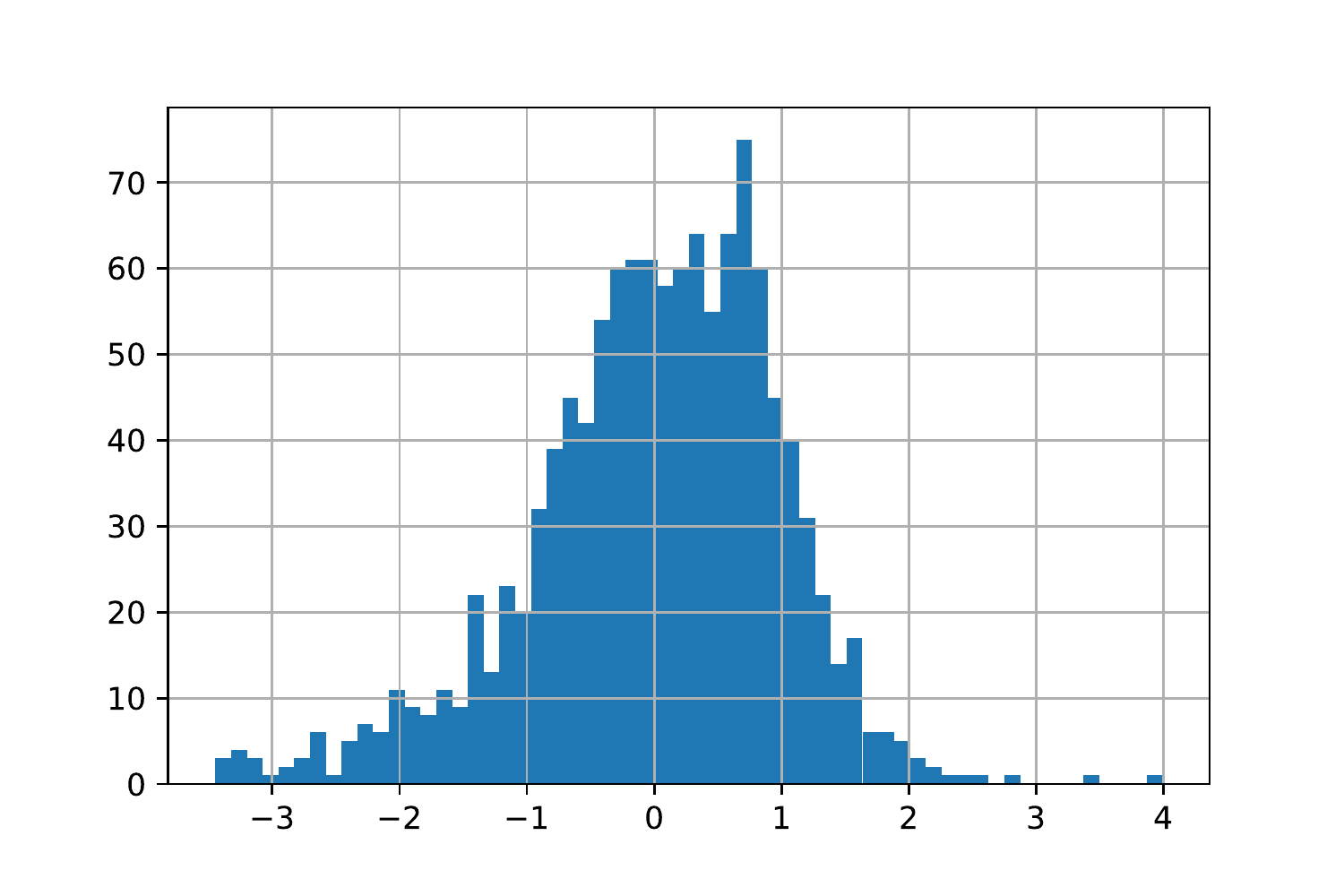}
\end{subfigure}
\caption{{\footnotesize Distribution of trend signal from the spot prices. On the left: all assets; in the middle: US index since 1900; on the right: Canadian index.}}\label{fig:histogram_trend} 
\end{figure*}

\begin{figure*}
\centering
\begin{subfigure}[t]{0.5\textwidth}
\centering 
\includegraphics[scale = 0.45]{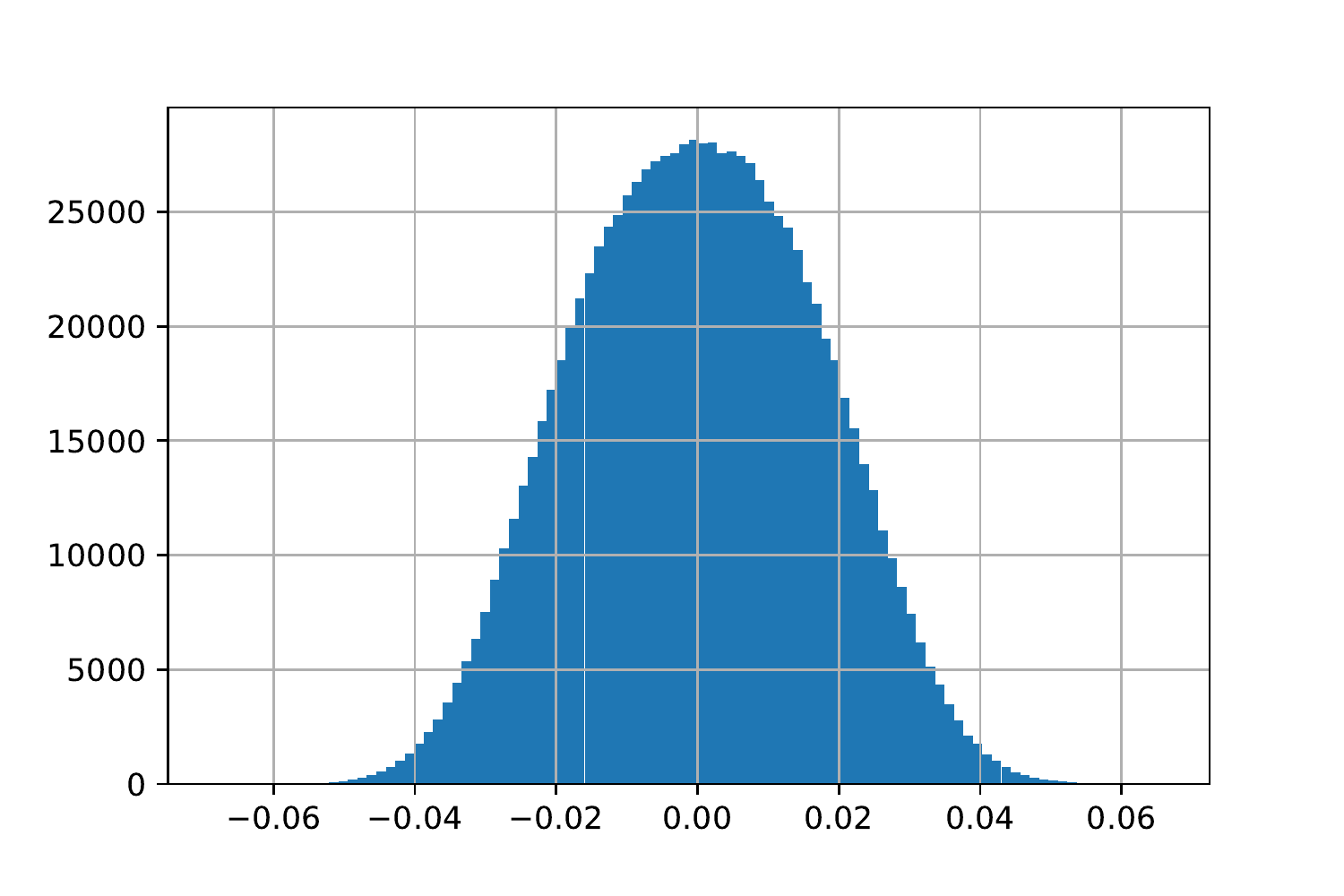}
 \end{subfigure}%
 ~
 \begin{subfigure}[t]{0.5\textwidth}
 \centering 
\includegraphics[scale=0.45]{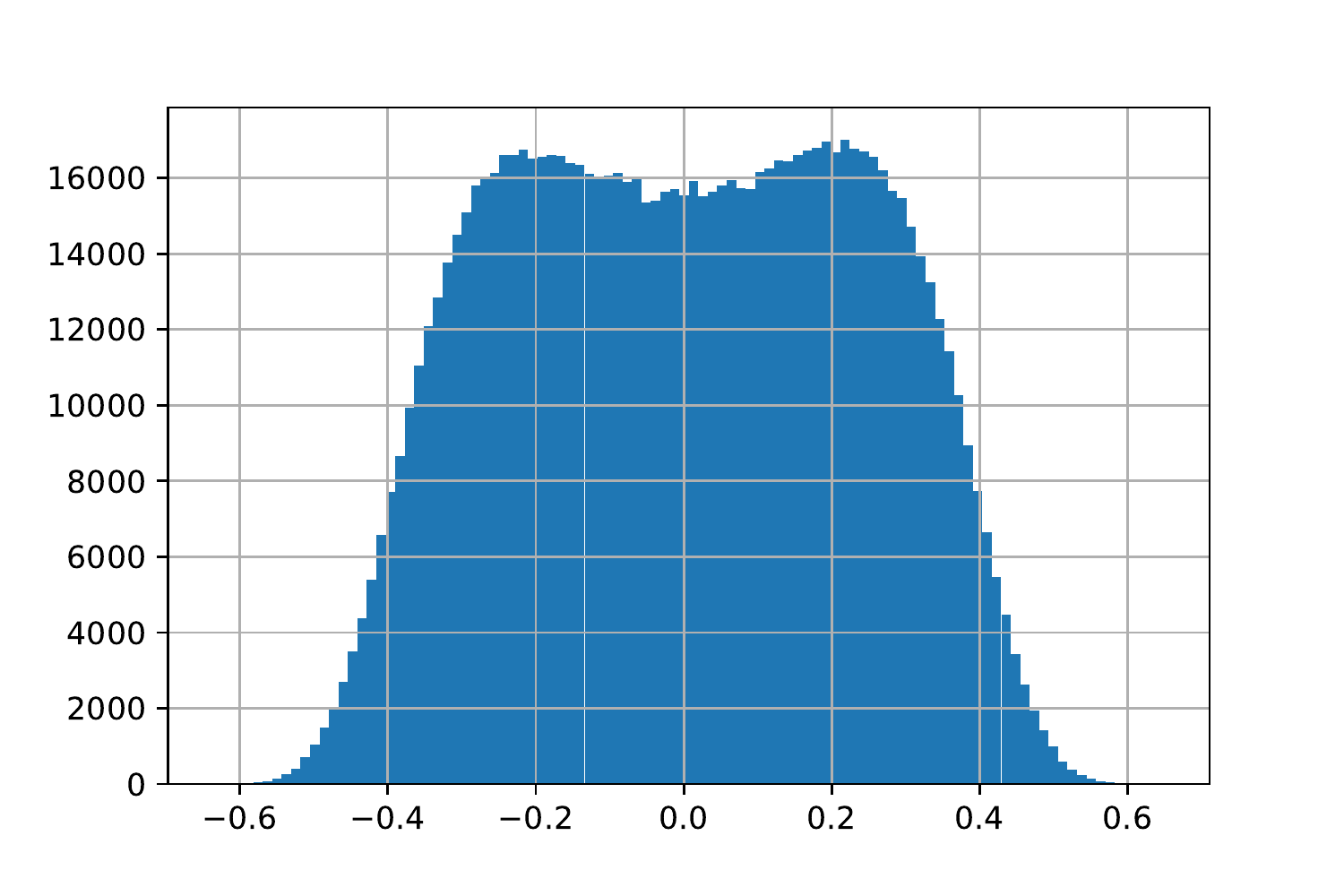}
\end{subfigure}
\caption{{\footnotesize Histogram of simulated trend signal (left) and price distortion (right) for model (\ref{eq:NonlinearModel}) with parameters: $\kappa = 0.0$, $\kappa_3 = 0.4$, $\beta = 0.03$, $\gamma = 50.0$, $\sigma_N =0.04$, $\sigma_V=0.02$, $g=0.001$, $v_0 = 5.0$.}}\label{fig:bimodality_theory} 
\end{figure*}

However, note that for the basic version of the model (\ref{eq:Model}), bimodality of mispricing always goes hand in hand with bimodality of the trend signal  (see \cite{CHWZ2008} and \cite{CHZ2011}). This co-existence of these two bimodalities is {\it not} observed in the data - see Figure \ref{fig:histogram_trend}. The extended model (\ref{eq:ModelNL}) does not have this limitation. When $\kappa < 0$ and $\kappa_3 > 0$ (like in the right panel of Figure \ref{fig:bimodality_sim}) one clearly finds a bimodal distribution of distortion and a unimodal distribution of trends. But even if both $\kappa$ and $\kappa_3$ are non-negative we can obtain bimodal distribution of price distortion together with unimodal distribution of trend signal (see Figure \ref{fig:bimodality_theory}). This is another reason for preferring the model with a non-linear demand of fundamentalists.

\section{Conclusions}\label{sec:Conclusions}

In this paper, we have reconsidered the coexistence of trend and value anomalies in financial markets from the point of view of Heterogeneous Agent Based Models. The field is already quite old, with, in particular, seminal papers by \cite{Chiarella1992} and \cite{LM1999}. Our major contribution is the estimation of a generalized Chiarella model on four asset classes (equity indices, commodities, FX rates and government bonds). In particular, our calibration procedure allows us to estimate the (unobservable) fundamental value of our different assets without having to rely on an external economic pricing model. In other words, the reference point around which the market price oscillates is an output of the calibration. 

Our extended Chiarella model comes in two versions: one where the fundamentalists' demand function is linear in the mispricing, and one where a non-linear (cubic) correction is considered. We find that, while capturing most of the phenomenology of the price dynamics, the linear model fails on several (related) counts: i) the linear model is not self-consistent as a clear non-linearity of the fundamental demand remains after calibration; ii) the fundamental value extracted from the linear model appears not to track closely enough the market price, a discrepancy partially corrected by a non-linear demand (although the variance of the mispricing is still large and compatible with Black's factor 2); iii) the distribution of mispricings is bimodal for some assets, as recently pointed out by \cite{SW2017}, while the distribution of trend signals remains unimodal. This feature is impossible to reproduce within a linear model but can be accommodated within the non-linear model. We have given several plausible, intuitive arguments for the existence of non-linearities in the fundamentalists' demand function, that lead to a bimodality in the distribution of price distortions. If confirmed by independent research, such bimodality would be a spectacular stigma of the failure of the efficient market hypothesis, since the most probable state of affairs would be that assets are either over-valued or under-valued by the market for long periods of time. 

The proposed model offers an intuitive explanation of mechanism producing the non-monotonic shape of trend effect observed by \cite{BCLMSS2017}. When we control trend effect for value (i.e. we condition price distortion to be zero) the mean-reversion of trend effect disappears. The interpretation of this result is that large trend signal goes hand in hand with large mispricing, which enhances the activity of fundamentalists who are wiping out the profits of trend-followers by taking the opposite position to trend.

Another noteworthy aspect of our paper concerns the estimation procedure, that is only slowly getting traction within the ABM literature (see \cite{Lux2017} and \cite{BMS2018} for other recent contributions). In fact, while the EM algorithm is perfectly adapted to calibrate the linear model, we advocate the use of the Unscented Kalman Filter to estimate its non-linear version.    

There are several follow-up research questions that would be well worth investigating. One is to extend the present analysis to single stocks, for which the dynamics is even more complex; in particular some short term mean-reversion is well known to take place before trend following effects set in. Second is to extend the Chiarella model to account for the time-evolution of the relative weight of the different groups of agents (trend followers, fundamentalists, noise-traders), possibly driven by the relative performance of these strategies -- much in the spirit of \cite{LUX1998,LM1999,GB2003}. Finally, it would be interesting to investigate different functional shapes for the non-linearity of the fundamentalists' demand (see e.g. footnote 8), and check the robustness of our conclusions with respect to different specifications. 

From a very practical point of view, our estimation methodology offers a general and efficient procedure to extract the fundamental value of a financial asset from price time series. We confirm the observation of \cite{BCLMSS2017} that universal trend following strategies should be supplemented by universal price-based value strategies in order to obtain the best investment results. Moreover, we emphasize the importance of controlling higher order terms in trends and price distortions when constructing such a combined strategy.

\newpage 
\bibliographystyle{elsarticle-harv} 
\bibliography{references}

\begin{thebibliography}{90}
\expandafter\ifx\csname natexlab\endcsname\relax\def\natexlab#1{#1}\fi
\expandafter\ifx\csname url\endcsname\relax
  \def\url#1{\texttt{#1}}\fi
\expandafter\ifx\csname urlprefix\endcsname\relax\def\urlprefix{URL }\fi

\bibitem[{Andrei and Cujean(2017)}]{AC2017}
Andrei, D., Cujean, J., 2017. Information percolation, momentum and reversal.
  {Journal of Financial Economics} 123~(3), 617--645.

\bibitem[{Asness et~al.(2013)Asness, Moskowitz, and Pedersen}]{AMP2013}
Asness, C.~S., Moskowitz, T.~J., Pedersen, L.~H., 2013. {Value and momentum
  everywhere}. {The Journal of Finance} 68~(3), 929--985.

\bibitem[{Barberis et~al.(1998)Barberis, Shleifer, and Vishny}]{BSV1998}
Barberis, N., Shleifer, A., Vishny, R., 1998. {A model of investor sentiment}.
  {Journal of Financial Economics} 49~(3), 307--343.

\bibitem[{Barde(2016)}]{Barde2016}
Barde, S., 2016. {Direct comparison of agent-based models of herding in
  financial markets}. {Journal of Economic Dynamics and Control} 73, 329--353.

\bibitem[{Barroso et~al.(2017)Barroso, Edelen, and Karehnke}]{BEK2017}
Barroso, P., Edelen, R.~M., Karehnke, P., 2017. {Crowding and the Moments of
  Momentum}. Tech. rep., working paper.

\bibitem[{Baum et~al.(1970)Baum, Petrie, Soules, and Weiss}]{BaumEtAl1970}
Baum, L.~E., Petrie, T., Soules, G., Weiss, N., 1970. {A maximization technique
  occurring in the statistical analysis of probabilistic functions of Markov
  chains}. {The Annals of Mathematical Statistics} 41~(1), 164--171.

\bibitem[{Beja and Goldman(1980)}]{BG1980}
Beja, A., Goldman, M.~B., 1980. {On the dynamic behavior of prices in
  disequilibrium}. {The Journal of Finance} 35~(2), 235--248.

\bibitem[{Bem(1965)}]{Bem1965}
Bem, D.~J., 1965. An experimental analysis of self-persuasion. {Journal of
  Experimental Social Psychology} 1~(3), 199--218.

\bibitem[{Bertschinger et~al.(2018)Bertschinger, Mozzhorin, and
  Sinha}]{BMS2018}
Bertschinger, N., Mozzhorin, I., Sinha, S., 2018. Reality-check for
  econophysics: Likelihood-based fitting of physics-inspired market models to
  empirical data. arXiv preprint arXiv:1803.03861.

\bibitem[{Black(1986)}]{Black1986}
Black, F., 1986. {Noise}. {The Journal of Finance} 41~(3), 528--543.

\bibitem[{Boswijk et~al.(2007)Boswijk, Hommes, and Manzan}]{BH2007}
Boswijk, H.~P., Hommes, C.~H., Manzan, S., 2007. {Behavioral heterogeneity in
  stock prices}. {Journal of Economic Dynamics and Control} 31~(6), 1938--1970.

\bibitem[{Bouchaud et~al.(2018)Bouchaud, Bonart, Donier, and Gould}]{BBDG2018}
Bouchaud, J.-P., Bonart, J., Donier, J., Gould, M., 2018. {Trades, quotes and
  prices: financial markets under the microscope}. {Cambridge University
  Press}.

\bibitem[{Bouchaud et~al.(2017)Bouchaud, Ciliberti, Lemperiere, Majewski,
  Seager, and Sin~Ronia}]{BCLMSS2017}
Bouchaud, J.-P., Ciliberti, S., Lemperiere, Y., Majewski, A., Seager, P.,
  Sin~Ronia, K., 2017. {Black was right: Price is within a factor 2 of Value}.

\bibitem[{Bouchaud and Cont(1998)}]{BC1998}
Bouchaud, J.-P., Cont, R., 1998. {A Langevin approach to stock market
  fluctuations and crashes}. {The European Physical Journal B-Condensed Matter
  and Complex Systems} 6~(4), 543--550.

\bibitem[{Brock and Hommes(1997)}]{BrH1997}
Brock, W.~A., Hommes, C.~H., 1997. A rational route to randomness.
  {Econometrica}, 1059--1095.

\bibitem[{Brock and Hommes(1998)}]{BH1998}
Brock, W.~A., Hommes, C.~H., 1998. {Heterogeneous beliefs and routes to chaos
  in a simple asset pricing model}. {Journal of Economic dynamics and Control}
  22~(8-9), 1235--1274.

\bibitem[{Campbell and Shiller(1988)}]{CS1988}
Campbell, J.~Y., Shiller, R.~J., 1988. Stock prices, earnings, and expected
  dividends. {The Journal of Finance} 43~(3), 661--676.

\bibitem[{Carhart(1997)}]{Carhart1997}
Carhart, M.~M., 1997. {On persistence in mutual fund performance}. {The Journal
  of Finance} 52~(1), 57--82.

\bibitem[{Challet et~al.(2013)Challet, Marsili, and Zhang}]{CMZ2013}
Challet, D., Marsili, M., Zhang, Y.-C., 2013. {Minority games: interacting
  agents in financial markets}. {OUP Catalogue}.

\bibitem[{Chen(1981)}]{Chen1981}
Chen, C.-F., 1981. {The EM approach to the multiple indicators and multiple
  causes model via the estimation of the latent variable}. Journal of the
  American Statistical Association 76~(375), 704--708.

\bibitem[{Chiarella(1992)}]{Chiarella1992}
Chiarella, C., 1992. {The dynamics of speculative behaviour}. {Annals of
  operations research} 37~(1), 101--123.

\bibitem[{Chiarella et~al.(2002)Chiarella, Dieci, and Gardini}]{CDG2002}
Chiarella, C., Dieci, R., Gardini, L., 2002. {Speculative behaviour and complex
  asset price dynamics: a global analysis}. {Journal of Economic Behavior \&
  Organization} 49~(2), 173--197.

\bibitem[{Chiarella et~al.(2006)Chiarella, Dieci, and Gardini}]{CDG2006}
Chiarella, C., Dieci, R., Gardini, L., 2006. {Asset price and wealth dynamics
  in a financial market with heterogeneous agents}. {Journal of Economic
  Dynamics and Control} 30~(9-10), 1755--1786.

\bibitem[{Chiarella et~al.(2009)Chiarella, Dieci, and He}]{CDH2009}
Chiarella, C., Dieci, R., He, X., 2009. {Heterogeneity, market mechanisms and
  asset price dynamics}. {Handbook of Financial Markets: Dynamics and
  Evolution}.

\bibitem[{Chiarella and He(2001)}]{CH2001}
Chiarella, C., He, X., 2001. {Asset price and wealth dynamics under
  heterogeneous expectations}. {Quantitative Finance} 1~(5), 509--526.

\bibitem[{Chiarella et~al.(2008)Chiarella, He, Wang, and Zheng}]{CHWZ2008}
Chiarella, C., He, X.-Z., Wang, D., Zheng, M., 2008. {The stochastic
  bifurcation behaviour of speculative financial markets}. {Physica A:
  Statistical Mechanics and its Applications} 387~(15), 3837--3846.

\bibitem[{Chiarella et~al.(2011)Chiarella, He, and Zheng}]{CHZ2011}
Chiarella, C., He, X.-Z., Zheng, M., 2011. {An analysis of the effect of noise
  in a heterogeneous agent financial market model}. {Journal of Economic
  Dynamics and Control} 35~(1), 148--162.

\bibitem[{Chiarella et~al.(2014)Chiarella, He, and Zwinkels}]{CHZ2014}
Chiarella, C., He, X.-Z., Zwinkels, R.~C., 2014. {Heterogeneous expectations in
  asset pricing: Empirical evidence from the S\&P500}. {Journal of Economic
  Behavior \& Organization} 105, 1--16.

\bibitem[{Daniel et~al.(1998)Daniel, Hirshleifer, and Subrahmanyam}]{DHS1998}
Daniel, K., Hirshleifer, D., Subrahmanyam, A., 1998. {Investor psychology and
  security market under-and overreactions}. {The Journal of Finance} 53~(6),
  1839--1885.

\bibitem[{Dao et~al.(2017)Dao, Nguyen, Deremble, Lemp{\'e}riere, Bouchaud, and
  Potters}]{DNDLBP2018}
Dao, T., Nguyen, T., Deremble, C., Lemp{\'e}riere, Y., Bouchaud, J., Potters,
  M., 2017. {Tail Protection for Long Investors: Trend Convexity at Work}.
  {Journal of Investment Strategies} 7~(1), 61--84.

\bibitem[{De~Bondt and Thaler(1985)}]{BWT1985}
De~Bondt, W.~F., Thaler, R., 1985. {Does the stock market overreact?} {The
  Journal of finance} 40~(3), 793--805.

\bibitem[{De~Long et~al.(1990)De~Long, Shleifer, Summers, and
  Waldmann}]{DSSW1990}
De~Long, J.~B., Shleifer, A., Summers, L.~H., Waldmann, R.~J., 1990. Noise
  trader risk in financial markets. Journal of Political Economy 98~(4),
  703--738.

\bibitem[{Dempster et~al.(1977)Dempster, Laird, and Rubin}]{DempsterEtAl1977}
Dempster, A.~P., Laird, N.~M., Rubin, D.~B., 1977. {Maximum likelihood from
  incomplete data via the EM algorithm}. {Journal of the Royal Statistical
  Society. Series B (methodological)}, 1--38.

\bibitem[{Dieci and He({2018})}]{DH2018}
Dieci, R., He, X.-Z., {2018}. {Heterogeneous Agent Models in Finance}. In:
  {Cars Hommes and Blake LeBaron} (Ed.), {Handbook of Computational Economics}.
  Vol.~{4}. {Elsevier}, pp. {257 -- 328}.

\bibitem[{Durbin and Koopman(2012)}]{DK2012}
Durbin, J., Koopman, S.~J., 2012. {Time Series Analysis by State Space
  Methods}. Vol.~38. {Oxford University Press}.

\bibitem[{Edwards(1968)}]{Edwards1968}
Edwards, W., 1968. Conservatism in human information processing. In: Kleinmutz,
  B. (Ed.), Formal Representation of Human Judgment. John Wiley and Sons, New
  York.

\bibitem[{Fama and French(1992)}]{FF1992}
Fama, E.~F., French, K.~R., 1992. The cross-section of expected stock returns.
  {the Journal of Finance} 47~(2), 427--465.

\bibitem[{Fama and French(2012)}]{FF2012}
Fama, E.~F., French, K.~R., 2012. {Size, value, and momentum in international
  stock returns}. {Journal of Financial Economics} 105~(3), 457--472.

\bibitem[{Fama and French(2016)}]{FF2016}
Fama, E.~F., French, K.~R., 2016. {Dissecting anomalies with a five-factor
  model}. {The Review of Financial Studies} 29~(1), 69--103.

\bibitem[{Franke and Westerhoff(2011)}]{FW2011}
Franke, R., Westerhoff, F., 2011. {Estimation of a structural stochastic
  volatility model of asset pricing}. {Computational Economics} 38~(1), 53--83.

\bibitem[{Franke and Westerhoff(2016)}]{FW2016}
Franke, R., Westerhoff, F., 2016. {Why a simple herding model may generate the
  stylized facts of daily returns: explanation and estimation}. {Journal of
  Economic Interaction and Coordination} 11~(1), 1--34.

\bibitem[{Ghonghadze and Lux(2016)}]{GL2016}
Ghonghadze, J., Lux, T., 2016. {Bringing an elementary agent-based model to the
  data: Estimation via GMM and an application to forecasting of asset price
  volatility}. {Journal of Empirical Finance} 37, 1--19.

\bibitem[{Giardina and Bouchaud(2003)}]{GB2003}
Giardina, I., Bouchaud, J.-P., 2003. Bubbles, crashes and intermittency in
  agent based market models. {The European Physical Journal B-Condensed Matter
  and Complex Systems} 31~(3), 421--437.

\bibitem[{Gordon(1962)}]{Gordon1962}
Gordon, M.~J., 1962. {The Investment, Financing, and Valuation of the
  Corporation}. {Homewood, Illinois: Richard D. Irwin, Inc.}

\bibitem[{Graham and Dodd(1934)}]{Graham1934}
Graham, B., Dodd, D.~L., 1934. {Security analysis}. {McGraw Hill Inc, New
  York}.

\bibitem[{Griffin et~al.(2003)Griffin, Ji, and Martin}]{GJS2003}
Griffin, J.~M., Ji, X., Martin, J.~S., 2003. {Momentum investing and business
  cycle risk: Evidence from pole to pole}. {The Journal of Finance} 58~(6),
  2515--2547.

\bibitem[{Hall and York(2001)}]{HY2001}
Hall, P., York, M., 2001. {On the calibration of Silverman's test for
  multimodality}. {Statistica Sinica}, 515--536.

\bibitem[{Hommes and LeBaron(2018)}]{HL2018}
Hommes, C., LeBaron, B. (Eds.), 2018. {Heterogeneous Agent Modeling}. Handbook
  of Computational Economics. {Elsevier Science}.

\bibitem[{Hommes(2006)}]{Hommes2006}
Hommes, C.~H., 2006. {Heterogeneous agent models in economics and finance}.
  {Handbook of Computational Economics} 2, 1109--1186.

\bibitem[{Hong and Stein(1999)}]{HS1999}
Hong, H., Stein, J.~C., 1999. {A unified theory of underreaction, momentum
  trading, and overreaction in asset markets}. {The Journal of Finance} 54~(6),
  2143--2184.

\bibitem[{Ide and Sornette(2002)}]{IS2002}
Ide, K., Sornette, D., 2002. {Oscillatory finite-time singularities in finance,
  population and rupture}. {Physica A: Statistical Mechanics and its
  Applications} 307~(1-2), 63--106.

\bibitem[{Jegadeesh(1990)}]{Jegadeesh1990}
Jegadeesh, N., 1990. {Evidence of predictable behavior of security returns}.
  {The Journal of Finance} 45~(3), 881--898.

\bibitem[{Jegadeesh and Titman(1993)}]{JT1993}
Jegadeesh, N., Titman, S., 1993. {Returns to buying winners and selling losers:
  Implications for stock market efficiency}. {The Journal of Finance} 48~(1),
  65--91.

\bibitem[{Julier and Uhlmann(1997)}]{JU1997}
Julier, S.~J., Uhlmann, J.~K., 1997. {New extension of the Kalman filter to
  nonlinear systems}. In: {Signal processing, sensor fusion, and target
  recognition VI}. Vol. 3068. {International Society for Optics and Photonics},
  pp. 182--194.

\bibitem[{Kahneman(2011)}]{Kahneman2011}
Kahneman, D., 2011. {Thinking Fast and Slow}. Penguin.

\bibitem[{Kahneman and Tversky(1973)}]{KT1973}
Kahneman, D., Tversky, A., 1973. On the psychology of prediction.
  {Psychological Review} 80~(4), 237.

\bibitem[{Kalman(1960)}]{Kalman1960}
Kalman, R.~E., 1960. {A new approach to linear filtering and prediction
  problems}. {Journal of Basic Engineering} 82~(1), 35--45.

\bibitem[{Kokkala et~al.(2014)Kokkala, Solin, and S{\"a}rkk{\"a}}]{KSS2014}
Kokkala, J., Solin, A., S{\"a}rkk{\"a}, S., 2014. {Expectation maximization
  based parameter estimation by sigma-point and particle smoothing}. In: {17th
  International Conference on Information Fusion 2014}. {IEEE}, pp. 1--8.

\bibitem[{Kokkala et~al.(2015)Kokkala, Solin, and S{\"a}rkk{\"a}}]{KSS2015}
Kokkala, J., Solin, A., S{\"a}rkk{\"a}, S., 2015. Sigma-point filtering and
  smoothing based parameter estimation in nonlinear dynamic systems. Submitted
  to Journal of Advances in Information Fusion.

\bibitem[{Kyle(1985)}]{Kyle1985}
Kyle, A.~S., 1985. {Continuous auctions and insider trading}. {Econometrica:
  Journal of the Econometric Society}, 1315--1335.

\bibitem[{Lakonishok et~al.(1994)Lakonishok, Shleifer, and Vishny}]{LSV1994}
Lakonishok, J., Shleifer, A., Vishny, R.~W., 1994. {Contrarian investment,
  extrapolation, and risk}. {The Journal of Finance} 49~(5), 1541--1578.

\bibitem[{Lamperti et~al.(2017)Lamperti, Roventini, and Sani}]{LRS2017}
Lamperti, F., Roventini, A., Sani, A., 2017. Agent-based model calibration
  using machine learning surrogates. arXiv preprint arXiv:1803.03861.

\bibitem[{Landier et~al.(2017)Landier, Ma, and Thesmar}]{LMT2017}
Landier, A., Ma, Y., Thesmar, D., 2017. {New Experimental Evidence on
  Expectations Formation}. Available at SSRN:
  https://ssrn.com/abstract=3046955.

\bibitem[{LeBaron(2006)}]{Lebaron2006}
LeBaron, B., 2006. {Agent-based computational finance}. {Handbook of
  Computational Economics} 2, 1187--1233.

\bibitem[{LeBaron et~al.(1999)LeBaron, Arthur, and Palmer}]{LAP1999}
LeBaron, B., Arthur, W.~B., Palmer, R., 1999. Time series properties of an
  artificial stock market. {Journal of Economic Dynamics and control}
  23~(9-10), 1487--1516.

\bibitem[{Lehmann(1990)}]{Lehmann1990}
Lehmann, B.~N., 1990. {Fads, martingales, and market efficiency}. {The
  Quarterly Journal of Economics} 105~(1), 1--28.

\bibitem[{Lemp{\'e}ri{\`e}re et~al.(2017)Lemp{\'e}ri{\`e}re, Deremble, Nguyen,
  Seager, Potters, and Bouchaud}]{LDNSPB2017}
Lemp{\'e}ri{\`e}re, Y., Deremble, C., Nguyen, T.-T., Seager, P., Potters, M.,
  Bouchaud, J.-P., 2017. {Risk premia: Asymmetric tail risks and excess
  returns}. {Quantitative Finance} 17~(1), 1--14.

\bibitem[{Lemp{\'e}ri{\`e}re et~al.(2014)Lemp{\'e}ri{\`e}re, Deremble, Seager,
  Potters, and Bouchaud}]{LDSPB2014}
Lemp{\'e}ri{\`e}re, Y., Deremble, C., Seager, P., Potters, M., Bouchaud, J.-P.,
  2014. Two centuries of trend following. Risk 3~(3), 41--61.

\bibitem[{Liu and Zhang(2008)}]{LZ2008}
Liu, L.~X., Zhang, L., 2008. {Momentum profits, factor pricing, and
  macroeconomic risk}. {The Review of Financial Studies} 21~(6), 2417--2448.

\bibitem[{Lux(1998)}]{LUX1998}
Lux, T., 1998. {The socio-economic dynamics of speculative markets: interacting
  agents, chaos, and the fat tails of return distributions}. {Journal of
  Economic Behavior \& Organization} 33~(2), 143--165.

\bibitem[{Lux(2017)}]{Lux2017}
Lux, T., 2017. {Estimation of agent-based models using sequential Monte Carlo
  methods}. Tech. rep., Economics Working Paper.

\bibitem[{Lux and Marchesi(1999)}]{LM1999}
Lux, T., Marchesi, M., 1999. Scaling and criticality in a stochastic
  multi-agent model of a financial market. Nature 397~(6719), 498.

\bibitem[{Moskowitz et~al.(2012)Moskowitz, Ooi, and Pedersen}]{MOP2012}
Moskowitz, T.~J., Ooi, Y.~H., Pedersen, L.~H., 2012. Time series momentum.
  {Journal of Financial Economics} 104~(2), 228--250.

\bibitem[{P{\'a}stor and Stambaugh(2003)}]{PS2003}
P{\'a}stor, L., Stambaugh, R.~F., 2003. {Liquidity risk and expected stock
  returns}. {Journal of Political Economy} 111~(3), 642--685.

\bibitem[{Rosenberg et~al.(1985)Rosenberg, Reid, and Lanstein}]{RRL1985}
Rosenberg, B., Reid, K., Lanstein, R., 1985. {Persuasive evidence of market
  inefficiency}. {The Journal of Portfolio Management} 11~(3), 9--16.

\bibitem[{Roweis and Ghahramani(2001)}]{RG2001}
Roweis, S., Ghahramani, Z., 2001. {Learning nonlinear dynamical systems using
  the expectation--maximization algorithm}. {Chapter in book: 'Kalman filtering
  and neural networks' edited by Simon Haykin}, 175--220.

\bibitem[{Sadka(2006)}]{Sadka2006}
Sadka, R., 2006. {Momentum and post-earnings-announcement drift anomalies: The
  role of liquidity risk}. {Journal of Financial Economics} 80~(2), 309--349.

\bibitem[{S{\"a}rkk{\"a}(2013)}]{Sarkka2013}
S{\"a}rkk{\"a}, S., 2013. {Bayesian filtering and smoothing}. Vol.~3.
  {Cambridge University Press}.

\bibitem[{Schmitt and Westerhoff(2017)}]{SW2017}
Schmitt, N., Westerhoff, F., 2017. {On the bimodality of the distribution of
  the S\&P 500's distortion: Empirical evidence and theoretical explanations}.
  {Journal of Economic Dynamics and Control} 80, 34--53.

\bibitem[{Shiller(1980)}]{Shiller1980}
Shiller, R.~J., 1980. Do stock prices move too much to be justified by
  subsequent changes in dividends? National Bureau of Economic Research
  Cambridge, Mass., USA.

\bibitem[{Shiller(1987)}]{Shiller1987}
Shiller, R.~J., 1987. {Investor behavior in the October 1987 stock market
  crash: Survey evidence}. {NBER working paper No. 2446, November 1987,
  published in Shiller, R.J., Market Volatility, MIT Press, Cambridge, 198,
  chapter 23}.

\bibitem[{Shiller(2000)}]{Shiller2000}
Shiller, R.~J., 2000. {Irrational exuberance}. {Princeton University Press}.

\bibitem[{Shumway and Stoffer(1982)}]{SS1982}
Shumway, R.~H., Stoffer, D.~S., 1982. {An approach to time series smoothing and
  forecasting using the EM algorithm}. Journal of time series analysis 3~(4),
  253--264.

\bibitem[{Silverman(1981)}]{Silverman1981}
Silverman, B.~W., 1981. {Using kernel density estimates to investigate
  multimodality}. {Journal of the Royal Statistical Society. Series B
  (Methodological)}, 97--99.

\bibitem[{Summers(1986)}]{Summers1986}
Summers, L.~H., 1986. {Does the stock market rationally reflect fundamental
  values?} {The Journal of Finance} 41~(3), 591--601.

\bibitem[{Thaler(1993)}]{Thaler1993}
Thaler, R.~H. (Ed.), 1993. {Advances in Behavioral Finance}. {Russell Sage
  Foundation}.

\bibitem[{Thaler(2005)}]{Thaler2005}
Thaler, R.~H. (Ed.), 2005. {Advances in Behavioral Finance. Volume 2}.
  {Princeton University Press}.

\bibitem[{Tversky and Kahneman(1974)}]{TK1974}
Tversky, A., Kahneman, D., 1974. {Judgment under Uncertainty: Heuristics and
  Biases}. {Science} 185~(4157), 1124--1131.

\bibitem[{Vayanos and Woolley(2013)}]{VW2013}
Vayanos, D., Woolley, P., 2013. {An institutional theory of momentum and
  reversal}. {The Review of Financial Studies} 26~(5), 1087--1145.

\bibitem[{Wyart and Bouchaud(2007)}]{WB2007}
Wyart, M., Bouchaud, J.-P., 2007. {Self-referential behaviour, overreaction and
  conventions in financial markets}. {Journal of Economic Behavior \&
  Organization} 63~(1), 1--24.

\end{thebibliography}

\newpage
\appendix
\section{Estimation results}\label{app:estimation_results}

\begin{table}[!h]
\begin{center}
{\footnotesize
\begin{tabular}{lrrrrrrrr}
\toprule
{} &  $\kappa$ &   $\beta$ &  $\gamma$ &  $\sigma_N$ &  $\sigma_V$ &  $g$ &  $v_0$ &     $\mathcal{L}$ \\
\midrule
\rowcolor{gray!30} \textbf{US     } &  0.015 &  0.015 &   36.7 &    0.043 &    0.018 &  0.0011 &   4.42 &  4596.4 \\
\textbf{UK     } &  0.015 &  0.015 &   40.9 &    0.036 &    0.018 &  0.0013 &   6.85 &  4920.7 \\
\rowcolor{gray!30} \textbf{AU     } &  0.015 &  0.015 &   40.8 &    0.039 &    0.018 &  0.0016 &   5.97 &  3053.9 \\
\textbf{CH     } &  0.015 &  0.015 &   32.6 &    0.044 &    0.018 &  0.0019 &   3.89 &  2037.9 \\
\rowcolor{gray!30} \textbf{JP     } &  0.015 &  0.015 &   25.1 &    0.059 &    0.018 &  0.0029 &   7.24 &  1611.7 \\
\textbf{CA     } &  0.015 &  0.015 &   34.5 &    0.045 &    0.018 &  0.0017 &   7.48 &  2011.1 \\
\rowcolor{gray!30} \textbf{DE     } &  0.015 &  0.015 &   32.2 &    0.044 &    0.018 &  0.0023 &  30.59 &  2481.2 \\
\toprule
\textbf{SUGAR  } &  0.094 &  0.020 &   24.1 &    0.074 &    0.036 & -0.0007 &   6.43 &  2945.0 \\
\rowcolor{gray!30} \textbf{CORN   } &  0.094 &  0.020 &   19.6 &    0.118 &    0.036 & -0.0008 &   2.75 &  1223.5 \\
\textbf{LCATTLE} &  0.094 &  0.020 &   33.7 &    0.062 &    0.036 &  0.0002 &   4.89 &  2363.3 \\
\rowcolor{gray!30} \textbf{WHEAT  } &  0.094 &  0.020 &   23.3 &    0.092 &    0.036 & -0.0009 &   3.28 &  1856.6 \\
\textbf{COPPER } &  0.094 &  0.020 &   30.6 &    0.062 &    0.036 & -0.0004 &   7.47 &  3317.9 \\
\rowcolor{gray!30} \textbf{NATGAS } &  0.094 &  0.020 &   14.8 &    0.182 &    0.036 &  0.0014 &   1.09 &    83.4 \\
\textbf{CRUDE  } &  0.094 &  0.020 &   14.1 &    0.109 &    0.036 & -0.0005 &   3.98 &  1011.6 \\
\toprule
\rowcolor{gray!30} \textbf{CHFUSD } &  0.017 &  0.005 &   49.5 &    0.034 &    0.010 &  0.0019 &  -0.84 &   965.9 \\
\textbf{JPYUSD } &  0.017 &  0.005 &   50.1 &    0.032 &    0.010 &  0.0022 &  -1.07 &   998.5 \\
\rowcolor{gray!30} \textbf{AUDUSD } &  0.017 &  0.005 &   54.2 &    0.033 &    0.010 &  0.0001 &  -0.25 &   990.5 \\
\textbf{GBPUSD } &  0.017 &  0.005 &   56.7 &    0.029 &    0.010 & -0.0005 &   0.68 &  1051.4 \\
\rowcolor{gray!30} \textbf{CADUSD } &  0.017 &  0.005 &   96.6 &    0.019 &    0.010 &  0.0013 &  -0.60 &  1251.7 \\
\textbf{EURUSD } &  0.017 &  0.005 &   51.1 &    0.032 &    0.010 &  0.0010 &  -0.21 &  1005.8 \\
\toprule
\rowcolor{gray!30} \textbf{USBND  } &  0.086 &  0.013 &   39.3 &    0.045 &    0.041 & -0.0006 &   1.43 &  1826.7 \\
\textbf{UKBND  } &  0.086 &  0.013 &   37.0 &    0.049 &    0.041 & -0.0015 &   2.01 &  1073.1 \\
\rowcolor{gray!30} \textbf{CHBND  } &  0.086 &  0.013 &   31.1 &    0.052 &    0.041 & -0.0017 &   1.63 &  1715.9 \\
\textbf{JPBND  } &  0.086 &  0.013 &   24.5 &    0.087 &    0.041 & -0.0075 &   2.20 &   419.8 \\
\rowcolor{gray!30} \textbf{AUBND  } &  0.086 &  0.013 &   38.5 &    0.044 &    0.041 & -0.0040 &   2.64 &   623.1 \\
\textbf{CABND  } &  0.086 &  0.013 &   47.7 &    0.036 &    0.041 & -0.0008 &   1.64 &  2103.8 \\
\rowcolor{gray!30} \textbf{DEBND  } &  0.086 &  0.013 &   35.1 &    0.046 &    0.041 & -0.0020 &   1.79 &  1251.7 \\
\bottomrule
\end{tabular}

\caption{{\footnotesize Estimation results for model (\ref{eq:DiscreteModel}). Predictive log-likelihood $\mathcal{L}$ is calculated according to (\ref{eq:PLL_KF}). CRUDE stands for crude oil, NATGAS for Henry Hub natural gas, LCATTLE for live cattle and abbreviation BND stands for government bonds.}}\label{tab:estimation_EM}}
\end{center}
\end{table}
\begin{table}
\begin{center}
{\footnotesize
\begin{tabular}{lrrrrrr}
\toprule
{} &     $\kappa$ &    $\beta$ &  $\sigma_N$ &  $\sigma_V$ &  $g$ &   $v_0$ \\
\midrule
\rowcolor{gray!30} \textbf{US     } &  4.459505 &  11.625 &   29.887 &    4.326 &   2.680 &  28.941 \\
\textbf{UK     } &  4.459505 &  11.625 &   19.069 &    4.326 &   2.797 &  28.488 \\
\rowcolor{gray!30} \textbf{AU     } &  4.459505 &  11.625 &   15.048 &    4.326 &   3.343 &  19.358 \\
\textbf{CH     } &  4.459505 &  11.625 &   25.885 &    4.326 &   3.286 &  14.578 \\
\rowcolor{gray!30} \textbf{JP     } &  4.459505 &  11.625 &   24.601 &    4.326 &   3.547 &  18.005 \\
\textbf{CA     } &  4.459505 &  11.625 &   25.887 &    4.326 &   4.576 &  64.456 \\
\rowcolor{gray!30} \textbf{DE     } &  4.459505 &  11.625 &   29.294 &    4.326 &   2.949 &  88.349 \\
\toprule
\textbf{SUGAR  } &  3.820824 &   4.379 &   30.009 &    7.957 &  -1.732 &  32.308 \\
\rowcolor{gray!30} \textbf{CORN   } &  3.820824 &   4.379 &    7.187 &    7.957 &  -2.359 &   5.760 \\
\textbf{LCATTLE} &  3.820824 &   4.379 &   26.617 &    7.957 &   0.851 &  19.140 \\
\rowcolor{gray!30} \textbf{WHEAT  } &  3.820824 &   4.379 &   34.577 &    7.957 &  -2.326 &  15.100 \\
\textbf{COPPER } &  3.820824 &   4.379 &   21.711 &    7.957 &  -0.624 &  31.440 \\
\rowcolor{gray!30} \textbf{NATGAS } &  3.820824 &   4.379 &   15.185 &    7.957 &   0.816 &  10.706 \\
\textbf{CRUDE  } &  3.820824 &   4.379 &   23.448 &    7.957 &  -0.665 &   7.889 \\
\toprule
\rowcolor{gray!30} \textbf{CHFUSD } &  3.747808 &   3.489 &   25.357 &    1.164 &   1.893 &  -3.930 \\
\textbf{JPYUSD } &  3.747808 &   3.489 &   24.584 &    1.164 &   2.222 &  -6.619 \\
\rowcolor{gray!30} \textbf{AUDUSD } &  3.747808 &   3.489 &   16.985 &    1.164 &   0.577 &  -0.798 \\
\textbf{GBPUSD } &  3.747808 &   3.489 &   22.620 &    1.164 &  -0.398 &   4.177 \\
\rowcolor{gray!30} \textbf{CADUSD } &  3.747808 &   3.489 &   15.658 &    1.164 &   0.531 &  -1.336 \\
\textbf{EURUSD } &  3.747808 &   3.489 &   26.910 &    1.164 &   0.935 &  -1.184 \\
\toprule
\rowcolor{gray!30} \textbf{USBND  } &  1.053739 &   1.212 &   19.007 &    2.115 &  -0.650 &  14.424 \\
\textbf{UKBND  } &  1.053739 &   1.212 &   20.205 &    2.115 &  -2.926 &  27.845 \\
\rowcolor{gray!30} \textbf{CHBND  } &  1.053739 &   1.212 &   13.438 &    2.115 &  -3.098 &  19.675 \\
\textbf{JPBND  } &  1.053739 &   1.212 &   12.280 &    2.115 &  -3.321 &  11.560 \\
\rowcolor{gray!30} \textbf{AUBND  } &  1.053739 &   1.212 &   17.907 &    2.115 &  -3.316 &  36.021 \\
\textbf{CABND  } &  1.053739 &   1.212 &   23.731 &    2.115 &  -1.517 &  23.660 \\
\rowcolor{gray!30} \textbf{DEBND  } &  1.053739 &   1.212 &   17.644 &    2.115 &  -3.143 &  21.471 \\
\bottomrule
\end{tabular}

\caption{{\footnotesize T-statistics of estimation for model (\ref{eq:DiscreteModel}). CRUDE stands for crude oil, NATGAS for Henry Hub natural gas, LCATTLE for live cattle and abbreviation BND stands for government bonds.}}\label{tab:tstat_EM}}
\end{center}
\end{table}

\begin{table}
\begin{center}
{\footnotesize
\begin{tabular}{lrrrrrrrrr}
\toprule
{} &  $\kappa$ &  $\kappa_3$ &   $\beta$ &  $\gamma$ &  $\sigma_N$ &  $\sigma_V$ &  $g$ &  $v_0$ &   $\hat{\mathcal{L}}$ \\
\midrule
\rowcolor{gray!30} \textbf{US     } &   -0.011 &    0.269 &  0.018 &   36.7 &    0.042 &    0.018 &  0.0011 &   4.41 &  4616.8 \\
\textbf{UK     } &   -0.011 &    0.269 &  0.018 &   40.9 &    0.034 &    0.018 &  0.0010 &   7.12 &  4932.8 \\
\rowcolor{gray!30} \textbf{AU     } &   -0.011 &    0.269 &  0.018 &   40.8 &    0.031 &    0.018 &  0.0014 &   5.71 &  2987.9 \\
\textbf{CH     } &   -0.011 &    0.269 &  0.018 &   32.6 &    0.043 &    0.018 &  0.0013 &   4.29 &  2019.6 \\
\rowcolor{gray!30} \textbf{JP     } &   -0.011 &    0.269 &  0.018 &   25.1 &    0.057 &    0.018 &  0.0024 &   7.94 &  1605.6 \\
\textbf{CA     } &   -0.011 &    0.269 &  0.018 &   34.5 &    0.045 &    0.018 &  0.0015 &   7.54 &  2016.6 \\
\rowcolor{gray!30} \textbf{DE     } &   -0.011 &    0.269 &  0.018 &   32.2 &    0.041 &    0.018 &  0.0023 &  31.03 &  2477.6 \\
\toprule
\textbf{SUGAR  } &   -0.065 &    0.536 &  0.011 &   24.1 &    0.074 &    0.036 & -0.0013 &   6.69 &  2949.7 \\
\rowcolor{gray!30} \textbf{CORN   } &   -0.065 &    0.536 &  0.011 &   19.6 &    0.105 &    0.036 & -0.0010 &   2.84 &  1425.1 \\
\textbf{LCATTLE} &   -0.065 &    0.536 &  0.011 &   33.7 &    0.061 &    0.036 &  0.0000 &   5.16 &  2385.7 \\
\rowcolor{gray!30} \textbf{WHEAT  } &   -0.065 &    0.536 &  0.011 &   23.3 &    0.090 &    0.036 & -0.0012 &   3.17 &  1887.7 \\
\textbf{COPPER } &   -0.065 &    0.536 &  0.011 &   30.6 &    0.057 &    0.036 & -0.0003 &   7.05 &  3370.6 \\
\rowcolor{gray!30} \textbf{NATGAS } &   -0.065 &    0.536 &  0.011 &   14.8 &    0.167 &    0.036 &  0.0018 &   0.99 &   102.8 \\
\textbf{CRUDE  } &   -0.065 &    0.536 &  0.011 &   14.1 &    0.085 &    0.036 & -0.0014 &  25.67 &   813.9 \\
\toprule
\rowcolor{gray!30} \textbf{CHFUSD } &   -0.019 &    0.581 &  0.006 &   49.5 &    0.034 &    0.010 &  0.0020 &  -1.27 &   962.7 \\
\textbf{JPYUSD } &   -0.019 &    0.581 &  0.006 &   50.1 &    0.032 &    0.010 &  0.0020 &  -1.10 &   996.7 \\
\rowcolor{gray!30} \textbf{AUDUSD } &   -0.019 &    0.581 &  0.006 &   54.2 &    0.029 &    0.010 & -0.0002 &   0.34 &   979.2 \\
\textbf{GBPUSD } &   -0.019 &    0.581 &  0.006 &   56.7 &    0.026 &    0.010 & -0.0010 &   0.76 &  1045.0 \\
\rowcolor{gray!30} \textbf{CADUSD } &   -0.019 &    0.581 &  0.006 &   96.6 &    0.019 &    0.010 & -0.0002 &  -0.02 &  1256.6 \\
\textbf{EURUSD } &   -0.019 &    0.581 &  0.006 &   51.1 &    0.030 &    0.010 &  0.0015 &  -0.55 &  1004.7 \\
\toprule
\rowcolor{gray!30} \textbf{USBND  } &   -0.179 &    7.435 &  0.016 &   39.3 &    0.042 &    0.041 &  0.0004 &   1.26 &  1874.8 \\
\textbf{UKBND  } &   -0.179 &    7.435 &  0.016 &   37.0 &    0.048 &    0.041 & -0.0014 &   2.00 &  1074.2 \\
\rowcolor{gray!30} \textbf{CHBND  } &   -0.179 &    7.435 &  0.016 &   31.1 &    0.050 &    0.041 & -0.0018 &   1.67 &  1758.0 \\
\textbf{JPBND  } &   -0.179 &    7.435 &  0.016 &   24.5 &    0.083 &    0.041 & -0.0082 &   2.35 &   399.2 \\
\rowcolor{gray!30} \textbf{AUBND  } &   -0.179 &    7.435 &  0.016 &   38.5 &    0.042 &    0.041 & -0.0017 &   2.66 &   628.4 \\
\textbf{CABND  } &   -0.179 &    7.435 &  0.016 &   47.7 &    0.033 &    0.041 &  0.0001 &   1.59 &  2134.6 \\
\rowcolor{gray!30} \textbf{DEBND  } &   -0.179 &    7.435 &  0.016 &   35.1 &    0.045 &    0.041 & -0.0016 &   1.69 &  1280.9 \\
\bottomrule
\end{tabular}
\caption{{\footnotesize Estimation results for model (\ref{eq:NonlinearModel}). Predictive log-likelihood $\hat{\mathcal{L}}$ is calculated according to (\ref{eq:PLL_UKF}). CRUDE stands for crude oil, NATGAS for Henry Hub natural gas, LCATTLE for live cattle and abbreviation BND stands for government bonds.}}\label{tab:estimation_direct}}
\end{center}
\end{table}

\begin{table}
\begin{center}
{\footnotesize
\begin{tabular}{lrrrrrr}
\toprule
{} &  $\kappa$ &  $\kappa_3$ &    $\beta$ &  $\sigma_N$ &  $g$ &    $v_0$ \\
\midrule
\rowcolor{gray!30} \textbf{US     } &   -2.289 &    2.150 &   8.705 &   27.377 &   5.270 &   76.319 \\
\textbf{UK     } &   -2.289 &    2.150 &   8.705 &   23.777 &   8.154 &  119.690 \\
\rowcolor{gray!30} \textbf{AU     } &   -2.289 &    2.150 &   8.705 &   27.986 &   3.411 &  572.320 \\
\textbf{CH     } &   -2.289 &    2.150 &   8.705 &   25.522 &   4.158 &   31.903 \\
\rowcolor{gray!30} \textbf{JP     } &   -2.289 &    2.150 &   8.705 &   21.722 &   4.362 &   29.014 \\
\textbf{CA     } &   -2.289 &    2.150 &   8.705 &   25.358 &   2.123 &  129.614 \\
\rowcolor{gray!30} \textbf{DE     } &   -2.289 &    2.150 &   8.705 &   24.464 &   4.721 &  319.856 \\
\toprule
\textbf{SUGAR  } &   -4.979 &    3.974 &   2.553 &   31.107 &  -2.945 &    4.577 \\
\rowcolor{gray!30} \textbf{CORN   } &   -4.979 &    3.974 &   2.553 &   11.510 &  -1.062 &   46.322 \\
\textbf{LCATTLE} &   -4.979 &    3.974 &   2.553 &   27.246 &   1.547 &   39.852 \\
\rowcolor{gray!30} \textbf{WHEAT  } &   -4.979 &    3.974 &   2.553 &   35.770 &  -2.630 &   20.007 \\
\textbf{COPPER } &   -4.979 &    3.974 &   2.553 &   17.763 &  -1.561 &  117.953 \\
\rowcolor{gray!30} \textbf{NATGAS } &   -4.979 &    3.974 &   2.553 &   15.579 &   1.545 &    7.006 \\
\textbf{CRUDE  } &   -4.979 &    3.974 &   2.553 &   12.699 &  -1.645 &   93.434 \\
\toprule
\rowcolor{gray!30} \textbf{CHFUSD } &   -1.386 &    1.369 &   3.418 &   24.778 &   3.873 &   -6.969 \\
\textbf{JPYUSD } &   -1.386 &    1.369 &   3.418 &   15.195 &   1.823 &   -2.970 \\
\rowcolor{gray!30} \textbf{AUDUSD } &   -1.386 &    1.369 &   3.418 &   12.293 &  -0.647 &   10.738 \\
\textbf{GBPUSD } &   -1.386 &    1.369 &   3.418 &   13.911 &  -1.975 &   22.620 \\
\rowcolor{gray!30} \textbf{CADUSD } &   -1.386 &    1.369 &   3.418 &   16.546 &  -0.407 &   -0.138 \\
\textbf{EURUSD } &   -1.386 &    1.369 &   3.418 &   22.244 &   8.167 &   -8.277 \\
\toprule
\rowcolor{gray!30} \textbf{USBND  } &   -9.591 &   17.931 &  13.270 &   21.581 &   2.524 &   34.161 \\
\textbf{UKBND  } &   -9.591 &   17.931 &  13.270 &   19.833 &  -2.111 &   69.679 \\
\rowcolor{gray!30} \textbf{CHBND  } &   -9.591 &   17.931 &  13.270 &   14.253 &  -3.006 &   18.736 \\
\textbf{JPBND  } &   -9.591 &   17.931 &  13.270 &    9.380 &  -1.336 &    1.791 \\
\rowcolor{gray!30} \textbf{AUBND  } &   -9.591 &   17.931 &  13.270 &   17.134 &  -3.410 &   88.784 \\
\textbf{CABND  } &   -9.591 &   17.931 &  13.270 &   29.540 &   1.124 &  323.950 \\
\rowcolor{gray!30} \textbf{DEBND  } &   -9.591 &   17.931 &  13.270 &   19.062 &  -4.249 &   94.968 \\
\bottomrule
\end{tabular}
\caption{{\footnotesize T-statistics of estimation for model (\ref{eq:NonlinearModel}). CRUDE stands for crude oil, NATGAS for Henry Hub natural gas, LCATTLE for live cattle and abbreviation BND stands for government bonds.}}\label{tab:tstat_direct}}
\end{center}
\end{table}

\begin{figure}
\begin{center}

 \includegraphics[scale=0.7]{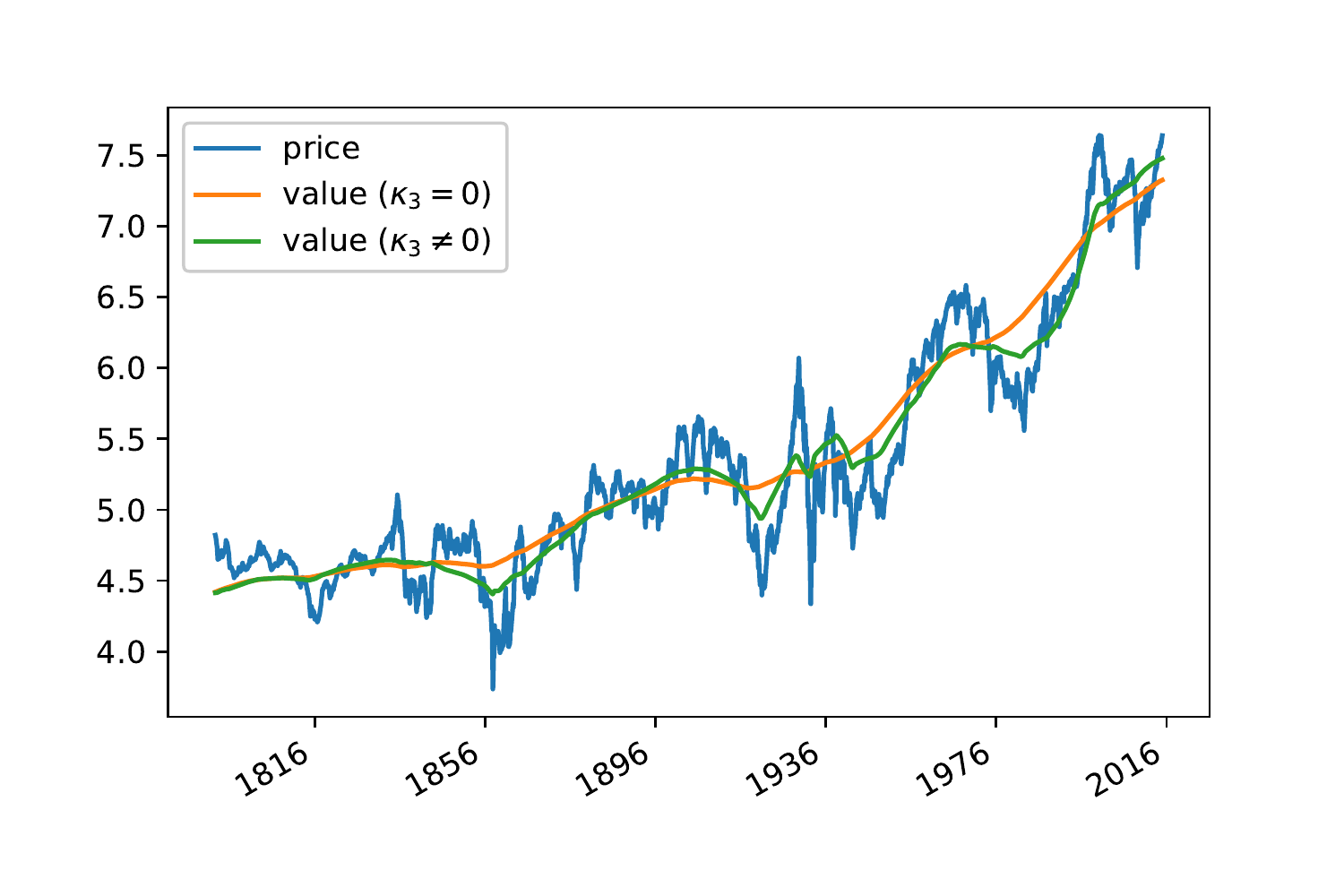} \caption{{\footnotesize Log-price of US stock index together with the smoothed value in model with linear demand of fundamentalists calculated by Kalman Smoother for parameters in Table \ref{tab:estimation_EM} and for model with non-linear demand of fundamentalists calculated by Unscented Kalman Smoother for parameters in Table \ref{tab:estimation_direct}.}}\label{fig:US} 
\end{center} 
\end{figure}
\begin{figure}
\begin{center}
\includegraphics[scale=0.7]{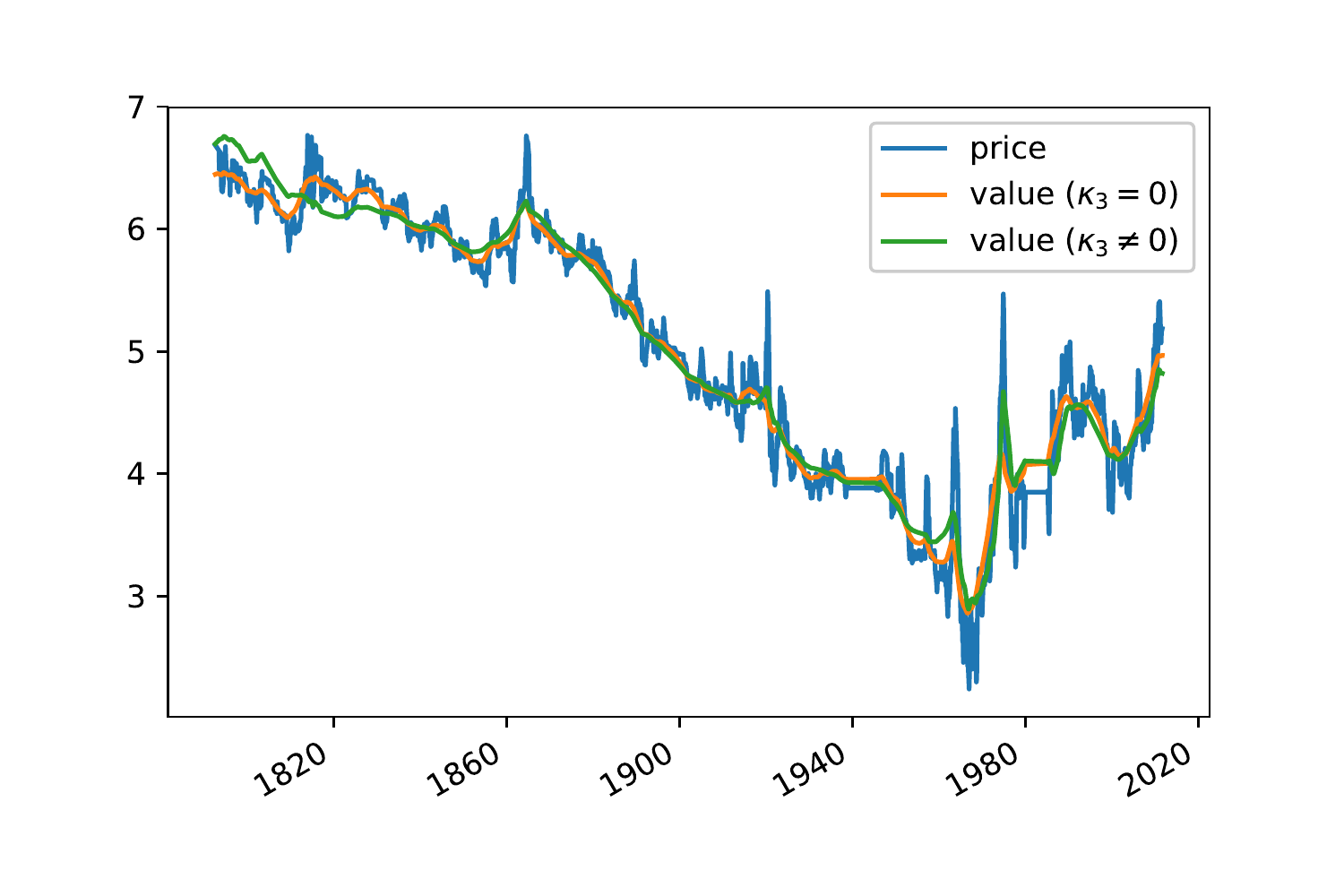} \caption{{\footnotesize Log-price of sugar together with the smoothed value in model with linear demand of fundamentalists calculated by Kalman Smoother for parameters in Table \ref{tab:estimation_EM} and for model with non-linear demand of fundamentalists calculated by Unscented Kalman Smoother for parameters in Table
\ref{tab:estimation_direct}.}}\label{fig:SUGAR} 
\end{center} 
\end{figure}

\begin{figure}
\begin{center}
\includegraphics[scale=0.7]{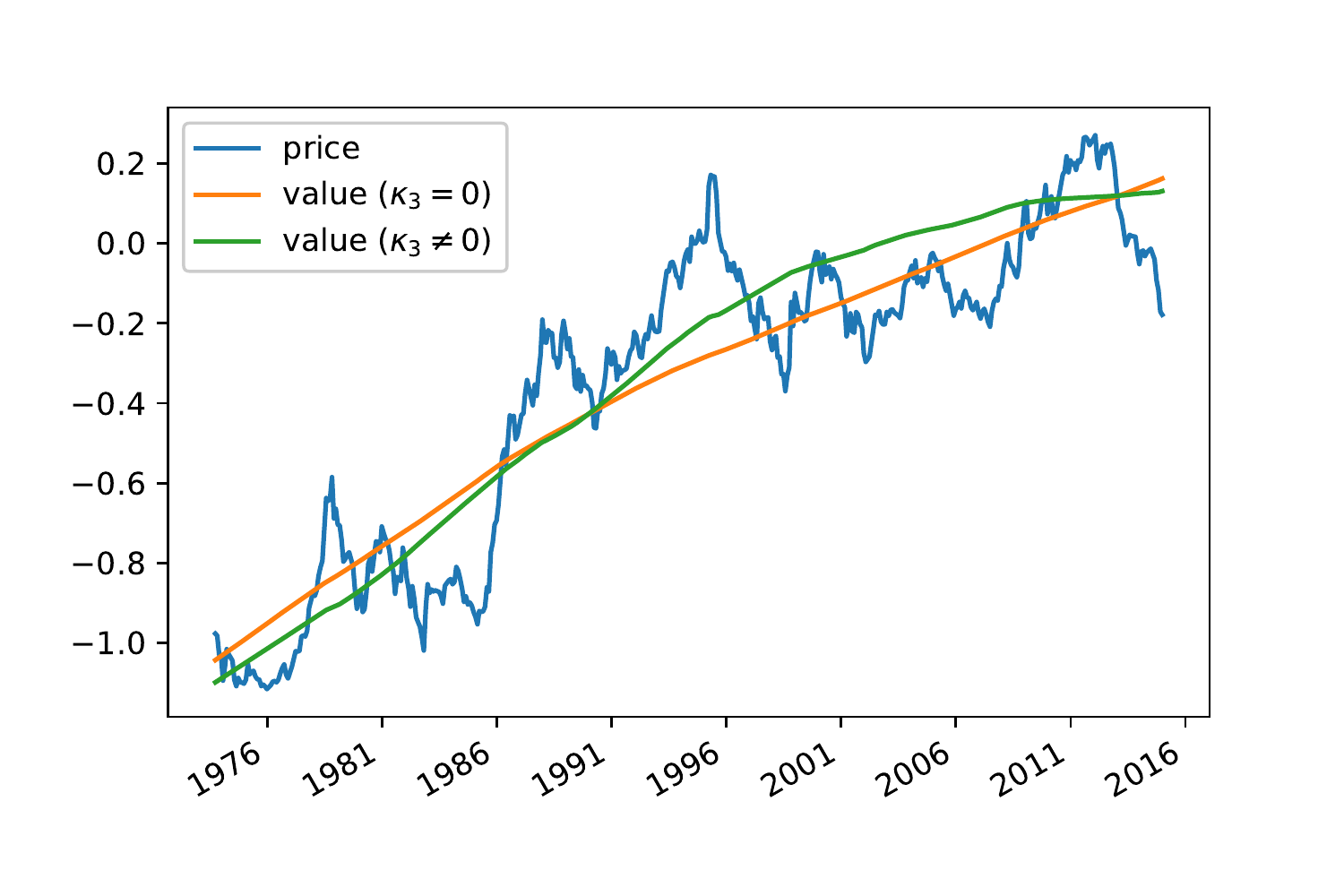} \caption{{\footnotesize Logarithm of JPY/USD exchange rate together with the smoothed value in model with linear demand of fundamentalists calculated by Kalman Smoother for parameters in Table \ref{tab:estimation_EM} and for model with non-linear demand of fundamentalists calculated by Unscented Kalman Smoother for parameters in Table \ref{tab:estimation_direct}.}}\label{fig:JPYUSD} 
\end{center} 
\end{figure}

\begin{figure}
\begin{center}
\includegraphics[scale=0.7]{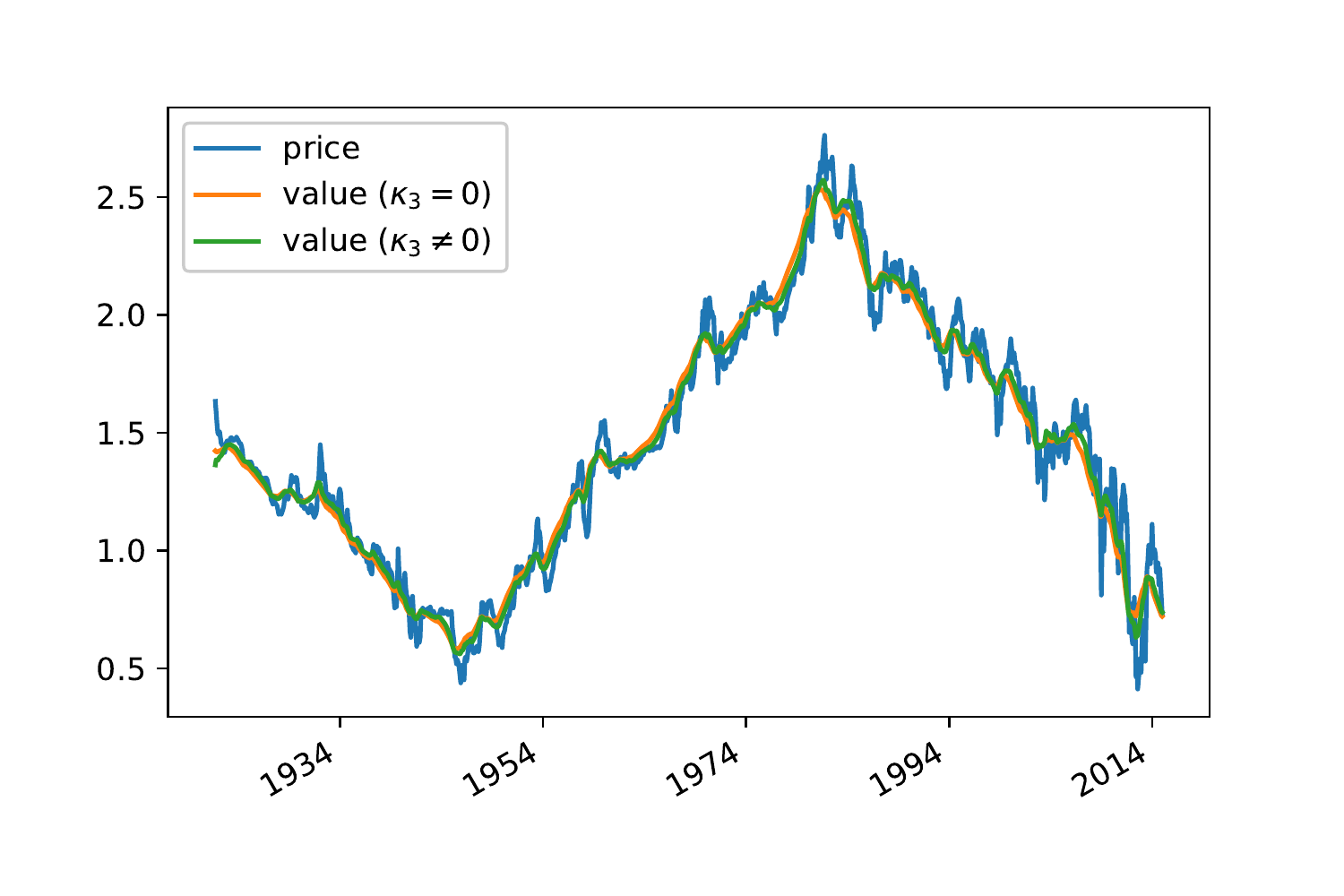} \caption{{\footnotesize Logarithm of yields on US sovereign bonds together with the smoothed value in model with linear demand of fundamentalists calculated by Kalman Smoother for parameters in Table \ref{tab:estimation_EM} and for model with non-linear demand of fundamentalists calculated by Unscented Kalman Smoother for parameters in Table \ref{tab:estimation_direct}. }}\label{fig:USBND} 
\end{center} 
\end{figure}

\newpage
\section{EM algorithm}\label{app:EMalgorithm}
\subsection{A description of the algorithm} 

The goal of EM algorithm is to find the set of parameters $\theta = \{ \kappa, \beta, \sigma_N, \sigma_V, g \}$ for which the model (\ref{eq:StateSpaceModel}) achieves the maximum likelihood. Since fundamental value $\tilde{v}_t$ is a hidden variable, it is hard to compute the likelihood function of system (\ref{eq:StateSpaceModel}) and it is hard to find its maximum directly, in a general case. On the other hand, we can rewrite the likelihood of the system as 
\begin{equation} 
L(\theta) = P\left( p_{1:T}| \theta \right) = \int P\left( p_{1:T}, \tilde{v}_{1:T} | \theta \right) d \tilde{v}_{1:T}. 
\end{equation} 
Then we can observe that the log-likelihood can be bounded from below by 
\begin{equation}\label{eq:inequality} 
\begin{split}
\LL{\theta} &= \ln \left( \int P\left( p_{1:T}, \tilde{v}_{1:T} | \theta \right) d \tilde{v}_{1:T} \right) \\ &= \ln \left( \int Q
\left( \tilde{v}_{1:T} \right) \frac{ P\left( p_{1:T},\tilde{v}_{1:T} | \theta \right) }{Q \left( \tilde{v}_{1:T} \right)} d \tilde{v}_{1:T} \right) \\ &\geq \int Q \left(
\tilde{v}_{1:T} \right) \ln \left( \frac{ P\left( p_{1:T}, \tilde{v}_{1:T} | \theta \right) }{Q \left( \tilde{v}_{1:T} \right)} \right) d \tilde{v}_{1:T} \eqdef \mathcal{F} \left( Q,
\theta\right), 
\end{split} 
\end{equation} 
where $Q$ is any distribution on $\tilde{v}_{1:T}$. Defining the energy of a global configuration $\left( p_{1:T}, \tilde{v}_{1:T} \right) $ to be $- \ln \left( p_{1:T}, \tilde{v}_{1:T} | \theta \right) $, the above quantity denoted by $\mathcal{F}$ can be rewritten as expected energy plus entropy 
\begin{equation}\label{eq:freeenergydef} 
\mathcal{F} \left( Q, \theta\right) = \int Q \left( \tilde{v}_{1:T} \right) \ln \left( P\left( p_{1:T}, \tilde{v}_{1:T} | \theta \right) \right) d \tilde{v}_{1:T} - \int Q \left( \tilde{v}_{1:T} \right) \ln \left( Q \left( \tilde{v}_{1:T} \right) \right) d \tilde{v}_{1:T} 
\end{equation} 
and it is called the free energy in statistical physics.

The idea of the EM algorithm is to maximize free energy $\mathcal{F}$ instead of maximizing likelihood $\mathcal{L}$. Algorithm reads 
\begin{enumerate} 
\item Choose $Q_0$ and $\theta_0$. 
\item For $n = \{0,1, 2, \ldots, N-1 \}$ repeat two-step procedure: 
\begin{itemize} 
\item \textbf{E-step}: $Q_{n+1} = \argmax_{Q} \mathcal{F} \left( Q, \theta_{n} \right)$ 
\item \textbf{M-step}: $\theta_{n+1} = \argmax_{\theta} \mathcal{F}
 \left( Q_{n+1}, \theta \right)$ 
\end{itemize} 
\end{enumerate}

Simple reasoning shows that the $\theta_n$ obtained by the above algorithm converges to the local maximum of log-likelihood $\LL{\theta}$. Lets rewrite the free energy as \begin{equation}\label{eq:KLdist} 
\begin{split} \mathcal{F} \left( Q, \theta\right) 
&= \int Q \left( \tilde{v}_{1:T} \right) \ln \left( \frac{ P\left( \tilde{v}_{1:T} | p_{1:T}, \theta \right) }{Q \left( \tilde{v}_{1:T} \right)} \right) d \tilde{v}_{1:T}\\ 
&+ \int Q \left( \tilde{v}_{1:T} \right) \ln \left( P\left( p_{1:T} | \theta \right) \right) d \tilde{v}_{1:T}\\ 
&= - D_{KL}(Q \left( \tilde{v}_{1:T} \right) || P\left( \tilde{v}_{1:T} | p_{1:T}, \theta \right) ) + \LL{ \theta }, 
\end{split} 
\end{equation} 
where $D_{KL}(Q||P)$ denotes Kullback–Leibler divergence from distribution $P$ to $Q$. From (\ref{eq:KLdist}) we immediately see that for $Q \left( \tilde{v}_{1:T} \right) = P\left( \tilde{v}_{1:T} | p_{1:T}, \theta \right)$ we obtain equality in (\ref{eq:inequality}). It means that in E-step we get $Q_{n+1} = P\left( \tilde{v}_{1:T} | p_{1:T}, \theta_n \right)$ and $\mathcal{F} \left( Q_{n+1}, \theta_{n} \right) = \LL{\theta_n}$. Then in M-step we search for $\theta_{n+1}$ such that $\mathcal{F} \left( Q_{n+1}, \theta_{n+1} \right) \geq \mathcal{F} \left( Q_{n+1}, \theta_{n} \right)$ and we can write it all together \begin{equation}\label{eq:convergence} \LL{\theta_{n+1}} \geq \mathcal{F} \left( Q_{n+1}, \theta_{n+1} \right) \geq \mathcal{F} \left( Q_{n+1}, \theta_{n} \right) = \LL{\theta_n}. \end{equation} we see that the $\theta_n$ converges to the local maximum of log-likelihood function.

The last equality in (\ref{eq:convergence}) holds only if $Q_{n+1}$ is exactly the distribution of $\tilde{v}_t$ conditioned on observed prices and parameters. We compute $P\left( \tilde{v}_{1:t} | p_{1:t}, \theta_n \right)$ by means of Bayesian filtering. However, only in the case of linear models we are able to compute it exactly and for non-linear models one has to use as $Q_{n+1}$ an approximation of $P\left( \tilde{v}_{1:T} | p_{1:T}, \theta_n \right)$, which will have non-zero Kullback–Leibler distance from the true distribution. In the cases when the distance is small we still can hope that EM algorithm will converge to the parameters maximizing likelihood. However, in general situation we cannot provide any guarantee and very often estimation with EM algorithm for models non-linear in fundamental value will provide results far away from the ones maximizing likelihood.

Going back to our linear in fundamental value version of our model, we know that $Q_{n+1} = P\left( \tilde{v}_{1:T} | p_{1:T}, \theta_n \right)$. Noticing that the second component of free energy in (\ref{eq:freeenergydef}) does not depend on $\theta$, we can simplify EM algorithm as follows: 
\begin{enumerate} 
\item Choose $\theta_0$. \item For $n = \{0, 1,2, \ldots, N-1 \}$ repeat two-step procedure: 
\begin{itemize} 
\item \textbf{E-step}: compute $\mathcal{G}(\theta, \theta_n) := E_{P\left( \tilde{v}_{1:T} | p_{1:T} , \theta_n \right) } \left[ \ln \left( P\left( p_{1:T}, \tilde{v}_{1:T} | \theta \right) \right) \right]$ 
\item \textbf{M-step}: $\theta_{n+1} = \argmax_{\theta} \mathcal{G}(\theta, \theta_n)$ 
\end{itemize} 
\end{enumerate}

\subsection{Computing E-step} 

In the E-step we compute the expectation of joint log-probability of $(p_{1:T}, \tilde{v}_{1:T})$ conditioned on $\theta$ with respect to probability of $\tilde{v}_{1:T}$ conditioned on $p_{1:T}$ and $\theta_n$. It is given by formula: 
\begin{equation}\label{eq:Gformula} 
\begin{split} \mathcal{G}(\theta, \theta_n) &= - \frac{1}{2\sigma_N^2} \sum_{t=1}^{T}E_n \left( p_t - p_{t-1} - \kappa \left( \tilde{v}_t - p_{t-1} \right) - \beta u_t \right)^2 - T \ln \sigma_N \\ 
&- \frac{1}{2\sigma_V^2}\sum_{t=1}^{T-1} E_n\left(\tilde{v}_{t+1} - \tilde{v}_{t} -g\right)^2 - (T-1) \ln \sigma_V - T \ln \left(2 \pi\right) \\ 
&- \frac{1}{2\sigma_0^2}\left( E_n\tilde{v}_1^2 + \tilde{v}_0^2 - 2\tilde{v}_0E_n\tilde{v}_1 + g^2 -2g\left( E_n\tilde{v}_{t+1} -\tilde{v}_{0} \right)\right) - \ln \sigma_0, 
\end{split} 
\end{equation} 
where all the expectations are taken with respect to distribution $P \left( \tilde{v}_{1:T} | p_{1:T} , \theta_n \right)$. Note also that $\tilde{v}_0$ is a parameter, not hidden variable. To calculate the above expression it is sufficient to know distributions $P(\tilde{v}_t | p_{1:T}, \theta_n)$ and $P(\tilde{v}_t, \tilde{v}_{t+1} | p_{1:T}, \theta_n)$ for $t = 1, \ldots, T-1$. The process of inferring the distribution of hidden variables given the parameters and the complete time series of observable variables of the system is called Bayesian smoothing. To proceed the smoothing one needs to proceed firstly Bayesian filtering $P(\tilde{v}_{t} | p_{1:t}, \theta_n)$ and predicting $P(\tilde{v}_{t+1} | p_{1:t}, \theta_n)$. Using Markov property of the process ($P(\tilde{v}_t | \tilde{v}_{t+1} , p_{1:t}) = P(\tilde{v}_t | \tilde{v}_{t+1} , p_{1:T})$), smoothing reads 
\begin{equation} 
\begin{split}
P(\tilde{v}_t, \tilde{v}_{t+1} | p_{1:T}, \theta_n) &= P(\tilde{v}_t | \tilde{v}_{t+1} , p_{1:t}, \theta_n) P(\tilde{v}_{t+1} | p_{1:T}, \theta_n) \\ 
&= P(\tilde{v}_{t} | p_{1:t}, \theta_n) \frac{P(\tilde{v}_{t+1} | \tilde{v}_{t} , \theta_n)}{P(\tilde{v}_{t+1} | p_{1:t}, \theta_n) } P(\tilde{v}_{t+1} | p_{1:T}, \theta_n) \end{split} 
\end{equation} 
and 
\begin{equation} 
\begin{split} 
P(\tilde{v}_t | p_{1:T}, \theta_n) &= \int P(\tilde{v}_t, \tilde{v}_{t+1} | p_{1:T}, \theta_n) d \tilde{v}_{t+1}\\ 
&= P(\tilde{v}_{t} | p_{1:t}, \theta_n) \int \frac{P(\tilde{v}_{t+1} | \tilde{v}_{t} , \theta_n) }{P\left(\tilde{v}_{t+1} | p_{1:t}, \theta_n\right) } P(\tilde{v}_{t+1} | p_{1:T}, \theta_n) d \tilde{v}_{t+1}. 
\end{split} 
\end{equation} 

Apart from filtering and predicting components on the right sides of both equations, we have the dynamics of value process $P(\tilde{v}_{t+1} | \tilde{v}_{t} , \theta_n) = \mathcal{N}(\tilde{v}_{t}, \sigma_V)$ and the smoothing one time step ahead $P(\tilde{v}_{t+1} | p_{1:T}, \theta_n)$. Therefore, we compute the smoothing distributions in backward recursions which we initialize by setting $P(\tilde{v}_{T} | p_{1:T}, \theta_n)$ equal to the filtered distribution.

To sum up, each iteration $n$ of E-step contains four parts: \begin{enumerate} \item Filter out distributions $P\left( \tilde{v}_t | p_{1:t},
 \theta_n \right)$ for $t = 1, \ldots, T$, \item Smooth out distributions $P\left( \tilde{v}_t | p_{1:T},
 \theta_n \right)$ for $t = T, \ldots, 1$, \item Smooth out distributions $P\left( \tilde{v}_t, \tilde{v}_{t+1} |
 p_{1:T}, \theta_n \right)$ for $t = T-1, \ldots, 1$, \item Compute $\mathcal{G}\left(\theta, \theta_n \right)$. \end{enumerate}

\subsection{Kalman filtering and smoothing}\label{sec:linearABM}

In the case of linear dynamic system the computations in E-step of EM algorithm are called Kalman filter and Kalman smoother. All the distributions obtained in this procedure (filtering, predicting and smoothing) are Gaussian: 

\begin{equation} 
\begin{split} 
P\left(\tilde{v}_{t} | p_{1:t}, \theta_n\right) &= \mathcal{N}\left( \tilde{v}_{t}^{t}, V_{t}^{t} \right), \\ 
P\left(\tilde{v}_{t+1} | p_{1:t}, \theta_n\right) &= \mathcal{N}\left( \tilde{v}_{t+1}^{t}, V_{t+1}^{t} \right), \\ 
P\left(\tilde{v}_{t} | p_{1:T}, \theta_n\right) &= \mathcal{N}\left( \tilde{v}_{t}^{T}, V_{t}^{T} \right). 
\end{split} 
\end{equation} 

The predicted and filtered fundamental values: $\tilde{v}_{t+1}^{t}$, $\tilde{v}_{t+1}^{t+1}$, and their variances: $V_{t+1}^{t}$, $V_{t+1}^{t+1}$ at $(n+1)$-th iteration of E-step are given by Kalman filter forward recursions\footnote{For esthetic reasons we write shortly $\tilde{v}_{t+1}^{t}$ instead of $(\tilde{v}_{t+1}^{t})^{(n)}$. The same holds true for $\tilde{v}_{t+1}^{t+1}$, $V_{t+1}^{t}$, $V_{t+1}^{t+1}$ and $K_{t+1}$.} 
\begin{equation}\label{eq:KalmanRecursions} 
\begin{split} 
\tilde{v}_{t+1}^{t} &= \tilde{v}_{t}^{t} + g^{(n)},\\ V_{t+1}^{t} &= \tilde{V}_{t}^{t} + \left(\sigma_V^{(n)}\right)^2,\\ 
K_{t+1} &= \frac{\kappa^{(n)} V_{t+1}^{t}}{\left(\kappa^{(n)}\right)^2 V_{t+1}^{t} + \left(\sigma_N^{(n)}\right)^2},\\ 
\tilde{v}_{t+1}^{t+1} &= \tilde{v}_{t+1}^{t} + K_{t+1} \left( p_t - p_{t-1} - \kappa^{(n)} \left( \tilde{v}_{t+1}^{t} - p_{t-1} \right) - \beta^{(n)} u_t \right),\\ 
V_{t+1}^{t+1} &= \tilde{V}_{t+1}^{t} - \kappa^{(n)} K_{t+1} \tilde{V}_{t+1}^{t}, 
\end{split} 
\end{equation} 
where $\tilde{v}_1^{0} = \tilde{v}_0^{(n)}$ and $V_{1}^{0} = \left(\sigma_0^{(n)}\right)^2$. The smoothed fundamental values $\tilde{v}_{t+1}^{T}$ and their variances $V_{t}^{T}$ at $(n+1)$-th iteration of E-step are given by Kalman smoother backward recursions\footnote{For esthetic reasons we write shortly
$\tilde{v}_{t+1}^{T}$ instead of $(\tilde{v}_{t-1}^{T})^{(n)}$. The same holds true for $V_{t-1}^T$, $C_{t-1,t}^T$ and $J_{t-1}$.} 
\begin{equation}\label{eq:SmoothingRecursion} 
\begin{split} J_{t-1} &= V_{t-1}^{t-1}/V_{t}^{t-1},\\ 
\tilde{v}_{t-1}^T &= \tilde{v}_{t-1}^{t-1} + J_{t-1} \left( \tilde{v}_{t}^T - \tilde{v}_{t-1}^{t-1} \right),\\ 
V_{t-1}^T &= V_{t-1}^{t-1} + (J_{t-1})^2 \left( V_{t}^T - V_{t-1}^{t} \right), 
\end{split} 
\end{equation} 
with $J_T = 0$. The recursions in Kalman smoothing procedure are often called Rauch–Tung–Striebel recursions. In order to compute E-step we also need the covariance of $\tilde{v}_{t-1}$ and $\tilde{v}_{t}$ conditioned on $p_{1:T}$ and $\theta^{(n)}$ and we denote this covariance by $C_{t-1,t}^T$: 
\begin{equation}\label{eq:covRecursion} 
C_{t-2,t-1}^T = V_{t-1}^{t-1}J_{t-2} + J_{t-1}\left(C_{t-1,t}^T -V_{t-1}^{t-1} \right)J_{t-2}, 
\end{equation} 
which is initialized as $C_{T-1,T}^T = \left( 1 - \kappa K_T \right) V_{T-1}^{T-1}$. The derivation of Kalman filter and smoothing recursions can be found for example in \cite{Sarkka2013}. The components in (\ref{eq:Gformula}) are given by 
\begin{equation} 
E _n \tilde{v}_{k} = \tilde{v}_k^T, \ \ \ \ \ E _n \tilde{v}_{k}^2 = V_{k}^T + (\tilde{v}_k^T)^2 \ \ \mbox{ and } \ \ E_n \tilde{v}_{k+1}\tilde{v}_{k} = C_{k-1,k}^T + \tilde{v}_k^T\tilde{v}_{k+1}^T. 
\end{equation}

 \subsection{Computing M-step}\label{sec:Mstep}
A set of parameters maximising $\mathcal{G}(\theta, \theta_n)$ can be found by solving set of equations $\frac{\partial}{\partial \theta}\mathcal{G}(\theta, \theta_n) = 0$. In this way we obtain: 

\begin{equation}
 \begin{bmatrix} \kappa^{(n+1)} \\ \beta^{(n+1)} \end{bmatrix} =
 A^{-1} b, 
\end{equation} 
where 
\begin{equation} A=
 \begin{bmatrix}
 \sum_{t=1}^T E_n\left[ (\tilde{v}_t - p_{t-1})^2 \right] &
 \sum_{t=1}^T E_n\left[ \tilde{v}_t - p_{t-1} \right] u_t
 \\ \sum_{t=1}^T E_n\left[ \tilde{v}_t - p_{t-1} \right] u_t &
 \sum_{t=1}^T u_t^2
 \end{bmatrix}
\end{equation} 
and 
\begin{equation} 
b=
 \begin{bmatrix}
 \sum_{t=1}^T E_n\left[ \tilde{v}_t - p_{t-1} \right] \left( p_t -p_{t-1} \right) \\
 \sum_{t=1}^T u_t \left( p_t - p_{t-1} \right)
\end{bmatrix}, 
\end{equation} 
where all conditional expectations are obtained via Kalman smoothing: $\tilde{v}_{t}^T$ and $V_{t}^T$ in (\ref{eq:SmoothingRecursion}). Moreover, 
\begin{equation} 
\sigma_N^{(n+1)} = \left( \frac{1}{T} \sum_{t=1}^{T}E_n \left( p_t - p_{t-1} - \kappa^{(n+1)} \left( \tilde{v}_t - p_{t-1} \right) - \beta^{(n+1)} u_t \right)^2 \right)^{1/2}, 
\end{equation} \begin{equation} 
g^{(n+1)} =\frac{1}{T-1} \sum_{k=1}^{T-1} \left( E_n\tilde{v}_{k+1} - E _n \tilde{v}_{k} \right),
\end{equation} 
\begin{equation} \sigma_{V}^{(n+1)} = \left( \frac{1}{T-1} \sum_{k=1}^{T-1} \left( E_n\tilde{v}_{k}^2 + E _n \tilde{v}_{k+1}^2 - 2E_n \tilde{v}_{k+1}\tilde{v}_{k} - \left( g^{(n+1)} \right)^2 \right) \right)^{1/2}, 
\end{equation} 
\begin{equation} 
\tilde{v}_0^{(n+1)} = E_n \tilde{v}_{1} - g^{(n+1)} 
\end{equation}
and 
\begin{equation} 
\sigma_{0}^{(n+1)} = \sqrt{E_n \tilde{v}_{1}^2 } - \left( \tilde{v}_0^{(n+1)} \right)^2. 
\end{equation}
The parameters obtained via EM algorithm are maximizing the predictive likelihood, which we will monitor:
\begin{equation}\label{eq:PLL_KF} 
\LL{\theta} = - \frac{T}{2} \ln 2 \pi -\frac{1}{2} \sum_{t=1}^{T}\left[ \ln\left( \kappa^2 V_t^{t-1} + \sigma_N^2 \right)
 -\frac{\left( p_{t} - p_{t-1} - \mu_t \right)^2}{2\left( \kappa^2 V_t^{t-1} + \sigma_N^2 \right)} \right], 
\end{equation} 
where $\mu_t = \kappa \left( \tilde{v}_{t}^{t-1} - p_{t-1} \right) + \beta u_{t} $. 

\section{Estimation of model \ref{eq:ModelNL} }\label{app:estNL}

We write the so-called predictive log-likelihood as 

\begin{equation}\label{eq:PLL_UKF} 
\begin{split} 
\LL{\theta} 
&= \ln P\left( p_{1:T}| \theta \right) = \sum_{t=1}^{T} \ln P\left( p_{t}| p_{1:t-1}, \theta \right)\\ 
&\approx - \frac{T}{2} \ln 2 \pi - \frac{1}{2} \sum_{t=1}^{T}\left[\ln S_t + \frac{\left(p_{t} - \hat{p}_{t} \right)^2}{S_t } \right] = \LLh{\theta},
\end{split} 
\end{equation} 
where we introduce the notation $p_{1:t} = \{p_1, p_2, \ldots, p_t \}$, $\hat{p}_{t}$ is the next-month price prediction with UKF and $S_t$ is the variance of that prediction given all price history and set of parameters $\theta$ (see formulas (\ref{eq:UKF_predictions})). Applying a standard optimizer\footnote{We use \texttt{minimize} from \texttt{scipy.optimize} package in Python using the Broyden-Fletcher-Goldfarb-Shanno (BFGS) method.} we search for the set of parameters that maximizes predictive log-likelihood, computed by means of UKF.

\subsection{Unscented Kalman Filter}\label{app:UKF}

 
Since model (\ref{eq:NonlinearModel}) has linear dynamics of fundamental value, the prediction step is the same as in Section \ref{sec:linearABM}. Therefore, the recursions for $\tilde{v}_{t+1}^t$, $V_{t+1}^t$ are the same as in (\ref{eq:KalmanRecursions}). To filter out the fundamental value we form sigma points 
\begin{equation} 
\begin{split} \mathcal{V}_{t+1}^{(0)} &= \tilde{v}_{t+1}^{t},\\ 
\mathcal{V}_{t+1}^{(1)} &= \tilde{v}_{t+1}^{t} + \sqrt{1 + \lambda}\sqrt{V_{t+1}^{t}},\\
 \mathcal{V}_{t+1}^{(2)} &= \tilde{v}_{t+1}^{t} - \sqrt{1 + \lambda}\sqrt{V_{t+1}^{t}}, 
\end{split} 
\end{equation} 
where $\lambda = a^2\left( 1 + k\right) -1$, $a$ and $k$ are parameters of Unscented Transform (we choose $a = 1.0$ and $k = 2.0$). We compute the expected price for each sigma point 
\begin{equation} 
\mathcal{P}_{t+1}^{(i)} = p_{t} + f \left( \mathcal{V}_{t+1}^{(i)} - p_t \right) + \beta u_{t+1}, \mbox{ for } i = 0,1,2. 
\end{equation}
The predicted price $\hat{p}_{t+1}$, the variance of price prediction $S_{t+1}$ , and the covariance of the fundamental value and the price $C_{t+1}$ are computed with the following forumlae: 
\begin{equation}\label{eq:UKF_predictions}
 \begin{split} 
\hat{p}_{t+1} &= \sum_{i=0}^2 W_i^{(m)} \mathcal{P}_{t+1}^{(i)}\\ 
S_{t+1} &= \sum_{i=0}^2 W_i^{(c)} \left( \mathcal{P}_{t+1}^{(i)} - \hat{p}_{t+1} \right)^2 + \sigma_N^2\\ 
C_{t+1} &= \sum_{i=0}^2 W_i^{(c)} \left( \mathcal{P}_{t+1}^{(i)} - \hat{p}_{t+1} \right) \left( \mathcal{V}_{t+1}^{(i)} - \tilde{v}_{t+1}^{t} \right),\\ 
\end{split} 
\end{equation} 
where weights of the Unscented Transform are given by 
\begin{equation}\label{eq:weights} 
\begin{split} 
W^{(m)}_0 &= \lambda/ \left( \lambda + 1 \right)\\ 
W^{(c)}_0 &= \lambda/ \left( \lambda + 1 \right) + (1 - a^2 + b)\\ 
W^{(m)}_i &= 1/ \left[ 2\left( \lambda + 1 \right) \right], \ \ i = 1, 2\\ 
W^{(c)}_i &= 1/ \left[ 2\left( \lambda + 1 \right) \right], \ \ i = 1, 2\\ 
\end{split} 
\end{equation} 
where $b$ is the parameter of Unscented Transform (we choose $b = 0.0$). The filter gain $K_{t+1}$, the filtered mean and variance of fundamental value, $\tilde{v}_{t+1}^{t+1}$ and $V_{t+1}^{t+1}$ are \begin{equation} 
\begin{split} K_{t+1} &= S_{t+1}/C_{t+1},\\ 
\tilde{v}_{t+1}^{t+1} &= \tilde{v}_{t+1}^{t} + K_{t+1} \left( p_{t+1} - \hat{p}_{t+1} \right),\\ 
V_{t+1}^{t+1} &= V_{t+1}^{t} - (K_{t+1})^2 S_{t+1}. 
\end{split} 
\end{equation} 
Then we approximate the predicting distribution by Gaussian and consequently the smoothing distributions are approximately Gaussian too. As a consequence, recursions for $\tilde{v}_{t+1}^{T}$, $V_{t-1}^T$, $C_{t-1,t}^T$ and $J_{t-1}$ are the same as in (\ref{eq:SmoothingRecursion}) and (\ref{eq:covRecursion}).

\end{document}